\documentclass[%
reprint,
superscriptaddress,
a4paper,
showpacs,preprintnumbers,
amsmath,amssymb,
aps,
prstab,
floatfix,
]{revtex4-1}

\usepackage{graphicx}
\usepackage[caption=false]{subfig}
\usepackage{hyperref}
\hypersetup{colorlinks=true,citecolor=blue}

\newcommand{\pinmrad}[0]{$\pi$~nm~rad}

\begin{document}
\title{Optics studies of a Muon Accumulator Ring based on FFA cells}
\author{O.R. Blanco-Garc\'ia}
\email{oblancog@lnf.infn.it}
\affiliation{INFN-LNF, Via E.~Fermi 40, 00044 Frascati, Rome, Italy}
\author{A. Ciarma}
\affiliation{INFN-LNF, Via E.~Fermi 40, 00044 Frascati, Rome, Italy}

\date{\today}

\begin{abstract}
  The production of an intense, high energy and low emittance muon beam is interesting for a possible muon collider. The Low EMittance Muon Accelerator~(LEMMA) team at the Istituto Nazionale di Fisica Nucleare~(INFN), in Italy, is studying the production of a 22.5~GeV low emittance muon beam from a high energy positron beam at twice the muon energy impinging on a fixed thin target.\par
  This muon source has two main advantages: the muon beam emittance is small thanks to the kinematics of the $e^+e^-$ annihilation process of a low emittance positron beam, and thanks to the asymmetric collision muons are produced with a relativistic gamma factor of ~200, extending their lifetime to 0.46~ms. The disadvantage of this scheme is the low production efficiency, which is in the order of $10^{-7}$ muon pairs per impinging positron in a target of 1\% of a radiation length, therefore, requiring a high positron rate.\par
  The LEMMA scheme proposes to perform the muon accumulation from multiple ($10^2$ to $10^3$) positron bunches to increase the population of a single muon bunch that is recirculated through the target using two small accumulator rings, one per muon species. The three beams ($\mu^+$ and $\mu^-$ at 22.5~GeV and $e^+$ at twice the muon energy) share the same phase space at the target on every positron bunch interaction, producing new muons inside the preserved beam emittance.\par  
  We study the requirements and optics design of the accumulator to recirculate the muons over the target using a Fixed Field Alternating Gradient~(FFA) arc.\par
  As a result, we achieve a compact 230~m long accumulator with two Interaction Points, energy acceptance of $\pm5$\%, low twiss beta function at the target $\beta^*_\mu=20$~cm, and a drift space 2$L^*$ of 20~cm enough to accommodate 1\% of a radiation length $X_0$ for several material options. These optics parameters are obtained with magnets similar to those foreseen for new colliders like FCC or CLIC, and could be extended further with new magnet designs.\par
  Simulations of the muon beam accumulation and target interaction show that the muon beam population increases by a factor 100 in the first few hundred turns, but, emittance grows due to multiple scattering with the target leading to particle losses. The final muon population and emittance is limited by the dynamic aperture of the machine.\par
  Although the FFA and Interaction Region designs are promising, it also points to some limitations.
  First, $\epsilon_N=5$~$\pi$~$\mu$m (or $200\times25$~$\pi$~nm) is the minimum normalized emittance obtained with the current value of $\beta^*_\mu$, and the production of lower $\beta^*_\mu$ values will require quadrupole gradients above 500~T/m.
  Second, the achieved low $\beta^*_\mu$ is not enough to mitigate the effect of multiple scattering with the target over a thousand turns.
  Third, the current Interaction Region for a low $\beta^*_\mu$ already reduces the energy acceptance of the machine to only $\pm5$\%.
  Fourth, the FFA cells are designed to correct low values of chromaticity, the arcs will need to be adapted consequently in length or strength when trying to push down the $\beta^*_\mu$ and/or increase the energy acceptance.\par
  Further efforts could be directed to design a second order apochromatic Interaction Region to increase the energy acceptance from $\pm5$\% up to $\pm$10\%, which would allow to accept a larger muon beam energy spread and thus increase the production efficiency. In addition, studies on magnets with good field regions larger than $\pm$2~cm will allow to increase dispersion and reduce the strength of sextupole magnets used for chromatic correction leading to larger dynamic aperture. Furthermore, quadrupole gradients beyond 500~T/m able to work in an Interaction Region that includes a target would allow to produce a smaller $\beta^*_\mu$ leading to a smaller muon beam emittance and mitigating the effect of multiple scattering over more than a few hundred turns.\par
  The current muon accumulation results will serve as input for beam combination studies.\par
\end{abstract}
\maketitle

\section{Introduction}\label{s:intro}
One of the many challenges to aboard in accelerator physics is the feasibility of a muon collider~\cite{updateofeuropeanstrategy}. Crucial parameters for this machine are the fast production and acceleration of a high quality muon beam measured in terms of high population, large lifetime and small emittance.\par
The Muon Accelerator Program~(MAP)~\cite{MAP} has already performed the initial studies for a possible muon collider reaching high luminosity in the TeV scale where positive and negative muon beams are produced as secondary particles, from the collision of a proton beam on a target. Due to the kinematics of this process the muon beams are produced with a very large transverse~(4D) and longitudinal~(+2D) emittance that is cooled down in later stages. Progress has been achieved in a small scale 4D emittance cooling test by the Muon Ionization Cooling Experiment~(MICE) group~\cite{cooling}, while 6D-cooling remains yet untested.\par 
The LEMMA~(Low Emittance Muon Accelerator)~\cite{NIM,Antonelli:IPAC16-TUPMY001,Boscolo:IPAC17-WEOBA3,eplusringopt,MaricaIPAC2019} team is studying the production of muon beams at about 22.5~GeV, with a bunch population of $10^9$ particles and extremely low normalized emittance of $\epsilon_N=0.040~\mu$m from the collision of a low emittance positron beam and a fixed thin target.\par
We note that a target is considered thin if the thickness of the material traversed by the positron beam is only a few percent of radiation length~$X_0$. In the case LEMMA concept the target material is Beryllium~(Be) and the target thickness is 3~mm, equivalent to 0.88\% of $X_0$.\par
\begin{figure}[thb]
  \includegraphics[width=0.48\textwidth]{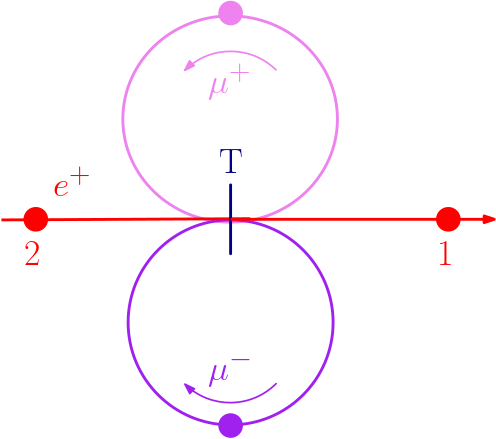}\caption{Schematic representation of the LEMMA accumulation scheme. Two muon bunches have been produced by the interaction of the first positron bunch with the target $T$. Muons are recirculated and synchronized with the arrival of a second positron bunch to increase the muon bunch population with minimal emittance growth. The cycle repeats over a thousand positron bunches.}\label{f:rings}
\end{figure}

The LEMMA scheme has two main advantages: no emittance cooling is required and muon lifetime is extended to 0.4~ms (more than 200 times the lifetime at rest or equivalently 120~km of run) before the muons decay. However, the combination of high bunch population and low emittance is challenging because the accumulation process requires the recirculation of the beam over the target more than a thousand times~(Fig.~\ref{f:rings}), leading to energy loss and emittance growth from multiple scattering with the target.\par
\begin{figure}[thb]
  \includegraphics[width=0.48\textwidth]{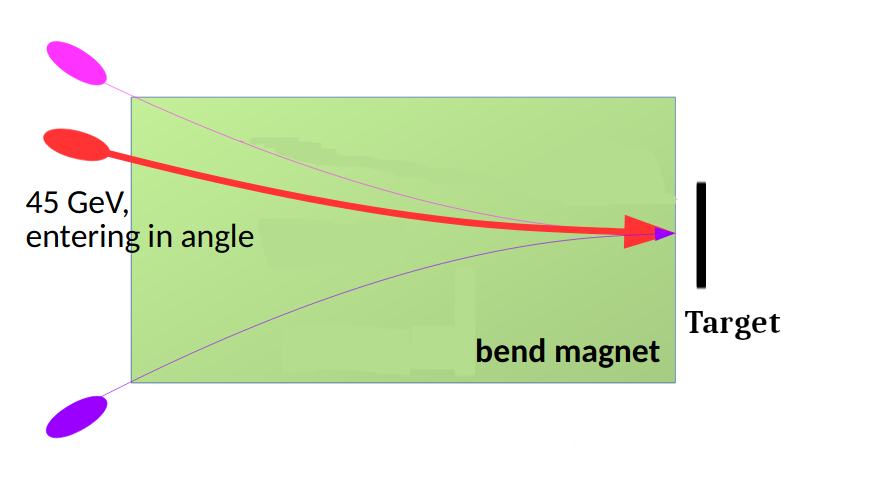}\\
  \includegraphics[width=0.48\textwidth]{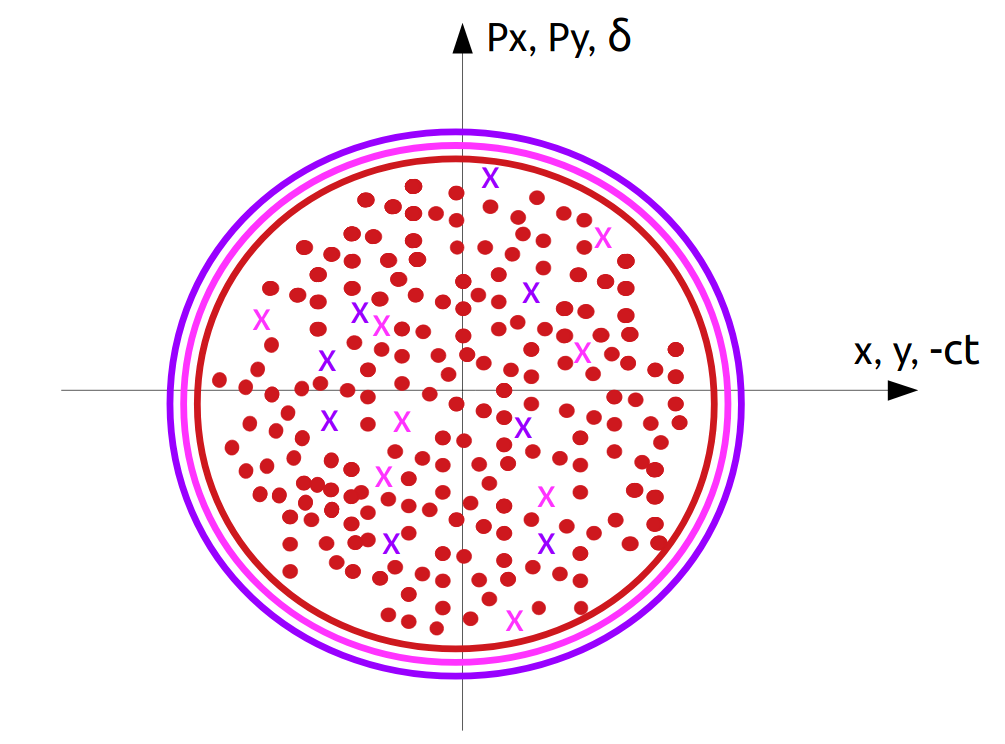}
  \caption{(TOP)  The three beams $e^+$ in red, $\mu^+$ in pink and $\mu^-$ in violet superimpose one another  at the target. (BOTTOM) Phase space representation at the thin target location. Beams are distributed over the same phase space, represented in colored circles. New muons are created inside the same phase space on every positron bunch passage and muon bunches recirculation.}\label{f:spacecharge}
\end{figure}

In the LEMMA scheme, three beams ($\mu^+$ and $\mu^-$ at 22.5~GeV and $e^+$ at twice the energy) occupy the same phase space at the target location in order to create new muons without emittance growth, see Fig.~\ref{f:spacecharge}.

In this article we concentrate in the possibility to create an accumulator design fulfilling the LEMMA specifications addressed theoretically in~\cite{muacc}. 
We list those requirements here:
\begin{itemize}
\item The accumulator ring should have a very short length in order to allow a large number of accumulation cycles.
\item The energy acceptance should be close to $\pm$20\% due to the muon pair production kinematic for 45~GeV positron beam. This is the minimum energy acceptance to get about $0.7\times10^{-7}$ muon pairs per impinging positron in a Beryllium target 3~mm thick.\par
\item The twiss $\beta^*_\mu$ at the beam--target Interaction Point~(IP) should be at most 1~cm over $\pm$20\% energy spread to create a low muon beam emittance with large divergence, comparable to the contribution to divergence from multiple scattering with the target over a thousand turns, mitigating the emittance growth.
\item The momentum compaction factor should be small over $\pm$20\% to preserve the bunch length in the order of millimeters given by the positron beam.
\end{itemize}
In addition, we would like to remark two points:
\begin{itemize}
\item We include in the requirements a radiofrequency~(RF) cavity to recover the beam energy loss in the target that becomes significative over a thousand turns, equivalent to traverse several radiation lengths of material.
\item At the end of the accumulation, the beam should be extracted. Therefore, we need to consider and extraction region.
\end{itemize}
A strong focusing accumulator optics and target material study~\cite{PhysRevAccelBeams.23.051001} have shown promising results on fulfilling these requirements.\par
In this article, we would like to extend the possibilities to the study of an alternative accumulator ring based in Fixed Field Alternating Gradient~(FFA) Cells because several publications show the possibility to design a small machine length and large energy acceptance~\cite{machida2012,sheehy2016,lagrange2016, garland2015}. Section~\ref{s:considerations} states the considerations that we have taken into account at the beginning of the accumulator study. Section~\ref{s:accsect} shows the accumulator sections. Section~\ref{s:ffacell} shows the FFA cell for high energy acceptance. Section~\ref{s:insert} shows the modification of the FFA cell to accommodate the insertions. Section~\ref{s:rfandkicker} shows the insertion optics for an RF and kicker. Section~\ref{s:beamseparation} shows the studies of beam separation. Section~\ref{ss:ir} shows the Interaction Region design. Section~\ref{s:accumulator} shows the accumulator. Section~\ref{s:beamsizeandtune} shows the beamsize and tune of the machine. Section~\ref{s:accumulation} shows the muon accumulation results for different beam energies and target materials. Section~\ref{s:further} shows further design improvements that have been foreseen during the design stage. Finally, in Section~\ref{s:conclusion} we conclude on the accumulation studies.

\section{Considerations before the design}\label{s:considerations}
There are some arbitrary assumptions we made in order to select among many possibilities the most promising optics designs with the current magnet technology before any accumulation study. We consider a good field region of an arc magnet in the order of $\pm$1 to $\pm$2~cm with a maximum peak magnetic field of 20~T. This should be a perfectly safe value for a good field region of any magnet because for a beam with $\pm$20\% energy spread and a horizontal dispersion $\eta_x$=0.1~m, the displacement of the beam to first order approximation corresponds to 2~cm~($0.1~\text{m}\times0.2$). Therefore, we have decided to pursue small dispersion along the arcs. \par
Sextupoles will need to be stronger in order to compensate the small horizontal dispersion for a given value of chromaticity per cell. We prefer to keep the dispersion small and assume that we have no limits on the achievable sextupole strength. It also implies that second order corrections are very sensitive and could limit the dynamic aperture of our design, but it did not seem to be a problem due to the small muon beam emittance.\par
As we will see, dynamic aperture puts a limit on the accumulation achievements due to emittance growth from multiple scattering with the target.\par
With respect to the optics design and as a starting point to understand the limitations of the FFA cell, we tried a second order model in MAD-X~\cite{MADX}. We recognize that it is not the best option for this design because FFA ring studies typically require a step-by-step integrator over well defined magnetic models that lead to better quantitative analysis of the particle trajectories with large difference in energy. However, we consider MAD-X and the MAD-X PTC~\cite{PTC} libraries  for the studies with $\pm$20\% of energy spread.\par
Keil in~\cite{emma} presents two possibilities on how to proceed:
\begin{itemize}
\item do a Taylor expansion of the magnetic field to second order involving dipole, quadrupole and sextupole components, or
\item displace the quad elements to feed down the dipole component and then add sextupoles.
\end{itemize}
We decided to start the design following the first option, i.e. do a Taylor expansion of the magnetic field that leads to superimposed independent dipole, quadrupole and sextupole components that in principle could be realized with a canted cosine theta type of magnet~\cite{cantedcosmagnet,blondel,caspi,goodzeit}.\par

\section{Accumulator sections}\label{s:accsect}
We divide the optics design in several minor sections in order to systematically approach the previously listed requirements. The initial subdivision includes:
\begin{itemize}
\item A high momentum acceptance arc cell for the arcs
\item A zero dispersion cell connecting the arcs with straight sections
\item An interaction region common to all three beams and the target
\item A radio frequency cavity region
\item An extraction region
\end{itemize}
In the following we explore different kinds of possible cells and subsections of the accumulator checking if they succeed in achieving the accumulator requirements.
\section{Large energy acceptance FFA cell}\label{s:ffacell}
The design shown in~\cite{bogomyagkov}, a new concept for synchrotron line sources at around 2 to 3~GeV, is composed by three canted-cosine magnets with dipole, quadrupole and sextupole components superimposed to achieve a compact cell optimized for $\pm40$\% momentum acceptance.\par
We have partially reproduced their results and continued to adjust the lattice to get the minimum possible circumference for a 22.5~GeV beam.\par
\begin{figure*}[hbt]
  \includegraphics[width=0.32\textwidth]{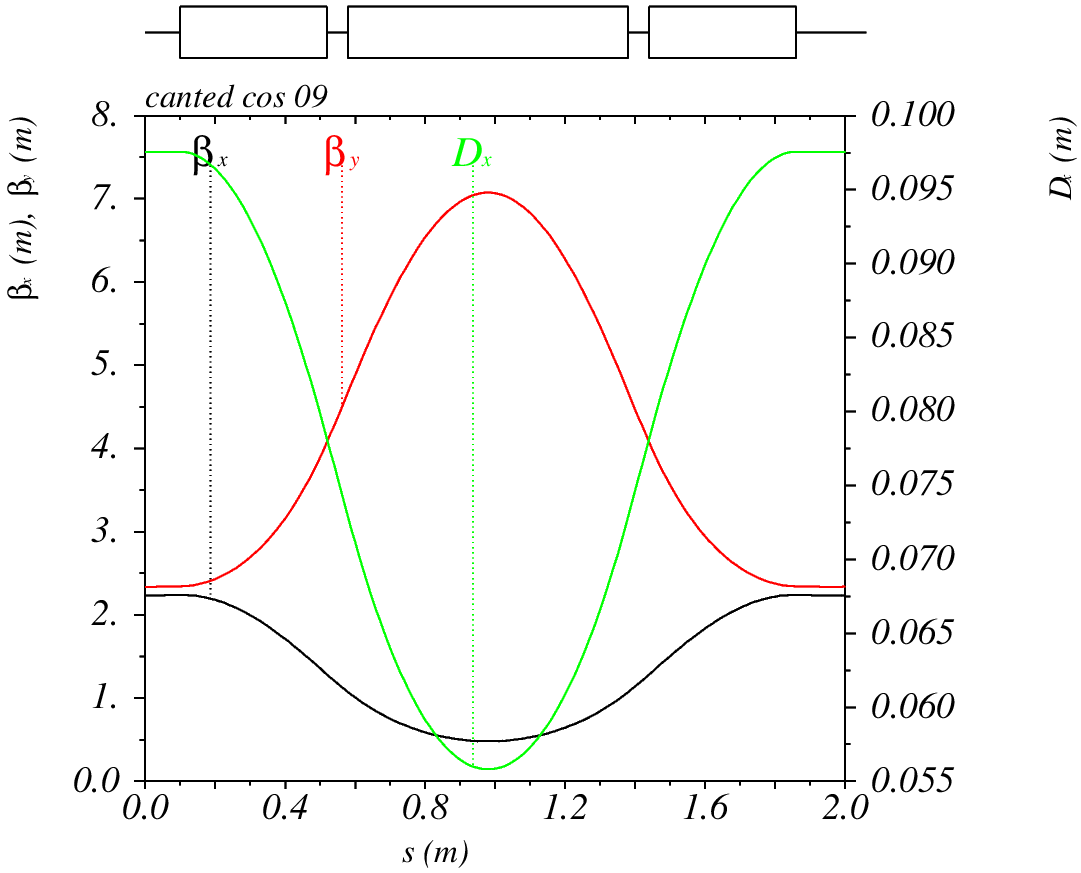}
  \includegraphics[width=0.32\textwidth]{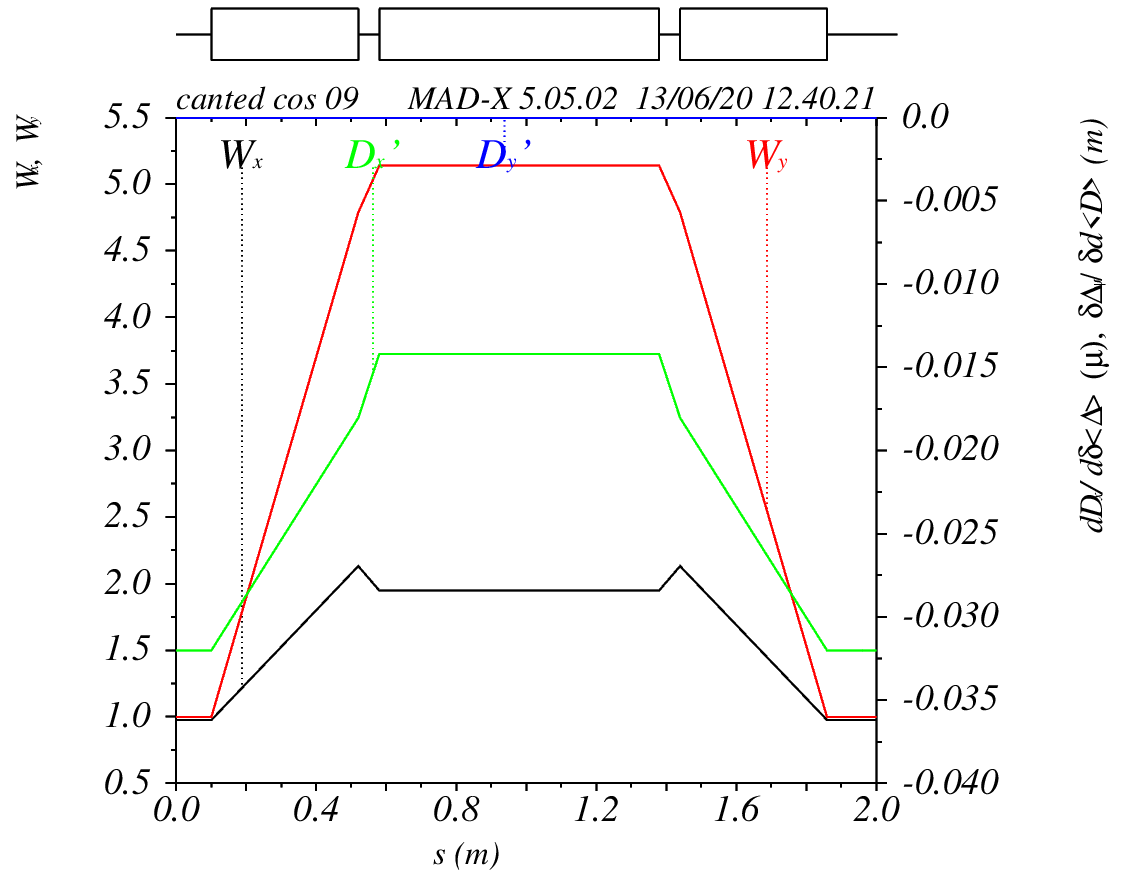}
  \includegraphics[width=0.32\textwidth]{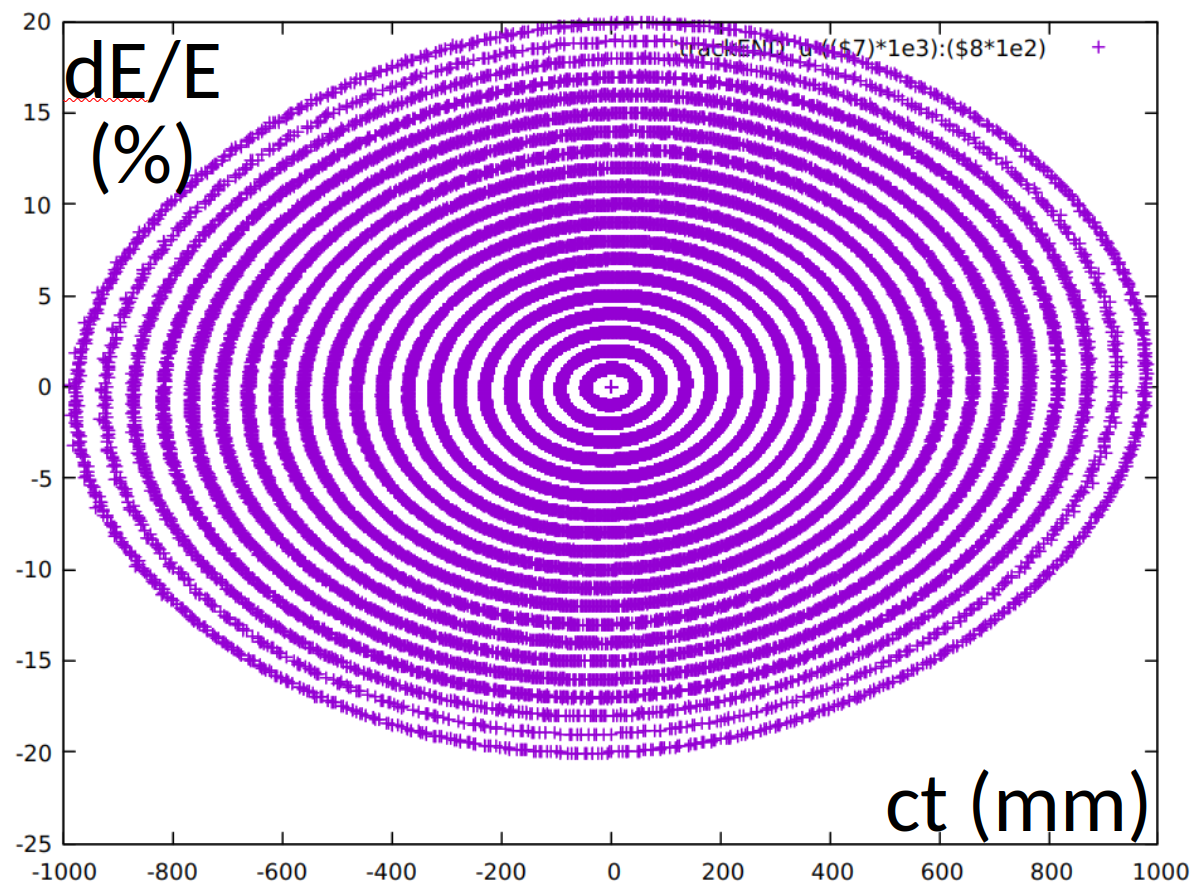}
  \caption{High energy acceptance FFA cell: (LEFT) Linear Optics. (CENTER) Second order optics functions $Wx$, $Wy$ and $DDX$ (or $DX'$ in the plot) as defined in MAD-X. (RIGHT) Longitudinal phase space resulting from particle tracking of a 98~m ring over one thousand turns starting with energy offset of $\pm$20\% in 1\% steps.}\label{f:cantedcos09}
\end{figure*}

The cell is shown in Fig.~\ref{f:cantedcos09}. It is composed by a Focusing-Defocusing-Focusing sequence where we have dipole B components superimposed~(BF-BD-BF). Chromaticity and dispersion were minimized together via multi-parameter match of phase advance in both planes varying the magnets lengths, dipole and quadrupole field components while checking the momentum acceptance from particle tracking in MAD-X PTC over $\pm$20\% energy offset. Chromaticity was corrected in a second step varying the sextupole components.\par
The separation between BF and BD magnets is 6~cm, while the distance between BFs in adjacent cells is 20~cm.
The magnets length, and the dipole, quadrupole and sextupole components are respectively:
\begin{itemize}
\item BF \;: 42~cm, 6.165~T, \;240.2~T/m, 2575~T/m$^2$
\item BD : 80~cm, 5.316~T, -182.7~T/m,-2931~T/m$^2$
\end{itemize}
The total length of the accumulator is 98~m consisting of 50 cells, 1.96~m long each. The maximum magnetic field at 2.5~cm moving radially outwards the reference orbit is just below 13~T, calculated as
\begin{equation}
  B = B_d + B_q x + \frac{1}{2}B_sx^2,
\end{equation}
where $B_d$ is the dipole field, $B_q$ is the quadrupole field, $B_s$ is the sextupole field, $B$ is the total sum and $x$ is the transverse horizontal displacement from the magnet center.\par
The natural chromaticity is $-14$ and $-17$ units in the horizontal and vertical planes respectively, which implies a very small chromaticity per cell of the order of -0.25 to -0.35. Sextupoles are tuned to correct the chromaticity on both planes.\par
The final momentum compaction factor is $5\times10^{-3}$, which is a quite large value for the needs of the muon accumulator ring. Using a cavity at 61~MHz and 400~MV, we get the required energy acceptance over more than $\pm20$\%, but, with a large bunch length of the order of one meter (see Fig.~\ref{f:cantedcos09} and Table~\ref{t:designs}).\par
We conclude from this cell design that the minimum length of an accumulator with $\pm20$\% energy acceptance is a little less that 100~m for magnets at about 13~T. The drawback is the large bunch length due to the momentum compaction factor to first order.\par
The next step is the reduction of the momentum compaction factor.\par

\subsection{FFA with reduced momentum compaction factor}
\begin{figure*}[hbt]
  \includegraphics[width=0.32\textwidth]{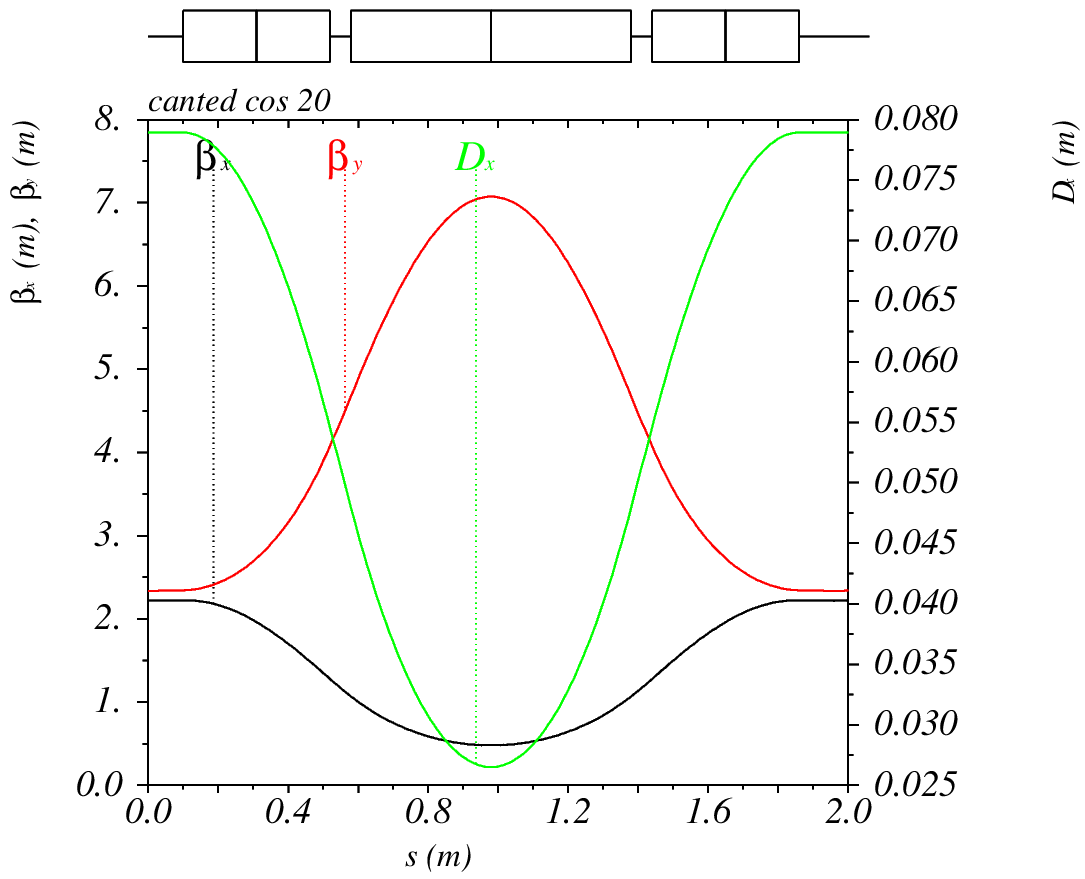}
  \includegraphics[width=0.32\textwidth]{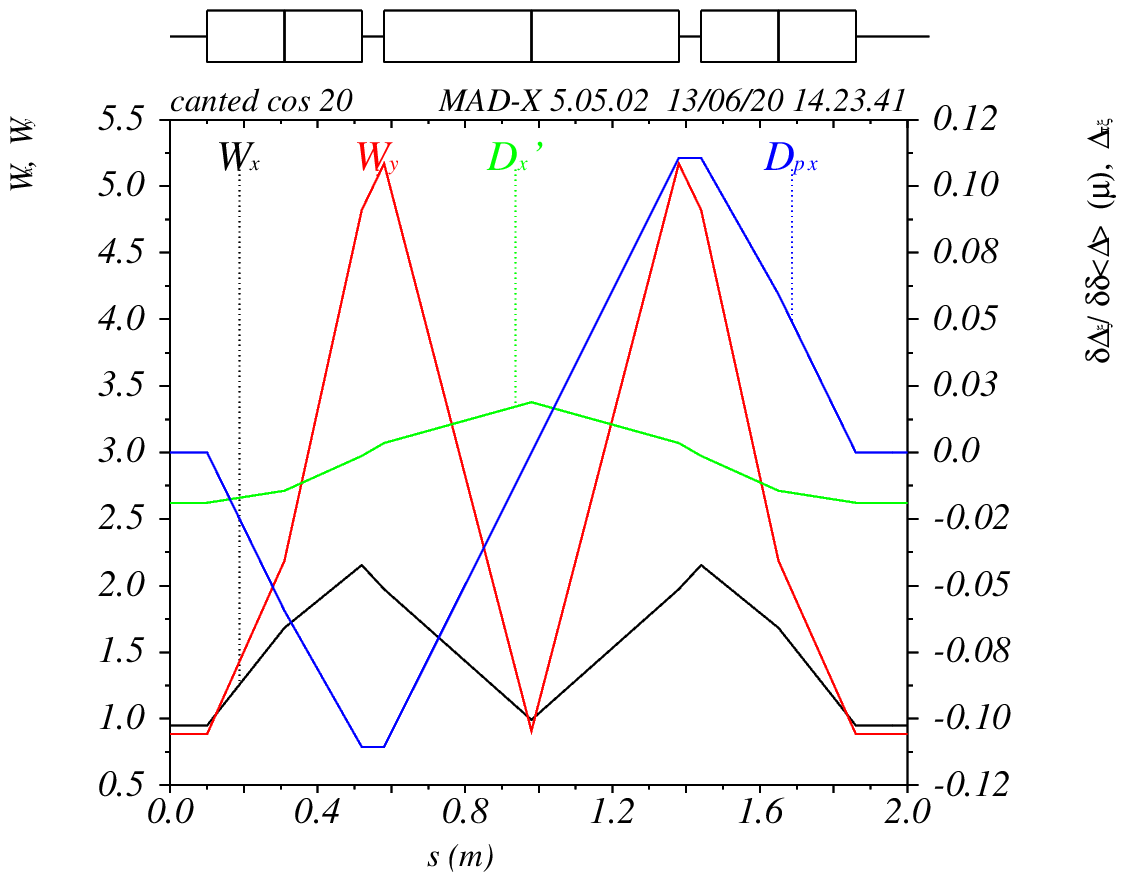}
  \includegraphics[width=0.32\textwidth]{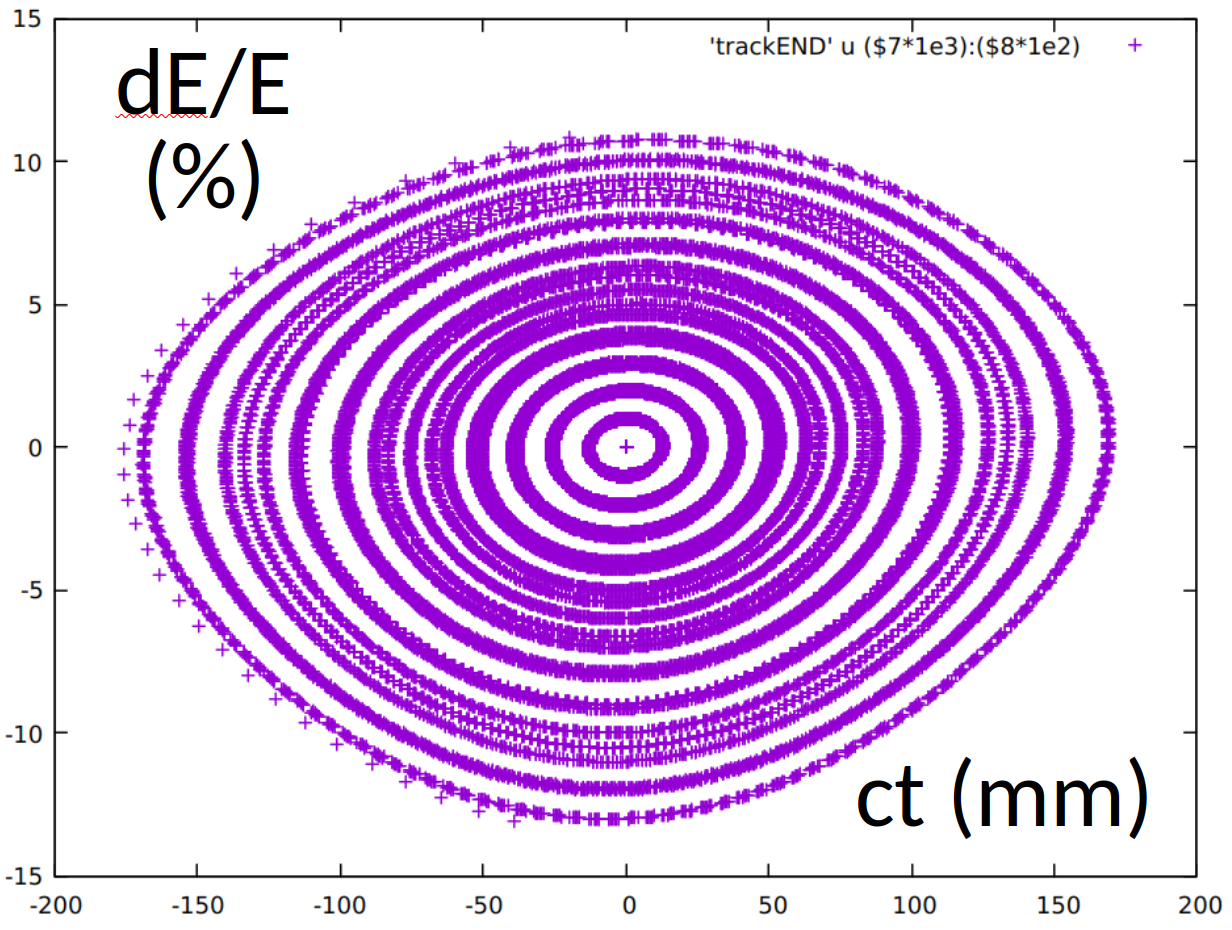}
  \caption{Reduced momentum compaction factor FFA cell: (LEFT) Linear Optics. (CENTER) Second order optics functions $Wx$, $Wy$, $DDX$ (or $DX'$ in the plot) and $DPX$ as defined in MAD-X. (RIGHT) Longitudinal phase space resulting from particle tracking of a 98~m long ring over one thousand turns starting with energy offset of $\pm$10\% in 1\% steps.}\label{f:cantedcos20}
\end{figure*}

The integral expression along the reference orbit to calculate the momentum compaction factor to first order is~\cite{wiedermann}
  \begin{equation}
    \alpha_{c1} = \frac{1}{C}\int_c\frac{\eta_x(s)}{\rho(s)}ds,\label{e:alfac1}
  \end{equation}
where $C$ is the accelerator circumference and $\rho$ is the local curvature of the reference trajectory $s$. Using the identity $\rho\theta=L$ valid for sector bend magnets, the integral in Eq.~(\ref{e:alfac1}) can be approximated for a fast calculation to the summation 
\begin{equation}
  \alpha_{c1} \approx \sum_{i=1}^{N} \eta_{xi}\theta_i
\end{equation}
over the $N$ cell elements, being $\eta_{xi}$ the horizontal dispersion and $\theta_i$ the bending angle at the $i$-th element.
Three possibilities to cancel $\alpha_{c1}$ arise:
\begin{itemize}
\item Cancel the horizontal dispersion, $\eta_x=0$. Which is possible in some particular types of cells called vertical FFA~(vFFA)~\cite{brooks}, and it is not addressed here.
\item Produce positive and negative values of $\eta_{xi}$, cancelling out the summation over the cell. This is a possible approach but produces large chromaticity due to the strong focusing to control the dispersion function. It has not been considered in this design because it has already been explored in~\cite{PhysRevAccelBeams.23.051001}.
  \item Negative bends, i.e. negative $\theta_i$. The disadvange is an increased length of the lattice, but, given the relatively small ring found before, we explored this idea.
\end{itemize}
Figure~\ref{f:cantedcos20} shows the result of the inclusion of anti bends in the design. We increased the magnetic field of the main dipoles to keep the length to 98~m. The cell superimposed magnetic components are:
\begin{itemize}
\item BF \;: 42~cm, -2.770~T,  \;240.2~T/m,  3.416~kT/m$^2$
\item BD : 80~cm, 14.697~T, -182.7~T/m, -5.519~kT/m$^2$
\end{itemize}
We minimized the dispersion function by putting the largest dipole field in the location of the vertically focusing magnet BD because $\eta_x$ follows the horizontal beta function~$\beta_x$, which has a minimum at the defocusing quadrupole.\par
Particle tracking shows a momentum acceptance of more than $\pm$10\%, and bunch length of 150~mm, from a reduction of $\alpha_{c1}$ to $1.56\times10^{-3}$ (see Fig.~\ref{f:cantedcos20} and Table~\ref{t:designs}).\par
\begin{figure}[hbt]
  \includegraphics[width=0.48\textwidth]{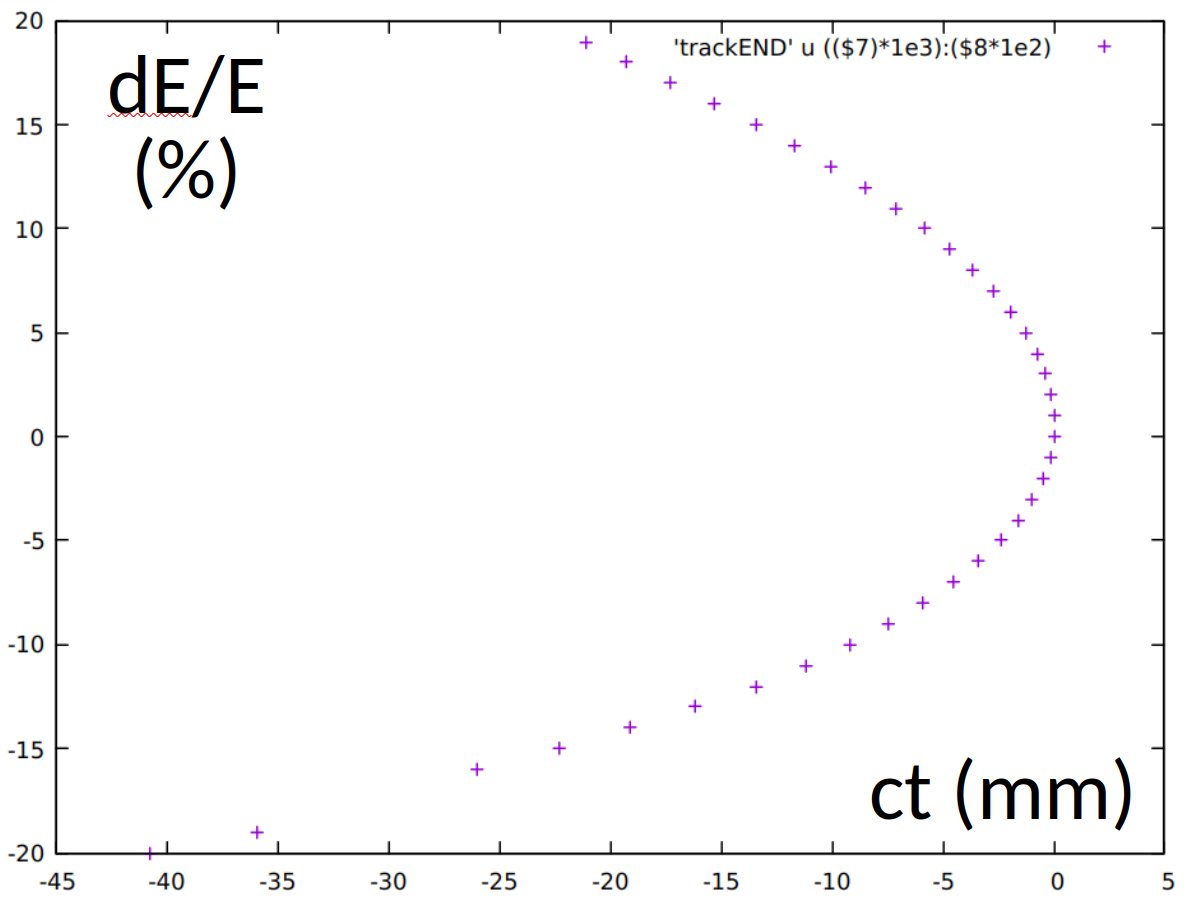}
  \caption{Longitudinal phase space at the end of one turn particle tracking for particles with energy offset between $\pm$20\% the nominal energy. The second order momentum compaction factor $\alpha_{c2}$ produces a non-linear longitudinal offset incompatible with energy gain in the RF section, limiting the energy acceptance.}\label{f:sndmom}
\end{figure}

While any arbitrarily small value of $\alpha_{c1}$ can be obtained, the second order momentum compaction factor~$\alpha_{c2}$ is not longer negligible and reduces the energy acceptance (See Fig.~\ref{f:sndmom}). We thus focus on the cancellation or reduction of $\alpha_{c2}$ to further increase the particle energy deviation $\delta$ acceptance.\par
We rewrite the momentum compaction factor as an expanded polynomial on energy as in~\cite{martin2011},
\begin{align}
  \alpha_c &= \frac{\Delta C/C }{\delta}\\
  \alpha_c(\delta) &= \alpha_{c1} + \alpha_{c2}\delta + \alpha_{c3}\delta^2 + O(\delta^3).
\end{align}
and we rewrite the integral
\begin{equation}
  \alpha_{c2} = \frac{1}{C}\int_c \left(\frac{\eta_{x}'^2}{2}+\frac{1}{2\rho}\frac{\partial \eta_{x}}{\partial \delta}\right)ds,
\end{equation}
valid for relativistic beams, where $\eta_x'$ is the derivative of dispersion with respect to $s$ and $1/2\cdot\partial \eta_x / \partial \delta$ is the second term in the expansion of position with respect to~$\delta$. It can be rewritten using the MAD-X notation as the summation
\begin{align}
  \alpha_{c2} \approx& \frac{1}{C}\left(\sum_{i=1}^N \frac{DPX^2_{i}}{2}+ \frac{DDX_i}{\rho_i}\right)\Delta s_i\\
  = & \frac{1}{C}\sum_{i=1}^N \frac{DPX^2_i}{2}\Delta s_i + \frac{1}{C}\sum_{i=1}^N\frac{DDX_i}{\rho_i}\Delta s_i\\
  = & \frac{DPX^2_{rms}}{2}+\frac{1}{C}\sum_{i=1}^{N}DDX_i\theta_i,
\end{align}
which can be evaluated over a single cell for a fast computation.\par
We note that $DPX^2_{rms}$ is always positive, therefore, the only way remaining to cancel the second order momentum compaction factor is to produce negative $DDX_i\theta_i$ in the cell.\par
The strongest dipole magnet BD creates the largest positive angle deflection $\theta\approx0.16$~rad. In a first attempt we produce negative $DDX$ at the center of the cell by rematching the sextupole strength. This proved to be effective but chromaticity was not longer cancelled, leaving the cell with zero momentum acceptance.\par
As a second alternative, we could vary independently the three sextupoles in the cell in order to cancel the horizontal and vertical chromaticities, $dq_x$ and $dq_y$ respectively, and $\alpha_{c2}$. However, all trials were unsuccessful because the three parameters are corrected at the expense of ten times stronger sextupoles that effectively kill the dynamic aperture.\par
A third approach was to create sextupole families combining sextupoles in two adjacent cells~\cite{robin1993}, however, it also resulted unsuccessful because the phase advance per cell~($\mu_x=0.1, \mu_y=0.3$) leaves a non orthogonal configuration. We remind that phase advance was optimized until we got a large energy acceptance but not orthogonality.\par
Taking into account the phase advance and following the ideas in~\cite{fartoukh}, we increased the number of sextupole families to 5 achieving a phase advance per super-cell of $\Delta\mu_x=0.5, \Delta\mu_y=1.5$. This allowed us to cancel out the geometrical errors introduced by the strong sextupoles while giving the lattice more flexibility to produce the negative DDX required to cancel $\alpha_{c2}$. However, several matching attempts showed no improvement in the dynamic aperture due to a large negative DDX peak produced at few sextupoles per super cell.\par
After these attempts we consider that a reduction of $\alpha_{c2}$ will require to minimize DPX. This is equivalent to increasing $\alpha_{c1}$ because the dispersion function remains almost constant.\par
In spite of the limitations in energy acceptance produced by the second order momentum compaction factor, we have shown that the minimum bunch length achievable with a simple FFA cell is in the order of 10~cm. With some optimization efforts using stronger gradients and dipole fields adding up to 20~T, we can minimize $DPX^2_{rms}$  reducing the bunch length to about 2~cm, and further work could be done to continue exploring other cell configurations.\par
\section{Cell for insertions}\label{s:insert}
The target location, the RF cavities and the beam extraction require dedicated insertions in the accumulator ring. These locations should have zero dispersion and therefore a matching section joining the insertions with the arc cell design is needed.\par
\begin{figure}[hbt]
  \includegraphics[width=0.48\textwidth]{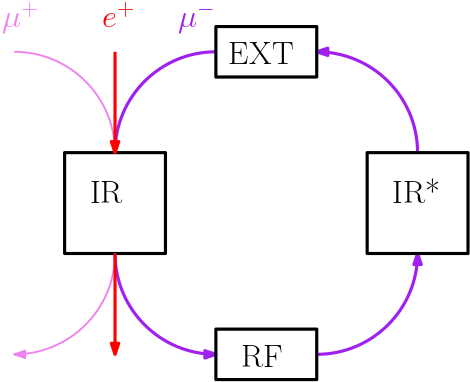}
  \caption{Muon Accumulator ring sections. Only one ring is schematically drawn, the other ring is a mirror reflection.}\label{f:ringsides}
\end{figure}

As a first step to include these regions we split the ring circumference in four equal arcs to preserve the symmetry~(see Fig.~\ref{f:ringsides}), each starting and ending with zero dispersion. Those four points will be occupied by two Interaction Points~(IPs) diametrically opposite, and two sections for the RF cavities and the beam extraction, located 90$^\circ$ away from the IPs. The survey layout is preserved as long as the RF cavity section and the beam extraction section are of the same length.\par
This simplified layout can be further optimized to have one only IP and a common region for the RF cavities and extraction kickers. However, the main goal is to study the overall parameters of a lattice design with FFA cells and further optimization can be left for a second step to reduce the total accumulator length in some few tens of meters.\par
As an open point we could still consider that the chromatic correction of two IPs is different to one IP. In this respect we remark that the interaction region should be designed to contribute little to chromaticity because the FFA cell energy acceptance is deteriorated when sextupole components are modified to correct not only the self magnet chromaticity but the total ring chromaticity. As a way to qualitatively measure this effect we will later refer to the natural chromaticity introduced by the arcs, and the single insertions.\par
\begin{figure}[htb]
  \includegraphics[width=0.48\textwidth]{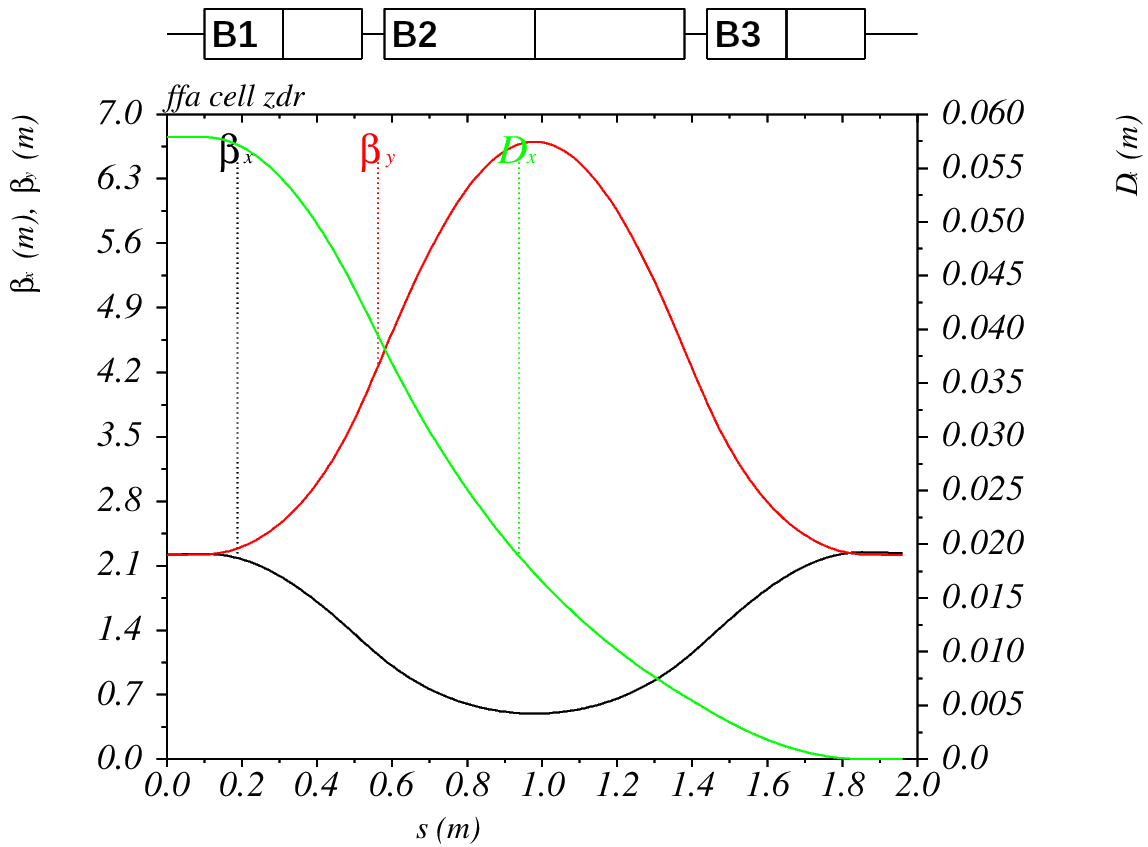}
  \includegraphics[width=0.48\textwidth]{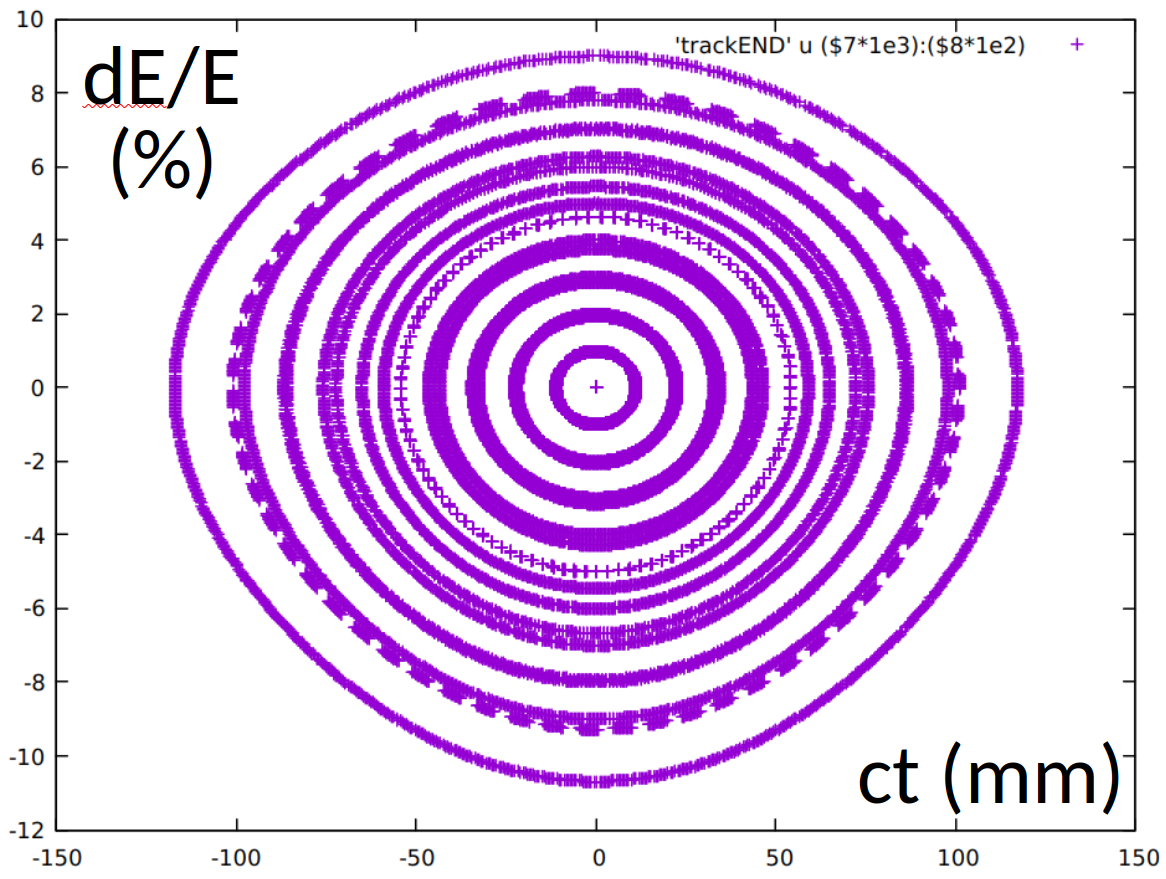}
  \caption{(TOP) FFA cell modified to get zero dispersion on one side. (BOTTOM) Longitudinal phase space resulting from particle tracking over a 98~m long ring over one thousand turns starting with energy offset of $\pm$9\% in steps of 1\%.}\label{f:cantedcos21}
\end{figure}

Several iterations were required in order to completely close the survey with a total angle of $2\pi$ in the horizontal plane while minimizing the ring circumference, dipole field and peak dispersion. The result is shown in Fig.~\ref{f:cantedcos21}.\par
The dispersion function at the end of the arc changed and sextupole components were rematched to achieve the chromatic correction. The effect is a reduction of the momentum acceptance from $\pm10$\% to $\pm$9\% which is not significative.\par
We list the magnet components obtained in this section for reference and include the parameter results in Table~\ref{t:designs}. However, they have been further modified when cancelling the additional chromaticity coming from the insertions.\par
\begin{itemize}
\item B1 : 42~cm, 0.0~T, \;\;240.2~T/m,    \;\;8.5~kT/m$^2$
\item B2 : 80~cm, 1.2~T, -182.7~T/m, -13.4~kT/m$^2$
\item B3 : 42~cm, 4.1~T, \;\;240.2~T/m,    \;\;0.0~kT/m$^2$
\end{itemize}
\section{Extraction Kickers and RF sections}\label{s:rfandkicker}
The muon beams lose few MeV when traversing a thin target~\cite{PhysRevAccelBeams.23.051001}. Although the exact amount depends on the target length and the effect is negligible for a fraction of a radiation length of material, the effect over hundreds or thousands of passages is relevant. This energy needs to be recovered during each accumulation cycle and therefore an RF cavity has been added to the design.\par
At the end of the accumulation cycle, the muon bunch should be extracted to continue with the next stages of the muon acceleration chain (not yet fully defined but a temptative proposal is shown in~\cite{blancomucoll}).\par
There are no particular restriction about the requirements for these sections, therefore we use twice the same optics design (one for the RF cavity and another for an extraction kicker section) where the main criteria has been to create a few meters long drift region for the RF cavities with large $\beta$ functions for the extraction kicker, and low chromaticity to only slightly increase the arc sextupoles in charge of chromatic correction.\par
Considering the RF cavities, the optimization process started with a 15~m long apochromatic lattice~\cite{blancoprab2020} that was modified to create a 4~m long drift, enough to allocate few RF cavities providing a total voltage in the order of a hundred MV.\par
For the extraction region the beta functions were inflated to a few meters in order to increase the effectiveness of a kick~$\Delta x'$, as in the expression
\begin{equation}
  \Delta x \propto \Delta x' \sqrt{\beta_k\beta_{ext}}\sin\Delta\phi_{x,ext}
\end{equation}
where $\Delta x$ is the displacement produced by the linear propagation of a kick at the kicker location and in a lattice with twiss optics functions $\beta_k$, $\beta_{ext}$ at the kicker and extraction points respectively, and a phase advance $\phi_{k,ext}$ between them.\par
In order to reduce the off-energy beta beating we opt for an apochromatic line. Magnet gradients are kept in the order of 100 to 200~T/m, comparable to that of the Future Circular Collider~(FCC)~\cite{fcchh}.\par
\begin{figure}[htb]
  \includegraphics[width=0.48\textwidth]{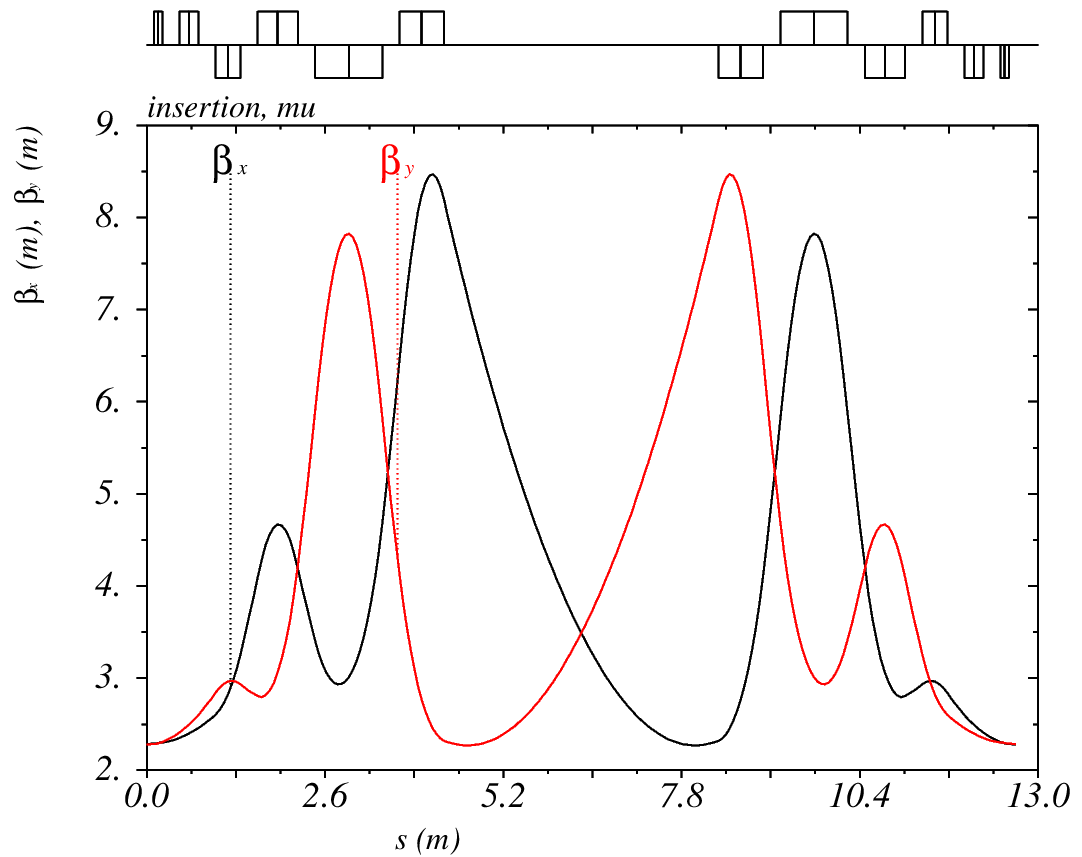}\\
  \includegraphics[width=0.48\textwidth]{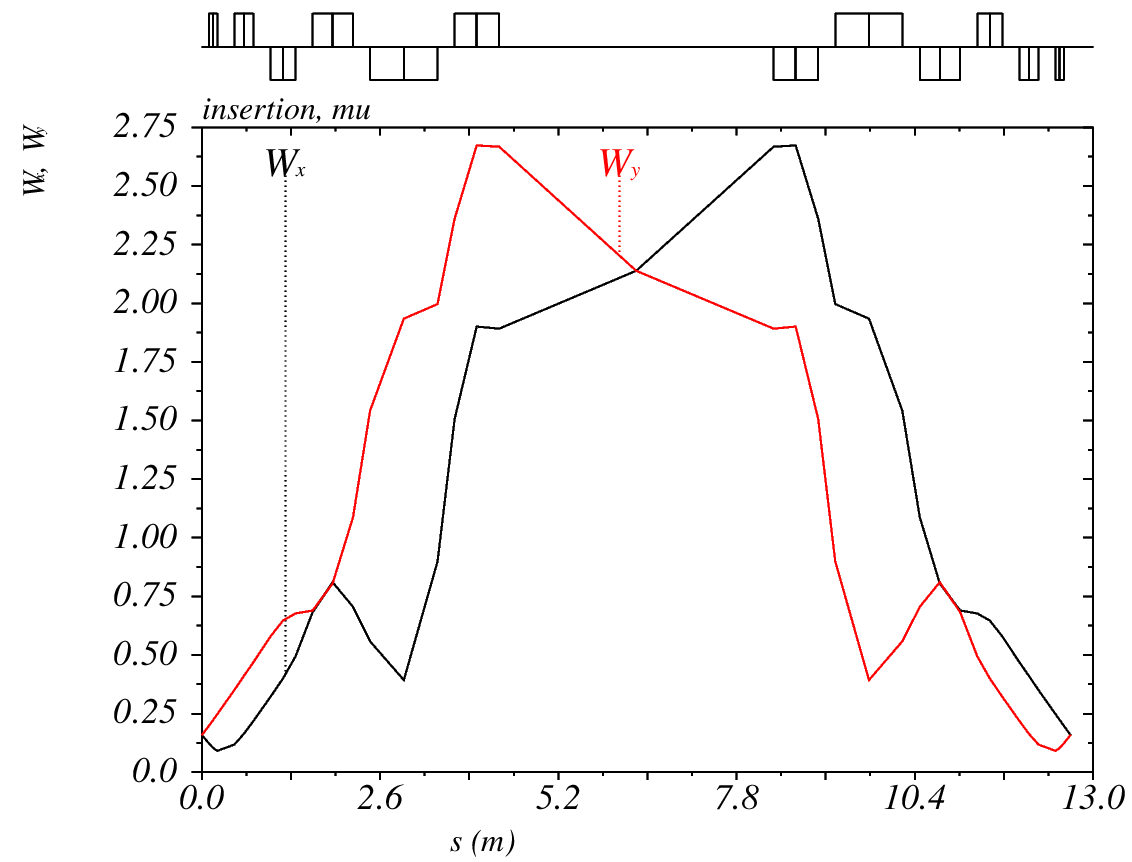}
  \caption{A 13~m long optics design for an apochromatic drift space 4~m long, with (TOP) twiss $\beta$ functions of few meters using FCC--like quadrupoles with gradients in the order of 100 to 200~T/m. (BOTTOM) The chromatic functions $W_x$ and $W_y$ are very small at the entry and exit point of the section. Two of this optics are used in the accumulator ring, one for the RF cavities and one for the extraction kicker.}\label{f:rf}
\end{figure}

The result of the optimization is a 13~m long region with a 4~m long drift and a $\beta$ function just below 10~m, shown in Fig.~\ref{f:rf}. It is for sure a preliminary result, because it does not match the best phase advance to extract the beam. However, it was included to take into account the possible increase of the accumulator ring length.\par

\section{Beam separation--combination}\label{s:beamseparation}
The region containing the target is common to three beams : $\mu^+\mu^-$ at about 22.5~GeV and $e^+$ at double the muon energy. In the LEMMA scheme (see Fig.~\ref{f:rings}) the interaction of the positron beam produces muon pairs that are transported and recirculated to pass through the target a few hundreds times. The muon beams therefore need to be separated and directed into the two muon rings, while a line/ring will transport the remaining positrons to the next stage in the positron chain.\par
When recirculated, the beams need to be recombined in order to increase the muon bunch population while mitigating the emittance growth.\par
\begin{figure}[htb]
  \includegraphics[width=0.48\textwidth]{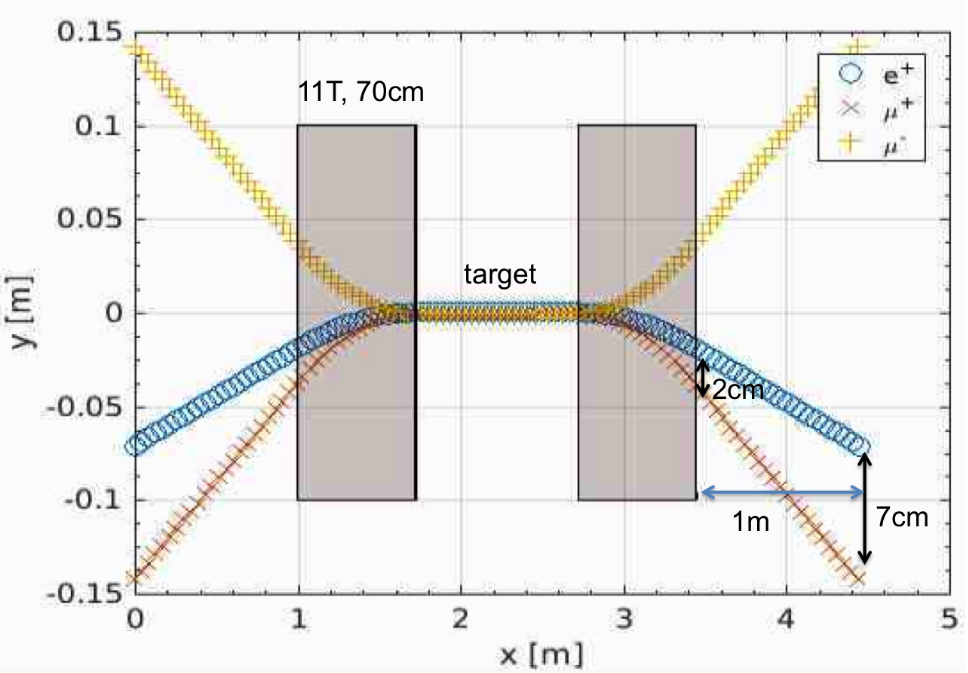}
  \caption{Beam separation with 11~T dipole magnets. The positron beam loses 0.2~GeV per passage, a value that could be too high. Image courtesy of Simone Liuzzo from ESRF.}\label{f:separationSimone}
\end{figure}

There are some aspects to take into account when considering the recombination of the muon beam:
\begin{itemize}
  \item The multiparticle interaction of different charges at different energies. We will not pursue any study about this subject on this article.
\item A second aspect is the effect of multiple scattering in the target on the final muon beam emittance. In the article~\cite{PhysRevAccelBeams.23.051001}, a $\beta^*_\mu$ from 1~m to 10~cm was explored for a monochromatic beam, but, proved not enough to mitigate the impact of a thousand turns through the target. We have achieved a $\beta^*_\mu=20$~cm over $\pm$5\% energy spread using quadrupole magnets with gradient of 500~T/m, and theoretically explored stronger focusing in order to determine the limits of such mitigation (see Section~\ref{ss:ir} for the Interaction Region design and Section~\ref{s:accumulation} for the accumulation results).
  \item The separation and combination of the beams will produce synchrotron radiation coming from the positron beam at high energy. We will take this into account when considering the strength of the dipole fields.
\end{itemize}
The beam separation and combination scheme has evolved with time. Initially, in collaboration with Susanna Guiducci from INFN and Simone Liuzzo from ESRF, a very strong magnet was foreseen to separate the beams in short space. Figure~\ref{f:separationSimone} shows the schematic diagram of three beams being combined, passed through the target and separated by 11~T 30~cm long dipole magnets. The expected energy loss of the muon beam is negligible for a single pass, while, the positron beam loses 0.2~GeV which is a considerable large amount of energy. The $e^+$ beam radiation could be high, therefore, they suggested to think in a lower B-Field. 

\subsection{Considerations due to radiation}
The positron beam energy loss expected to come as synchrotron radiation when positrons traverse strong dipole magnets forces the separation and combination design to consider normal conducting magnets on the order of 2~T.\par
\begin{figure}[htb]
  \includegraphics[width=0.48\textwidth]{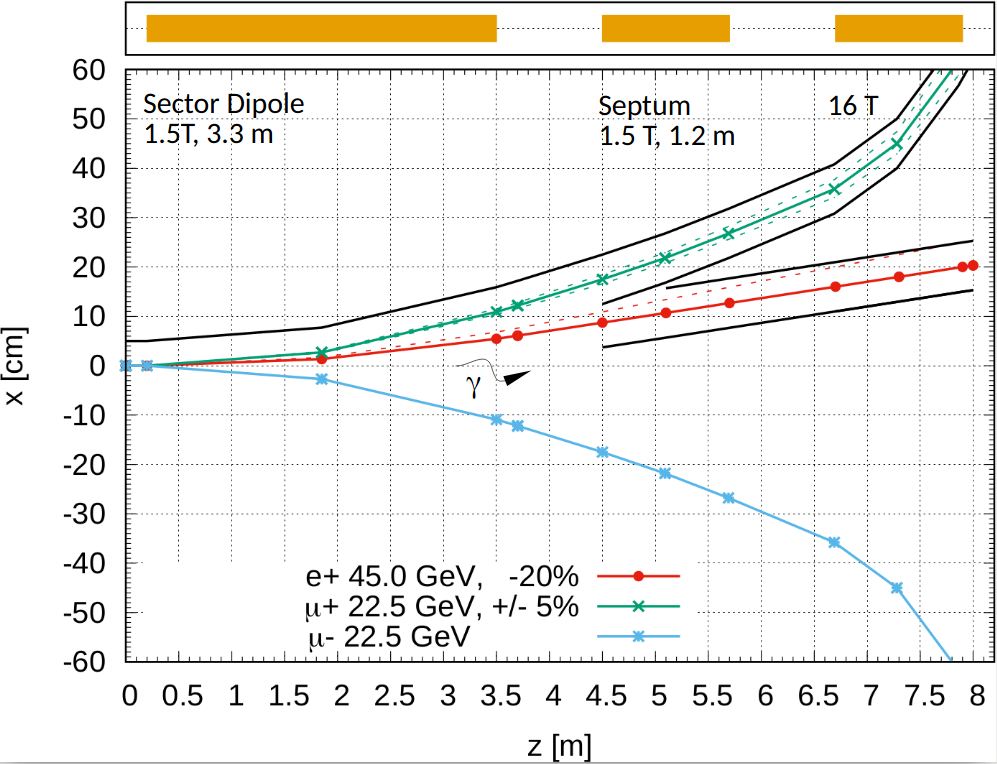}
  \caption{Beam separation with magnets below 2~T. A dipole field separates the three beams while a septum directs the muon beams into the rings letting the positron beam pass through. The positron beam loses few tens of MeVs while the photon critical energy corresponds to strong X-rays at 2~MeV. The separation region is less than 10~m long in total. The 16~T magnets correspond to the beginning of the accumulator ring.}\label{f:beamseparation2019}
\end{figure}

We continued the design with a 1.5~T 3.3~m long dipole magnet leading to a 1.5~T and 1.2~m long septum to leave the positron beam escape the interaction region while the muon beams are separated into assumed 16~T dipole magnets of independent accumulator rings, see Fig.~\ref{f:beamseparation2019}. The aperture has been schematically drawn as black lines at $\pm$5~cm around the $e^+$ and $\mu^+\mu^-$ trajectories.\par
From particle tracking in MAD-X PTC it was clear that the aperture will limit the energy acceptance. A muon beam with large energy spread would occupy the whole aperture when reaching the septum, so only $\pm$5\% energy spread lines have been tracked around the nominal energy line. We also verified that positrons with energy as low as 36~GeV or -20\% energy offset pass through the separation region. At the entrance in the accumulator, beam pipes are separated by about 10~cm, which should be a safe value to consider separated magnets.\par
The positron beam loses 20~MeV in the 3.3 m long 1.5~T magnet, which is small and in the same order of magnitude of energy loss due to the interaction with a thin target. Photons are emitted from the interaction of the beam and the dipole field with a critical energy of 2~MeV, which is a value comparable with the FCC--ee Interaction Region expectations.\par
Given the acceptable critical energy of the irradiated photons and the small energy loss of the positron beam, the design of the Interaction Region will include magnets on the other of 1.5 to 2~T. The optics will be matched so that the beam behavior does not perturb the arc.\par
\subsection{Exploring the idea of a separation aided by a sextupole field}
\begin{figure}[htb]
  \includegraphics[width=0.48\textwidth]{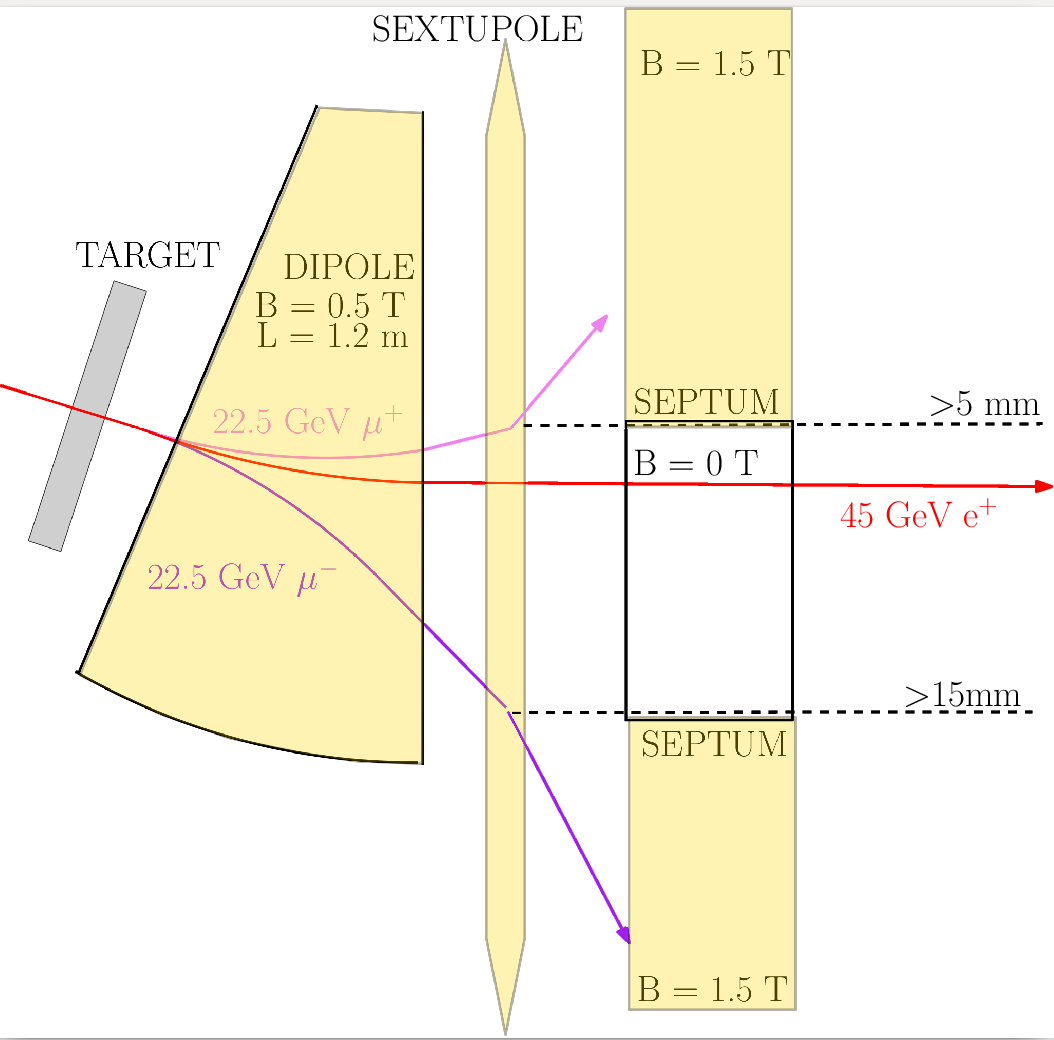}
  \caption{Beams separation aided by a large aperture sextupole magnet. The three beam coming out the target are initially separated by a low dipole field. The positron beam will continue to cross a sextupole magnet near the axis which leaves it almost unperturbed, while, the two muon beams are kicked away from the sextupole axis in opposite directions increasing the beam separation before the septum.}\label{f:sepSext}
\end{figure}

As an alternative to the single dipole, we have considered the usage of a sextupole magnet. Figure~\ref{f:sepSext} shows the schematic diagram of a beam separation assisted by a large aperture sextupole.\par
\begin{figure*}[htb]
  \includegraphics[width=0.98\textwidth]{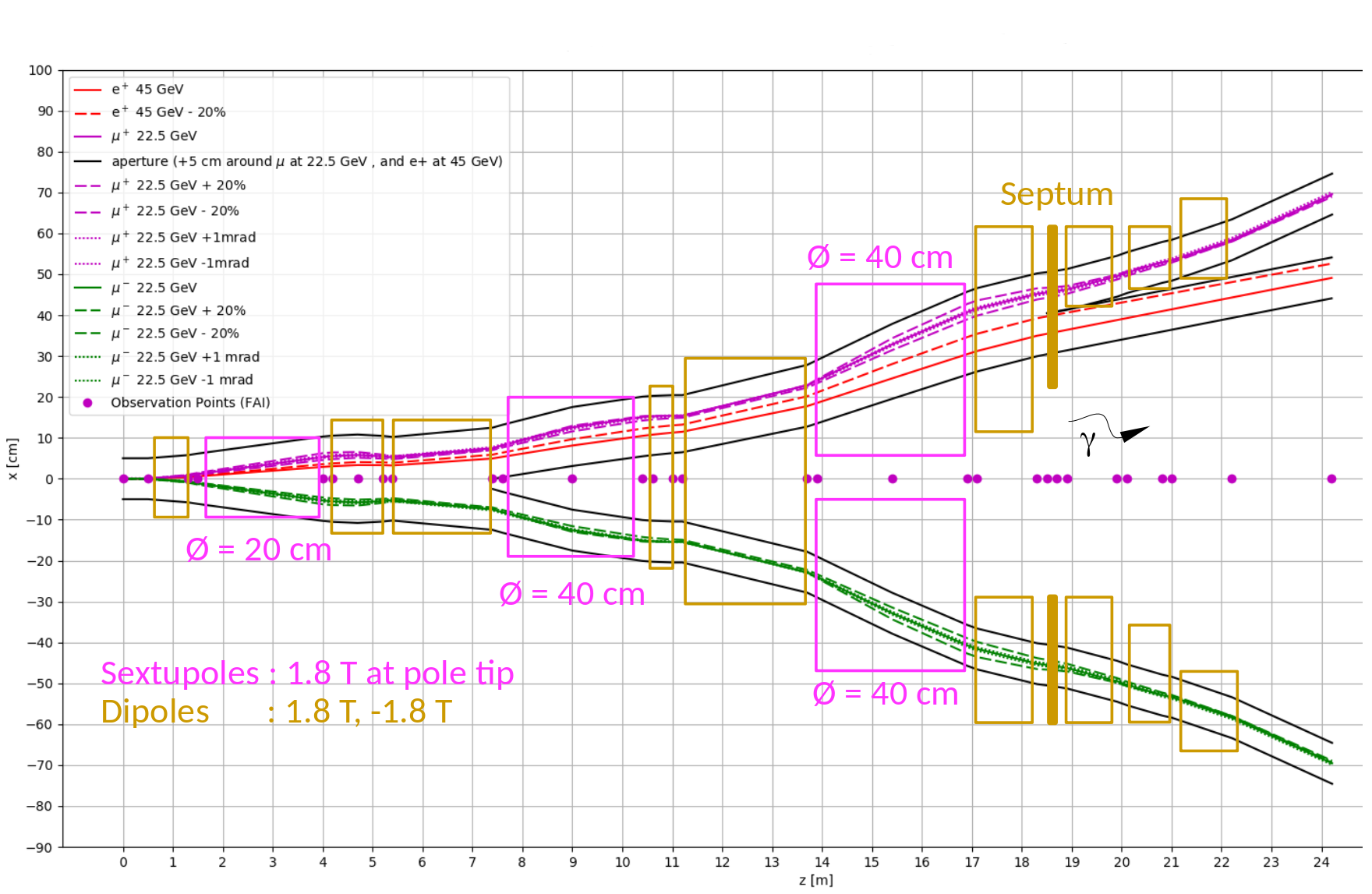}
  \caption{Beam separation design using a combination of dipole and sextupole magnets below 2~T peak field in order to reduce the amount of synchrotron radiation coming from the high energy positron beam. Three beams with energy offset of $\pm$20\% have been tracked to show the energy acceptance of the separation region. The total length is less than 30~m long.}\label{f:separation3}
\end{figure*}
Three beams exit the target and enter a relatively low dipole field of 0.5~T and 1.2~m in length, after which the three beams are separated. In order to give an extra separation kick to the muon beams while reducing the radiation of the positron beam, a sextupole is located among the septum and the dipole. The sextupole will kick the two muon beams in opposite directions because the kick sign will depend on the muon charge irrespective of the crossing side, as shown by the square term in the equations
\begin{align}
  x' &= K_2L (x^2+y^2)\\
  &= \frac{1}{B_\rho}\left(\frac{\partial^2 B_y}{\partial x^2}\right)Lx^2 ; \text{\;\;(assuming }y=0) \\
  &= \frac{q}{P}\left(\frac{\partial^2 B_y}{\partial x^2}\right)Lx^2
\end{align}
where $x'$ is the angle change in the trajectory of the particle due to a normal focusing sextupole, $B_\rho$ is the particle magnetic rigidity, $q$ is the charge of the particle, $x$ and $y$ are the transverse particle coordinates when crossing a thin sextupole, $L$ is the length of the sextupole, and
\begin{equation}
  \frac{\partial^2 B_y}{\partial x^2}=B_\rho K_2.
\end{equation}
We have considered a gradient of 800~T/m$^2$, a diameter of aperture equal to 76~mm, a length of 1.4~m, a magnetic peak field of 1.6~T and a good field region of 10~mm (as the one in the FCC--ee ARC). This gives a $k_2=10.8$/m$^3$ and a $k_2L=15.1$/m$^2$. Using this parameters we can estimate an additional kick of 1.5~mrad, which is small, but could lead to a reduction in the level of radiation.\par
The design was continued tracking particles in pyzugoubi~\cite{pyzgoubi}, a python~\cite{python} interface to the tracking code Zgoubi~\cite{zgoubi}, allowing to estimate a 20~m long separation region taking into account the tracks of all beams with an energy spread of $\pm$20\%, see Fig.~\ref{f:separation3}. It consist in a composition of dipoles and sextupoles with reduced peak magnetic field that effectively separates the three beams.\par
The total length required to separate the beams is slightly less than 30~m long, and therefore is too long for our design. Section~\ref{ss:ir} is dedicated to the Interaction Region design and we integrate the results of these studies with the muon beam optics.\par
\section{Interaction Region}\label{ss:ir}
The interaction region of the muon accumulator ring in the LEMMA scheme is one of the most critical parts because it is common to the three particles species at two different energy levels ($\mu^+$ and $\mu^- $ at about 22.5~GeV and $e^+$ at double the muon energy), and must mitigate the effect of multiple scattering of beams with the target located at the Interaction Point~(IP).\par
Multiple scattering with the target will increase the beam divergence per passage by an amount $\Delta\sigma'\propto \sqrt{X_0}/E$, where $E$ is the particle energy and $X_0$ is the thickness of the material in radiation length units. We expect that the divergence growth due multiple scattering will be more significant for the muon beam because the muon beam has only half of the positron beam energy, and in addition the muon beams pass a large number of times through the target (equivalent to a large $X_0$).\par
As a way to mitigate the effect one could try to focus the beams so that the divergence is much larger than the contribution from multiple scattering over some number of passages~$N$. The contribution from multiple scattering is uncorrelated, thus, we require a beam divergence
\begin{equation}
  \sigma' \geq \sqrt{N}\Delta \sigma'
\end{equation}
with $N\approx100$ for the positron beam, $N\approx10^3$ for the muon beam, in the initial LEMMA proposal. This forces the interaction region design to have a very small $\beta^*$ at the IP, able to produce a focal point with large divergence from a low emittance beam~$\epsilon$, according to $\sigma' = \sqrt{\epsilon/\beta}$.\par
In~\cite{eplusringopt} it was found that a $\beta^*_{e^+} = 50$~cm was enough to keep low the positron beam emittance over more than 50 turns through out a thin target of 0.89\%$X_0$, i.e. almost half a radiation length of material when adding all passages. In the case of the muon beam we estimated that a $\beta^*_{\mu}=1$~cm will be required to avoid the emittance growth over a thousand turns through out the material.\par
Due to the large muon beam energy spread coming from the kinematics of the collision, we note that the Interaction Region should not only have a small $\beta^*$ at the nominal energy, but also over the whole muon beam energy range.\par
\begin{figure}[htb]
  \includegraphics[width=0.48\textwidth]{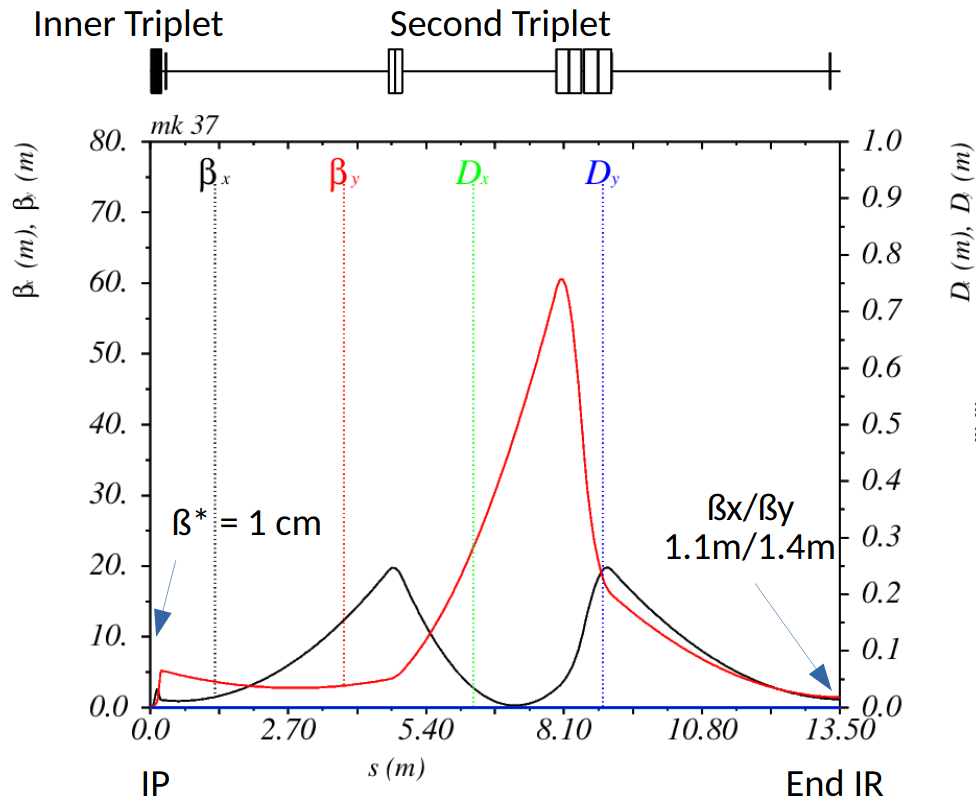}\\
    \includegraphics[width=0.48\textwidth]{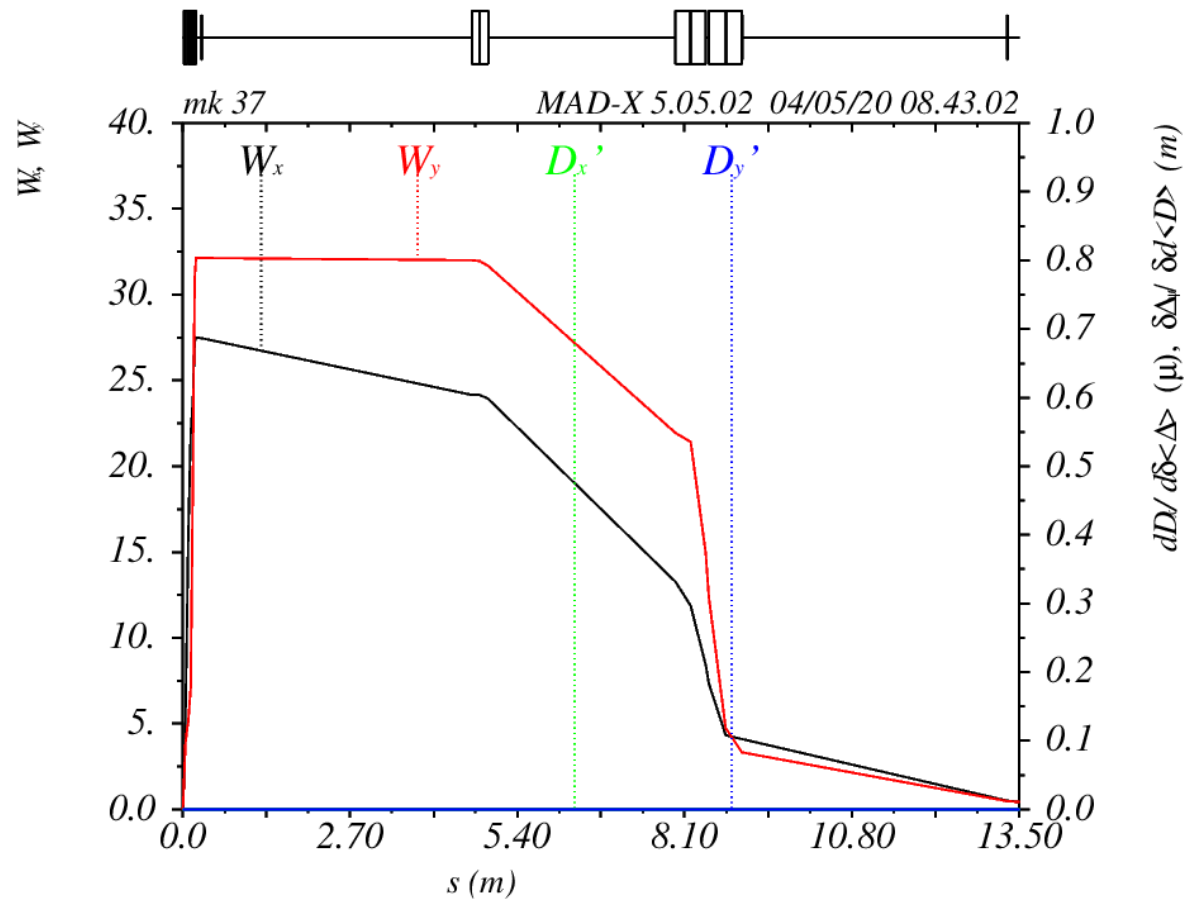}
  \caption{Theoretical evaluation of an Interaction Region design to produce (TOP) a $\beta^*_\mu=1$~cm over $\pm2$\% energy spread, for $L^*=1~$cm. The design consists in two triplets, the nearest to the IP focalize the beam, while the second produces opposite chromaticity and brings the $\beta$ functions to some meters (possible values of a zero dispersion cell to be connected with). (BOTTOM) Apochromaticity of the design achieved by cancelling the $W_x$ and $W_y$ chromatic functions at the IP and at the end points.}\label{f:betamustar1cm}
\end{figure}

 
We have explored the magnet requirements to get a $\beta^*_\mu=1$~cm as shown in Fig.~\ref{f:betamustar1cm}. The design consists in two triplets, the first one reduces the beta functions to 1~cm in both planes while the second triplet matches the $\beta$ functions to some arbitrary value in the order of a meter and produces the opposite chromaticity to cancel the chromatic functions at the end point.\par
The space between the IP and the nearest quadrupoles is $L^*=1$~cm. This means that only very thin targets could be considered.\par
On top of the difficulty to put a target in 2~cm ($2L^*$) of space, the quadrupole magnets in the inner triplet are expected to have a gradient of 20~kT/m, i.e. be at 20~T peak with 1~mm of aperture radius. This is a gradient that has not been yet achieved with any type of magnet technology. Quadrupole gradients in the order of 200 to 400~T/m are foreseen in the FCC design. The CLIC QD0 magnet pushes the gradient to a bit more than 500~T/m~\cite{Modena:1427609,testQD0}. We require at least a factor ten higher gradients.\par
\begin{figure}[htb]
  \includegraphics[width=0.48\textwidth]{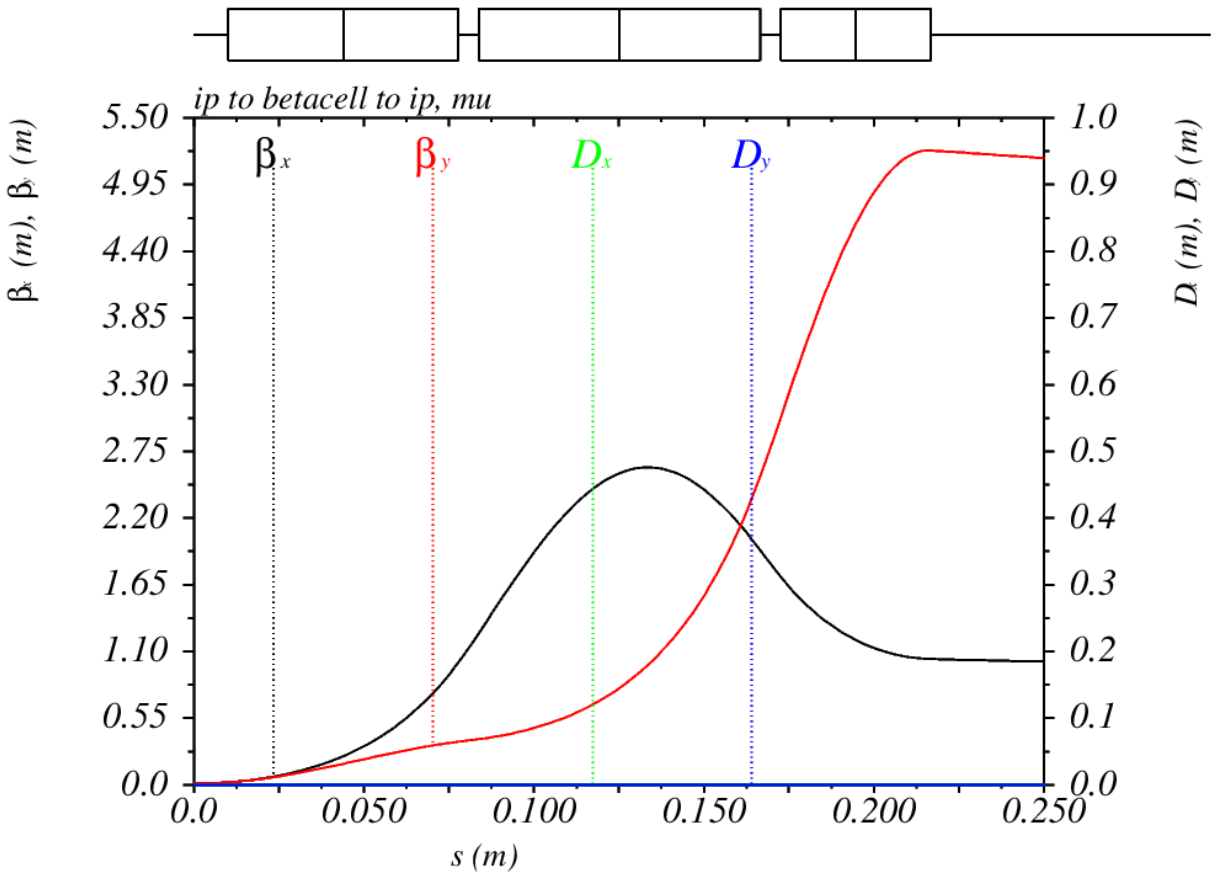}\\
  \includegraphics[width=0.48\textwidth]{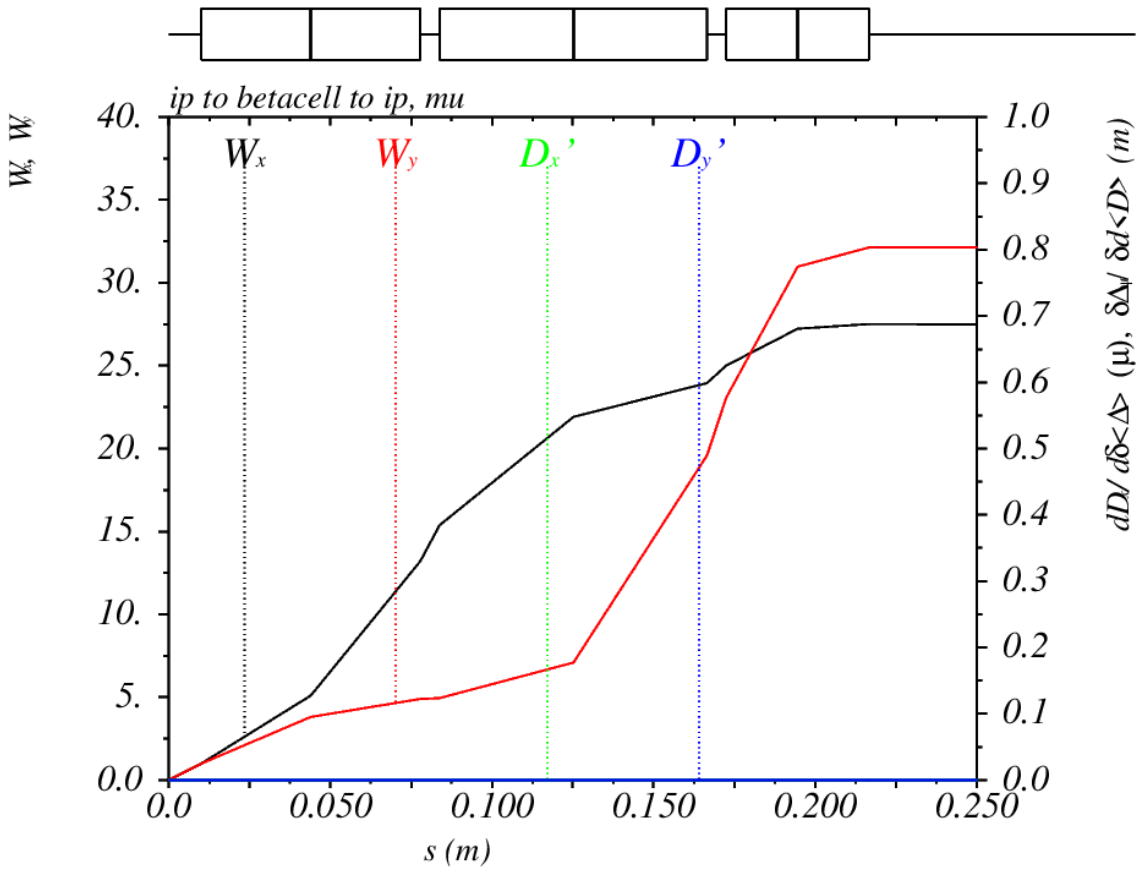}
  \caption{Zoom into the inner triplet (TOP) $\beta$ functions and (BOTTOM) chromatic functions for a $\beta^*_\mu=1$~cm and $L^*=1$~cm}\label{f:innertriplet20k}
\end{figure}

The main reason to use such an extreme gradient is shown in Fig.~\ref{f:innertriplet20k}. The optics twiss $\beta$ and chromatic $W$ functions around the IP grow fast near the IP due to the low $\beta^*$, and as a result, the dynamic aperture of the IR is reduced because of the off-momentum beta--beating. In order to give an order of magnitude of the $\beta$ variation with energy, we rewrite here the expression of the chromatic functions
\begin{equation}
  W_{X,Y}=\sqrt{\left(\frac{\partial \alpha_{x,y}}{\partial p_t}-\frac{\alpha_{x,y}}{\beta_{x,y}}\frac{\partial \beta_{x,y}}{\partial p_t}\right)^2+\left(\frac{1}{\beta_{x,y}}\frac{\partial \beta_{x,y}}{\partial p_t}\right)^2}.
\end{equation}
Assuming that $\alpha$ and its derivative with respect to momentum $p_t$ cancel at some location in the triplet, we can write
\begin{equation}
\beta_{x,y}W_{x,y}\Delta p_t\approx \Delta\beta_{x,y},
\end{equation}
which in our inner triplet evaluates to $W_{x,y}=40$, $\Delta p_t\approx0.2$ and $\beta_{x,y}=5$~m, giving $\Delta \beta_{x,y}=40$~m, which is a very large $\beta$ excursion for a small aperture triplet, effectively limiting the energy acceptance.\par
\begin{figure}[htb]
  \includegraphics[width=0.48\textwidth]{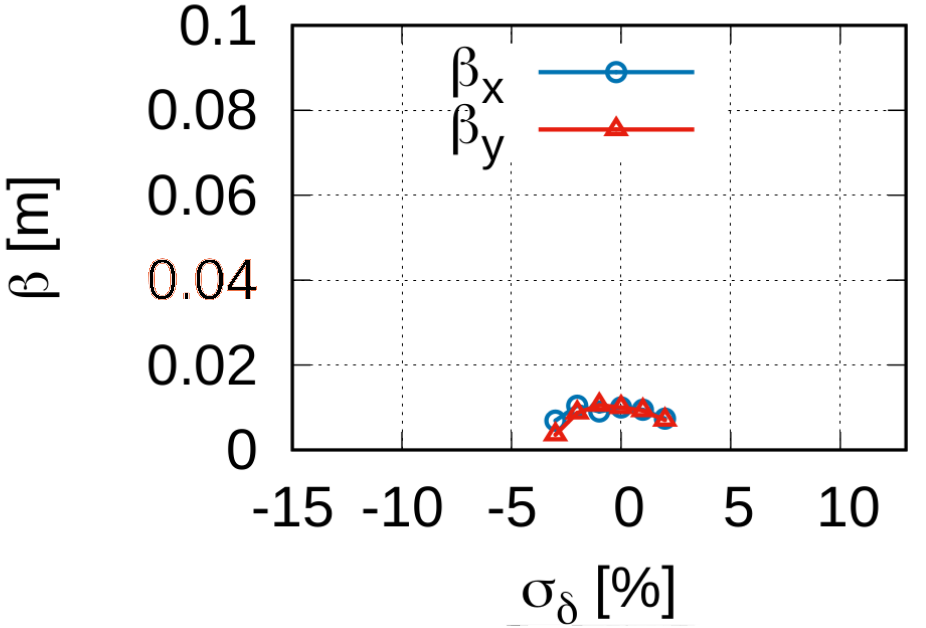}\\
  \includegraphics[width=0.48\textwidth]{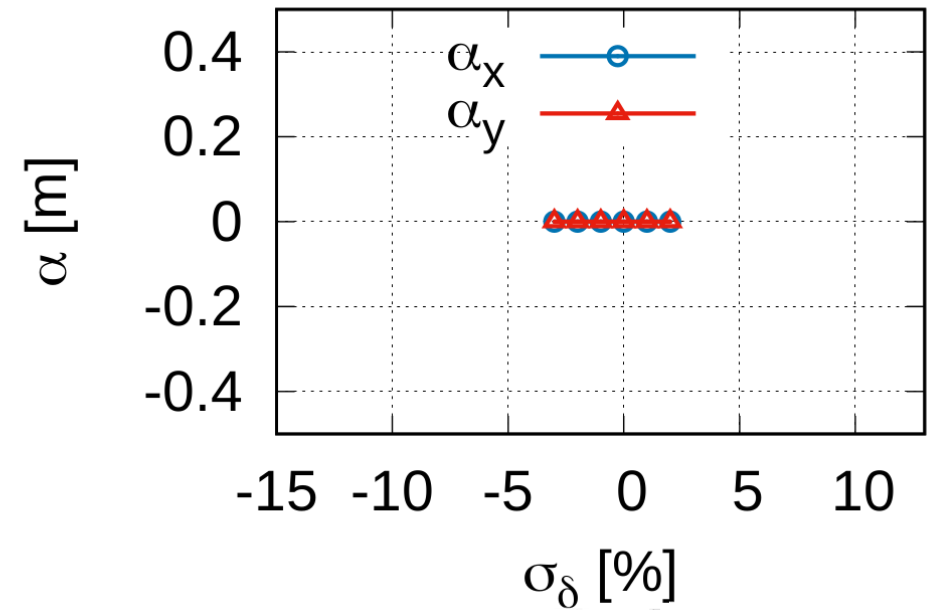}
  \caption{Twiss functions $\beta^*$ and $\alpha^*$ at the IP of a theoretical interaction region with a triplet at 20~kT/m, and $L^*=1$~cm for a range of energy spread. The twiss computation succeeds in the energy range of $\pm2$\% the muon beam energy.}\label{f:betamustar1cmeoffset}
\end{figure}

In spite of the problems with the magnets apertures, one could be interested in the theoretical model behavior at different energies. Figure~\ref{f:betamustar1cmeoffset} shows the twiss calculation for several energy offsets in steps of 1\% the nominal energy where we attach one interaction region with a mirrored interaction region to get the twiss results.\par
The design is stable in the energy range of $\pm$2\% the muon beam energy. This value however is very small for the requirements of accumulation, as it would only accept muons from a positron beam energy just above the muon pair production threshold reducing the production efficiency to almost $10^{-8}$ muon pairs per positron, see Table~\ref{t:posEmusigmadelta}.\par
From all previous aspects, we do not pursue a design achieving a $\beta^*_\mu=1$~cm any further.\par
In order to reduce the $\beta^*$ over a large energy range we have started from an apochromatic transport line able to focus the positron and muon beams~\cite{blancoprab2020}. The transport line was designed to have a $\beta^*_\mu=20$~cm over $\pm$5\% of energy spread, and a $\beta^*_{e+}=50$~cm at twice the muon beam energy.\par
\begin{figure}[htb]
  \includegraphics[width=0.48\textwidth]{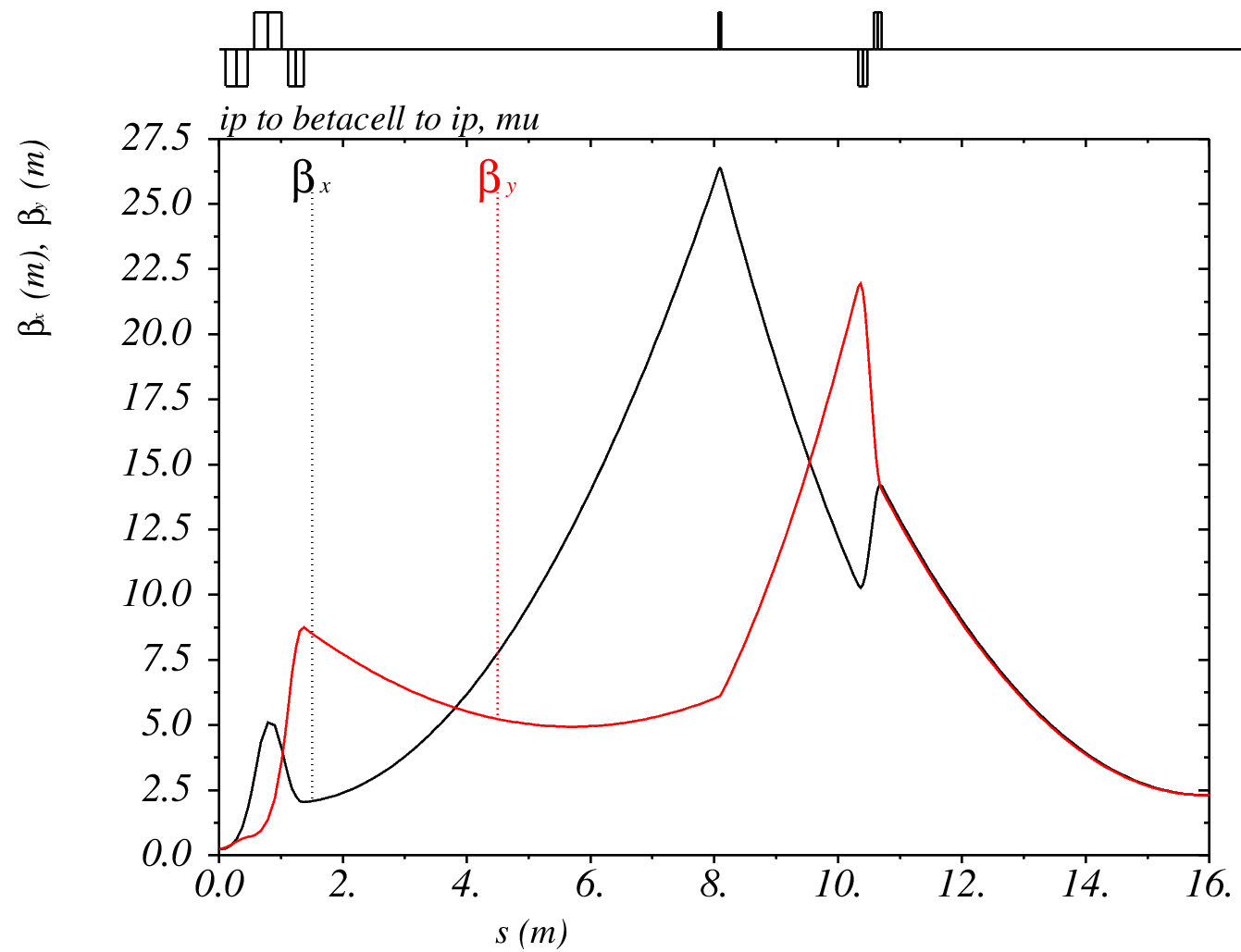}\\
  \includegraphics[width=0.48\textwidth]{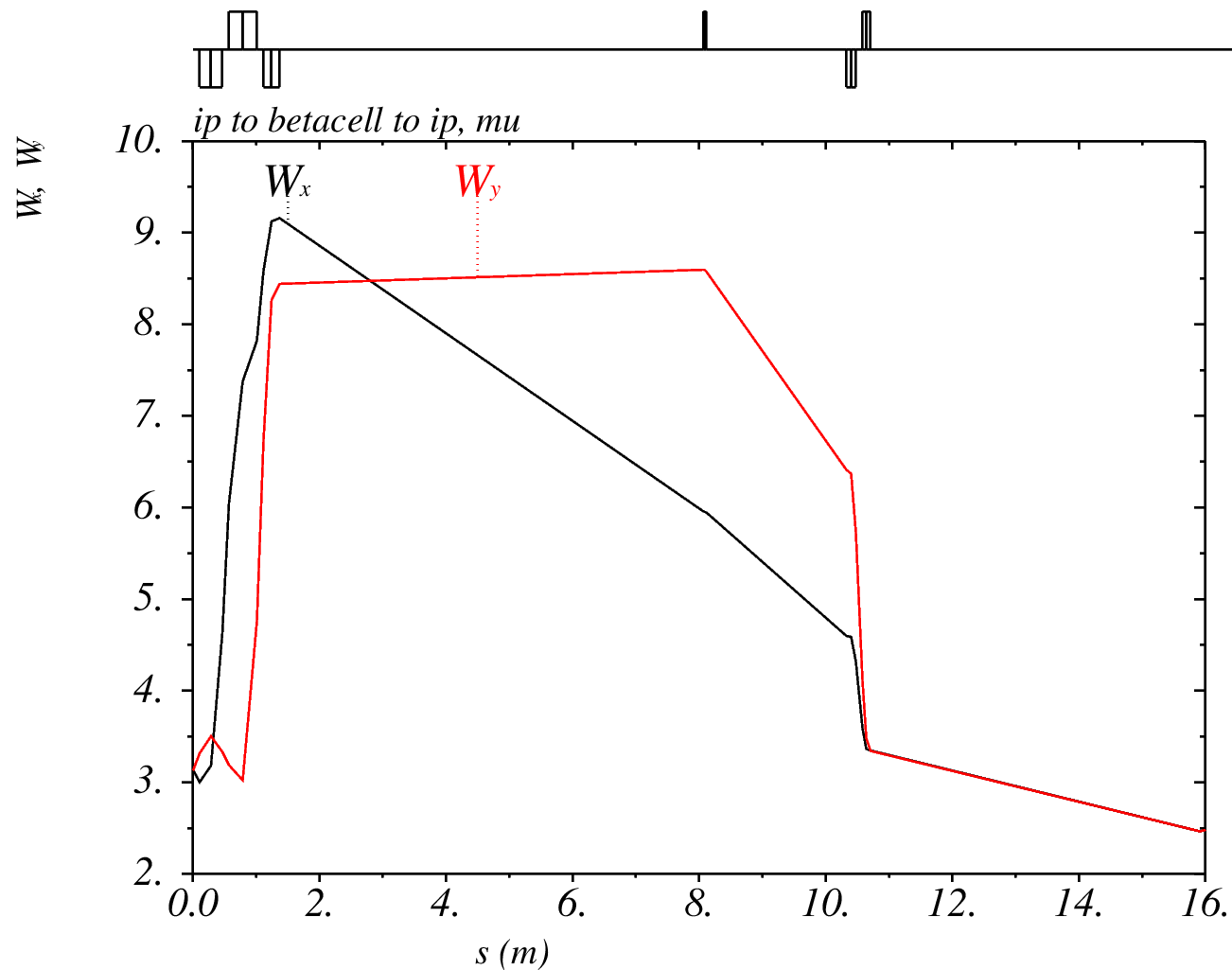}
  \caption{Half of the Interaction Region design to produce (TOP) a $\beta^*_\mu=20$~cm over $\pm5$\% energy spread, for $L^*=10~$cm. The design consists in two triplets: the nearest to the IP, at $s=0$, focalizes the beam using a gradient of about 500~T/m, while the second produces opposite chromaticity and brings the $\beta$ functions to some meters to match the zero dispersion cell with magnets at 100 to 200~T/m. (BOTTOM) The low chromaticity of the design is achieved by partially cancelling the $W_x$ and $W_y$ chromatic functions at the IP and at the end points. The chromaticity introduced by the line is below 1.5 units on each plane.}\label{f:ir}
\end{figure}

Figure~\ref{f:ir} shows the result of the antisymmetric design after rematching the transport line to the $\beta$ functions at zero dispersion cell entry side. We see a lattice where the beta functions in horizontal and vertical plane start with a value of 20~cm and they propagate over more than 15~m to get a beta value  close to 2.2~m.\par
The optimization also minimized the chromatic functions $Wx$ and $Wy$ which will allow for a smaller off-energy beta-beating in the arc and therefore a better chromatic correction. The chromaticity added by this section is less than 1.5~units per plane which could be corrected by sextupoles in a quarter of the arc running at an additional 50\% strength (1.5/12 cells is 50\%$\times$0.25, the arc cell chromaticity).\par

\subsection{Including the beam separation into the Interaction Region Design}
Profiting from the long drifts to reduce the beta from 2~m in the arc to 20~cm at the Interaction Region, we have included three \emph{vertical dipoles} with reduced magnetic fields in the range of 1 to 2~T, mitigating the positron synchrotron radiation, as explained in Section~\ref{s:beamseparation}. The combination of the three vertical dipoles produces an early and smooth separation/combination of the three beams while cancelling the vertical dispersion $\eta_y$ and its derivative with respect to $s$, $\eta'_y$.\par
\begin{figure}[htb]
  \includegraphics[width=0.48\textwidth]{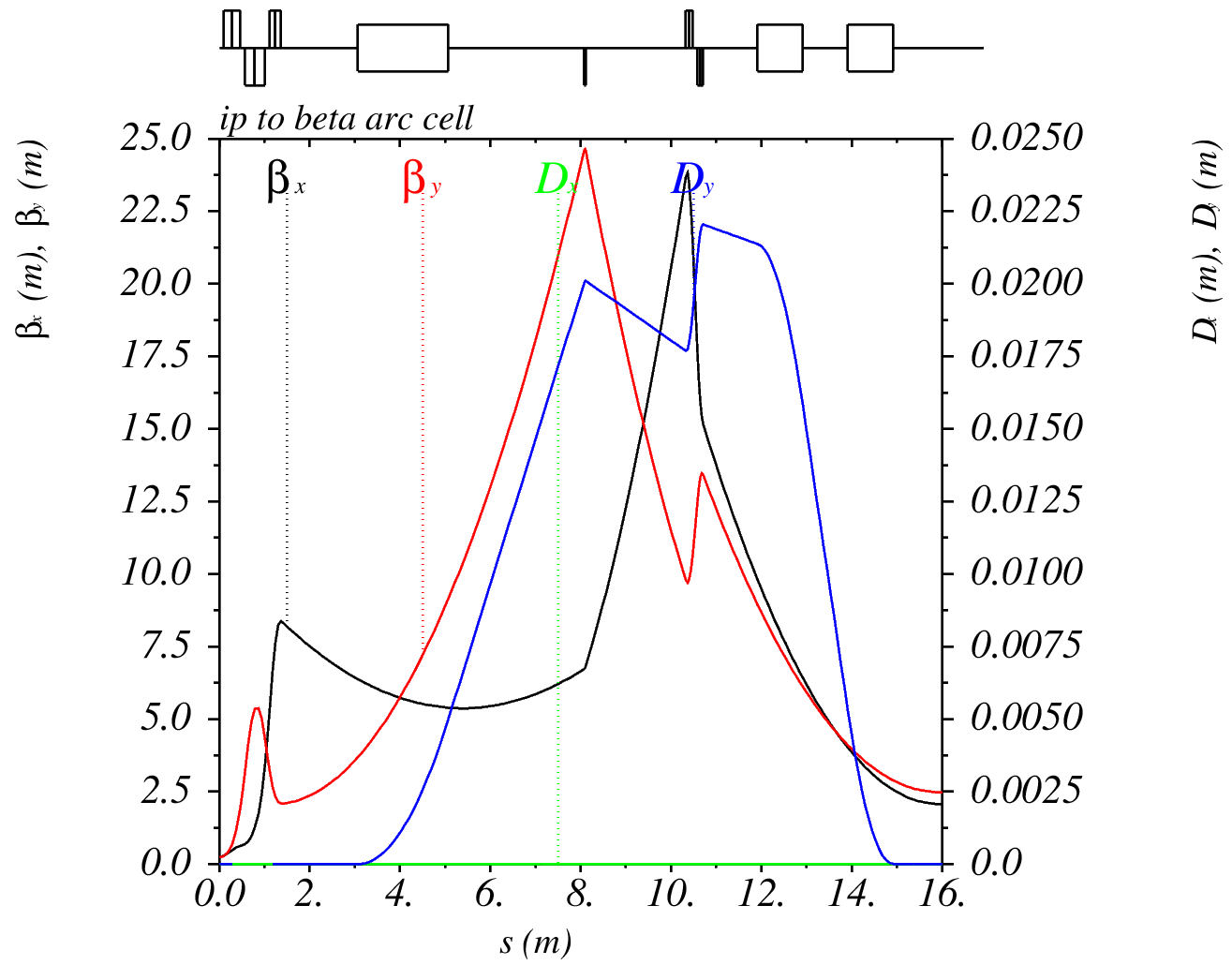}
  \caption{Half of the Interaction Region design with IP at $s=0$~m and the beginning of the zero dispersion arc cell at $s=16$~m. Three vertical dipoles, with a field lower than 2~T, separate the beams while cancelling $\eta_y$ and $\eta'_y$. The first triplet achieves a $\beta^*_\mu=20$~cm over $\pm5$\% energy spread for $L^*=10$~cm using 500~T/m gradients, while the second triplet matches the beta functions with the arc. All quads length, strength and in-between drifts have been matched to minimize chromaticity as in the symmetric apochromatic design.}\label{f:verticalIR}
\end{figure}

In order to keep the symmetry of the vertical dipole configuration at the combination~(i.e. before the IP) and the separation~(i.e. after the IP), the apochromatic Interaction Region has been rematched for a symmetric reflection instead of the asymmetric solution previously described (in the asymmetric solution the mirrored magnets polarity is inversed, while this is not the case in the symmetric solution). The optimization process matched the $\beta^*_\mu$, the $\beta$ at the arc cell entrance and the chromaticity leaving a small beta-beat. It seems possible to improve the matching if additional quadrupoles are included in the simplex minimization, but as a first approximation it doesn't represent a large problem on the design. Figure~\ref{f:verticalIR} shows the result of the modifications in the initial concept of the interaction region.\par
\begin{figure}[htb]
  \includegraphics[width=0.48\textwidth]{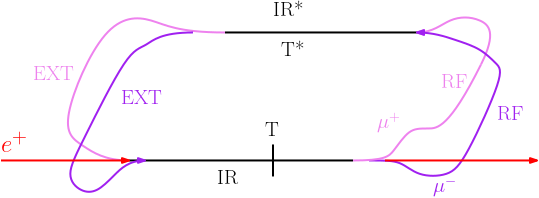}
  \caption{Rings with vertical separation can be one on top of the other. It also allows to use a second target $T^*$ given a second positron beam.}\label{f:vertical}
\end{figure}

The beams reach the arcs completely separated as a consequence of having low magnetic field vertical dipoles in the IR. This scheme also allows to consider that both rings will bend in the same direction with one on top of the other, therefore, minimize the footprint of the machine, see Fig.~\ref{f:vertical}, and allowing to use a second Interaction Region~(IR*) opposite to the location of the first one, that could be also considered a muon source in the case of a second positron source and target.\par
The length of the machine would benefit from having multiple IPs as it will reduce the \textit{length-over-IP-number} ratio, leading to an improved muon accumulation. In the case of this particular design we have 231~m/2~IPs for magnets at about 14~T, but the combination arc--IR could be repeated as many times as required because it reduces the peak magnet dipole field as the circumference grows.\par

\section{the accumulator}\label{s:accumulator}
The muon accumulator ring design, as described before, corresponds to the connection of the interaction region(s), the arcs, the RF cavity section and the extraction section. We use all previous optics designs and rematch some elements to properly adapt the twiss optics functions and correct the total chromaticity.\par
\begin{figure}[htb]
  \includegraphics[width=0.48\textwidth]{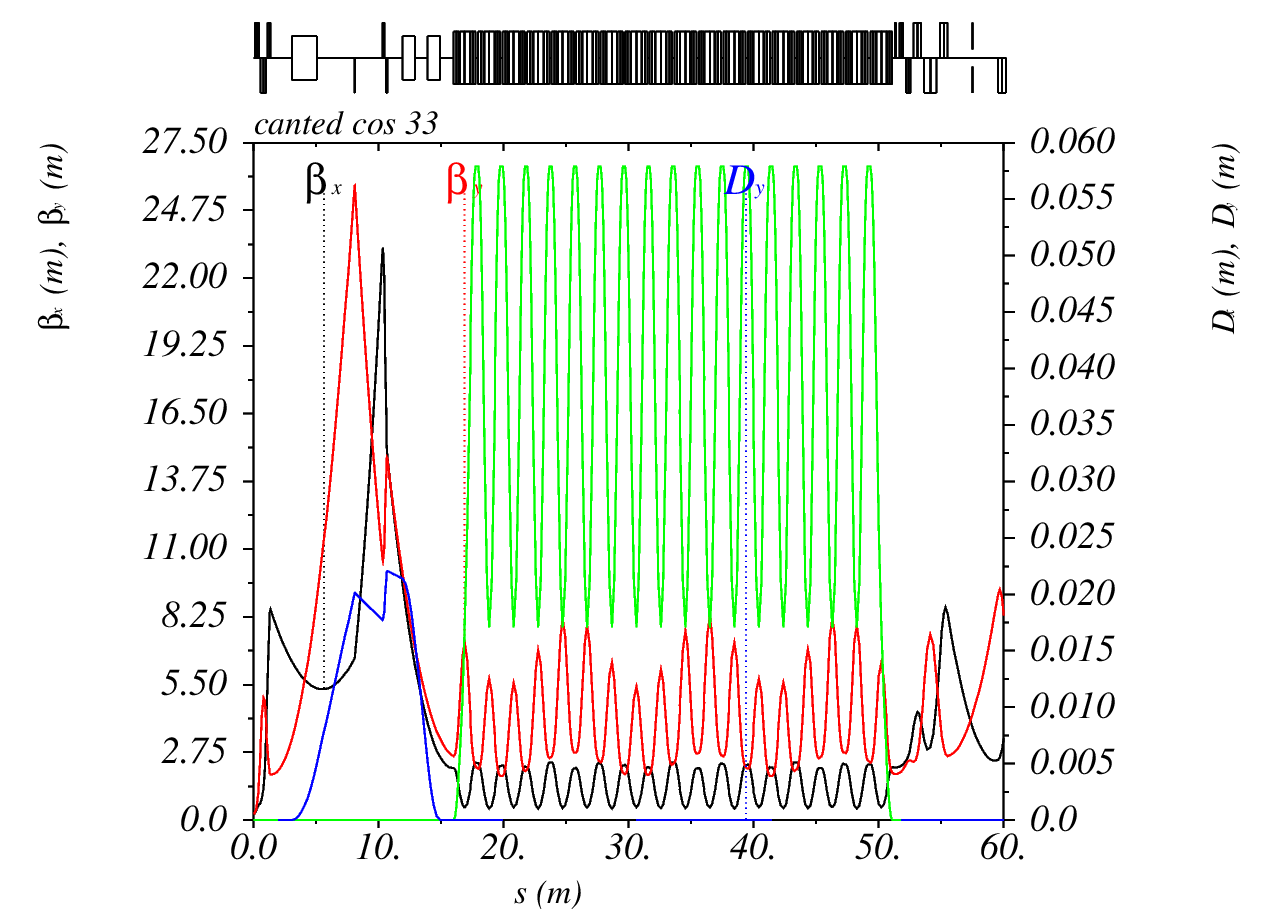}
  \caption{Half of the muon accumulator ring (IP, arc and one straight section used for RF cavities).}\label{f:hcantedcos33}
\end{figure}
\begin{figure*}[htb]
  \includegraphics[width=0.98\textwidth]{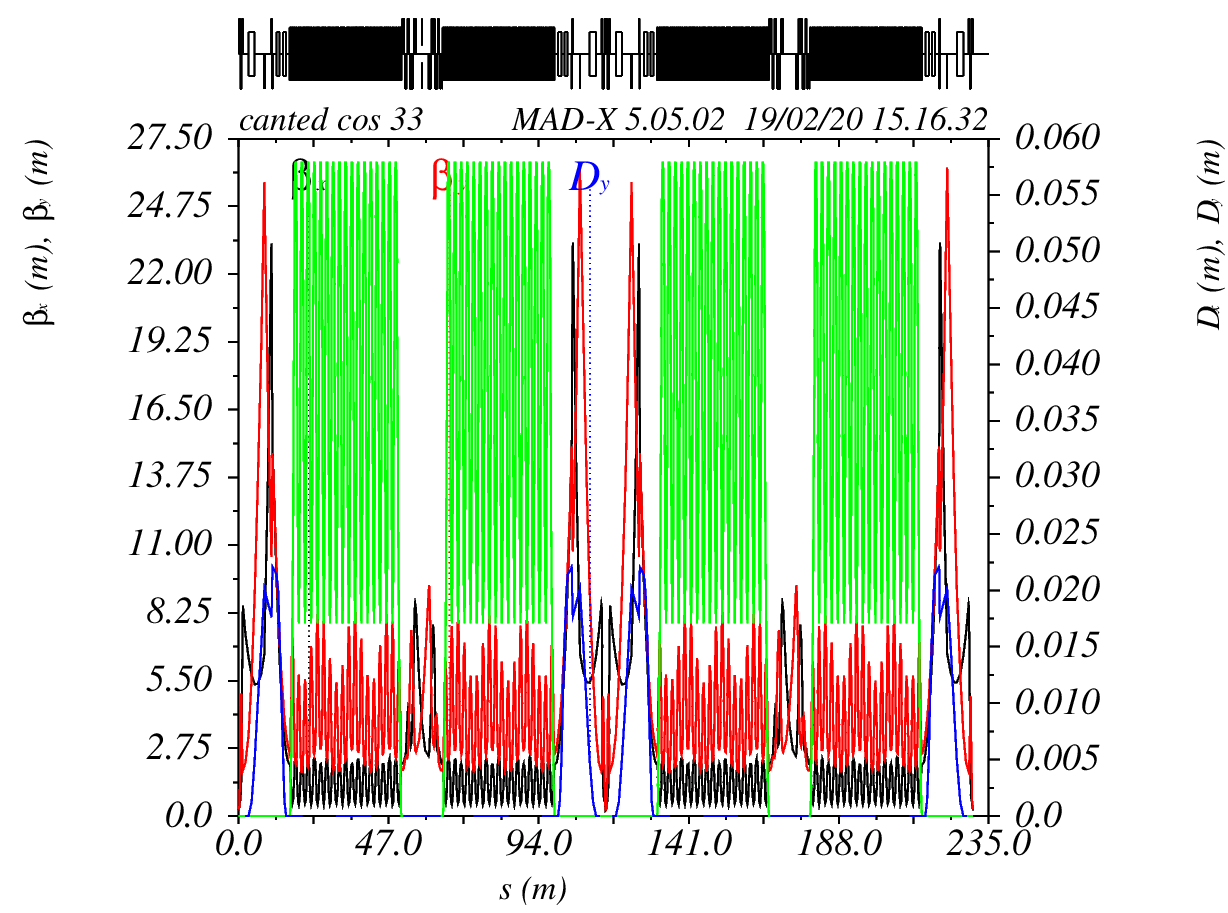}
  \caption{Accumulator Ring with arcs based on large energy acceptance FFA cells. The first and second IPs are at $s=0$~m and $s=115$~m, respectively, where the twiss beta functions are minimized to $\beta^*_\mu=20$~cm over $\pm$5\% energy spread for an $L^*=10$~cm. The RF section is located at $s=58$~m while the extraction section is located at $s=173$~m. Four FFA arcs join the insertions using combined function magnets at 15~T peak, foreseen to be of canted cosine type.}\label{f:cantedcos33}
\end{figure*}

Figure~\ref{f:cantedcos33} shows the twiss functions over the entire length of the accumulator ring while Table~\ref{t:muacc} shows the main parameters obtained with this optics. Table~\ref{t:designs} shows only the most relevant parameters in comparison to previous design stages. Figure~\ref{f:hcantedcos33} shows half of the accumulator ring.\par
\begin{figure}[htb]
  \includegraphics[width=0.48\textwidth]{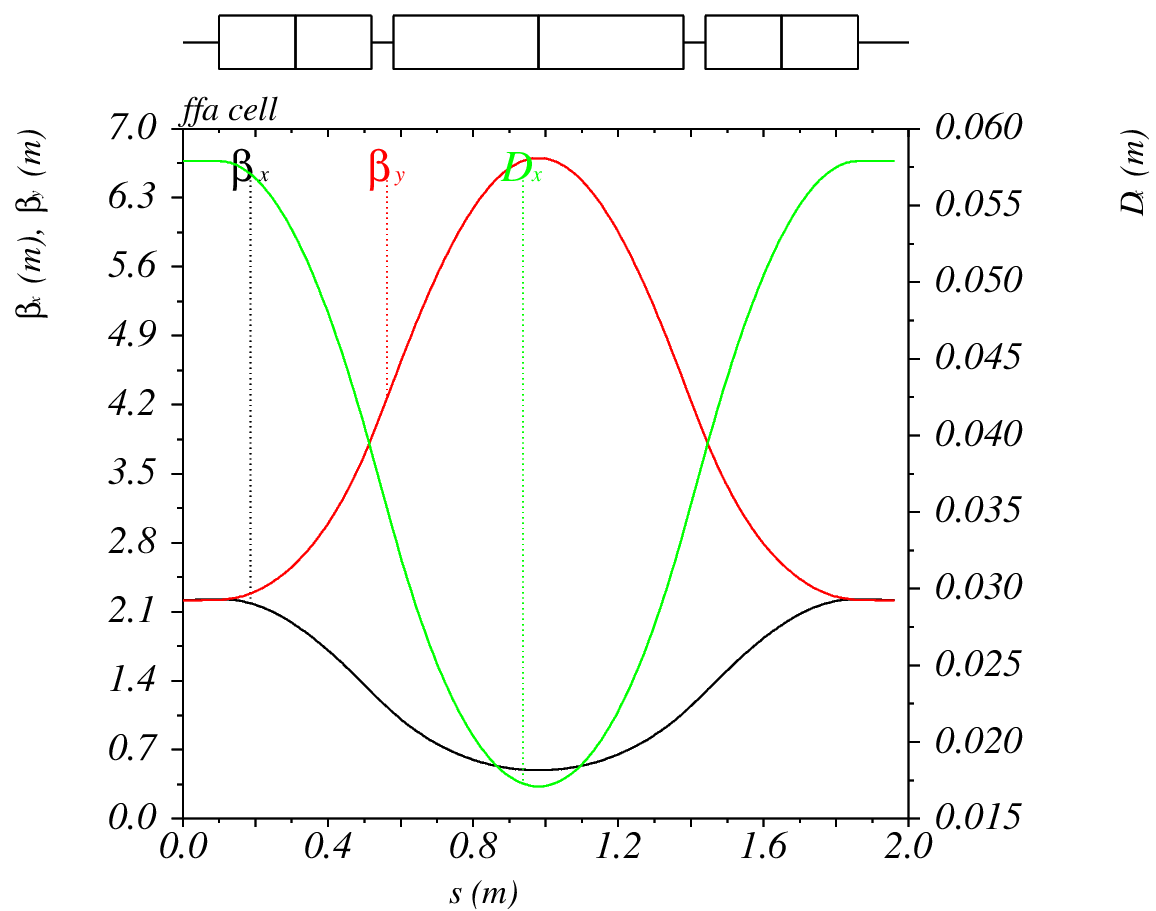}
  \caption{Arc cell resulting from the rematch of the accumulator ring with insertions and chromatic correction.}\label{f:cellaffa}
\end{figure}

 In Fig.~\ref{f:cellaffa} we show the arc cell after rematching dipole, quadrupole and sextupole fields.  For a 1~cm good field region the magnets are below 15~T, while the individual components are\par
\begin{itemize}
\item BF \;: 42~cm, -3.1~T,\;\;  238~T/m,\;\;\;   6.0~kT/m$^2$
\item BD : 80~cm, 12.0~T, -183~T/m, -10.3~kT/m$^2$
\end{itemize}
This cell sequence is composed by three BF--BD--BF magnets, where the distances between BD and BF magnets is 6~cm while the distance between BF magnets in adjacent cells is 20~cm.\par
The zero dispersion cell is composed by the magnets:
\begin{itemize}
\item B1 : 42~cm, 0.0~T,  \;238~T/m,      \;\;\;9.6~kT/m$^2$
\item B2 : 80~cm, 1.2~T, -183~T/m, -12.9~kT/m$^2$
\item B3 : 42~cm, 4.1~T,  \;238~T/m,      \;\;\;0.0~kT/m$^2$
\end{itemize}
\subsection{Survey}
\begin{figure*}[thb]
  \includegraphics[width=0.60\textwidth]{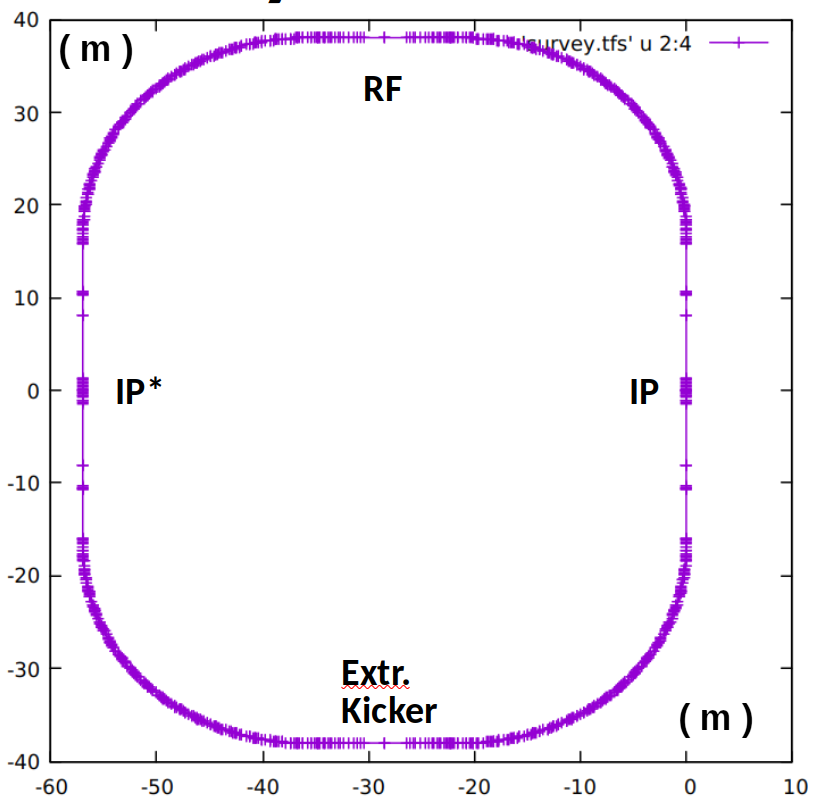}\\
  \includegraphics[width=0.60\textwidth]{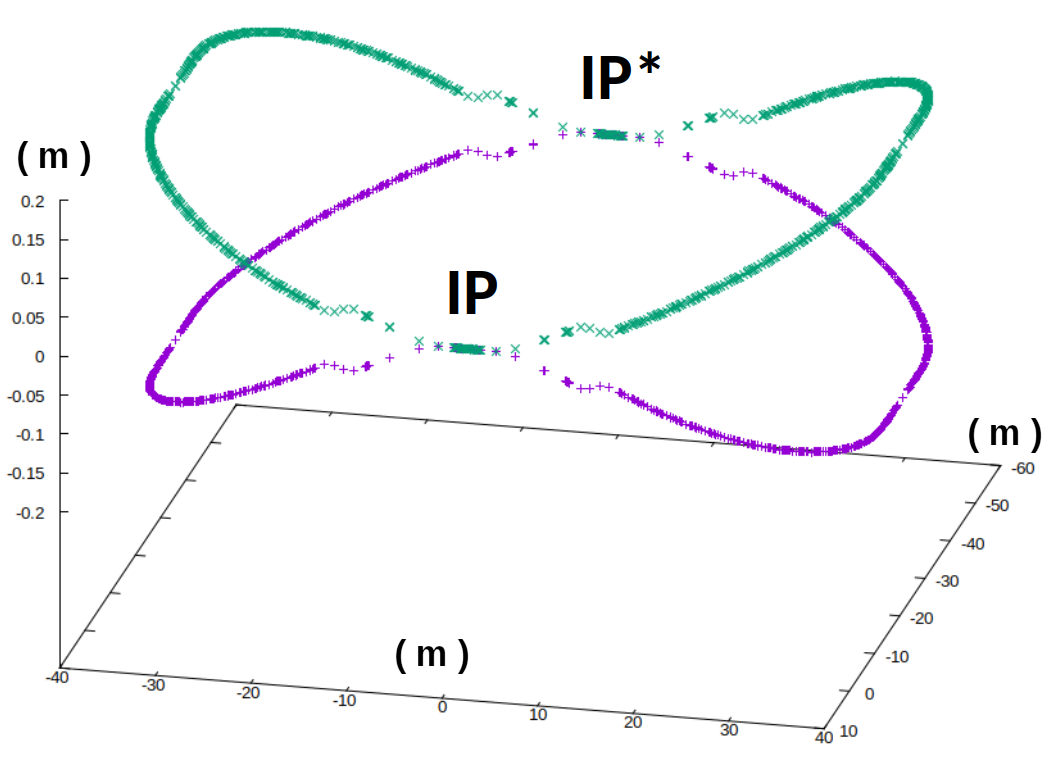}\\
  \includegraphics[width=0.60\textwidth]{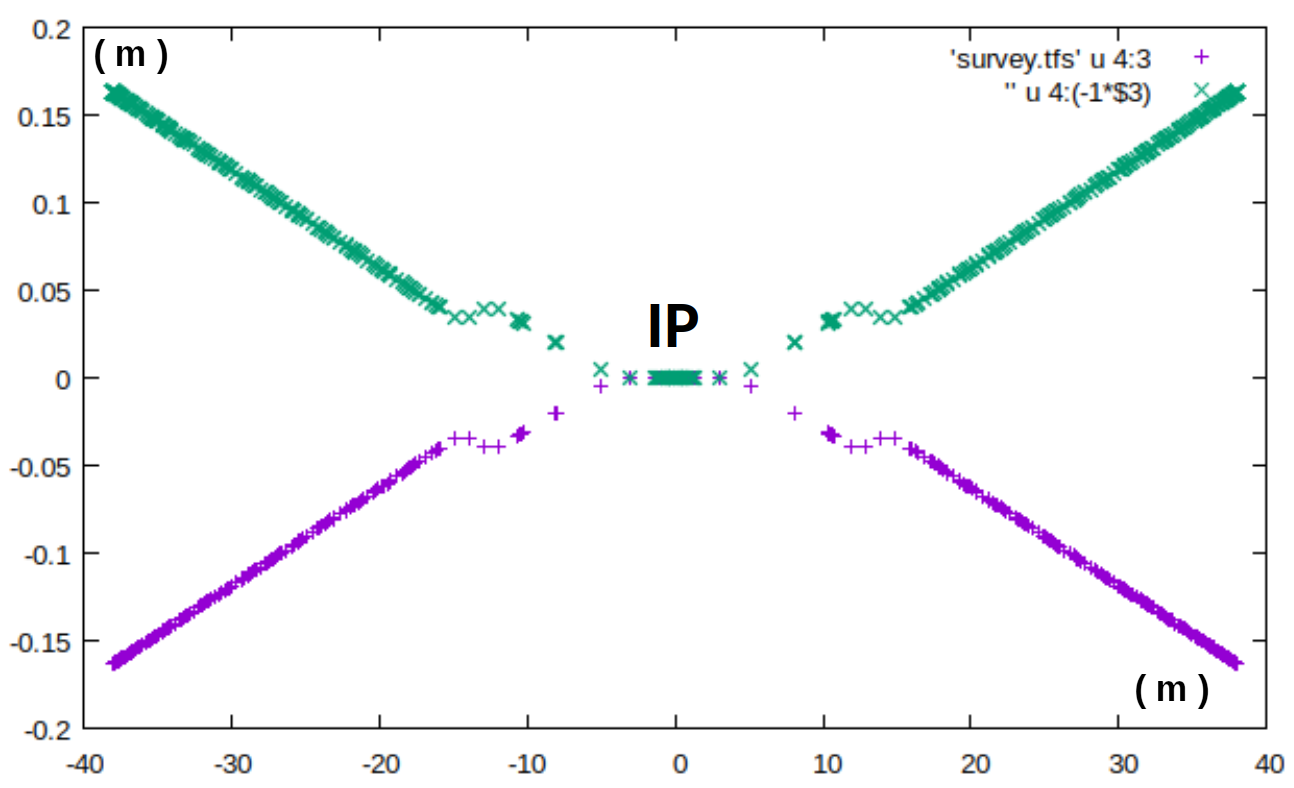}
  \caption{Accumulator ring survey, units in meters. (TOP) Top view, IP is at coordinate (0,0) (CENTER) Three dimensional representation of the Survey map of the two rings obtained from MAD-X and plotted with Gnuplot~\cite{Gnuplot}. (BOTTOM) Side view of the positive and negative muon rings separation.}\label{f:survey}
\end{figure*}

Figure~\ref{f:survey} shows the survey of the accumulator ring. The arcs are entirely separated for muon species while the interaction region is common to all three beams. The positron beam enters on one side of the IP, crosses the IP and passes in-between the two muon arcs.
\section{Beam size and tune footprint}\label{s:beamsizeandtune}
\begin{figure}[htb]
  \includegraphics[width=0.48\textwidth]{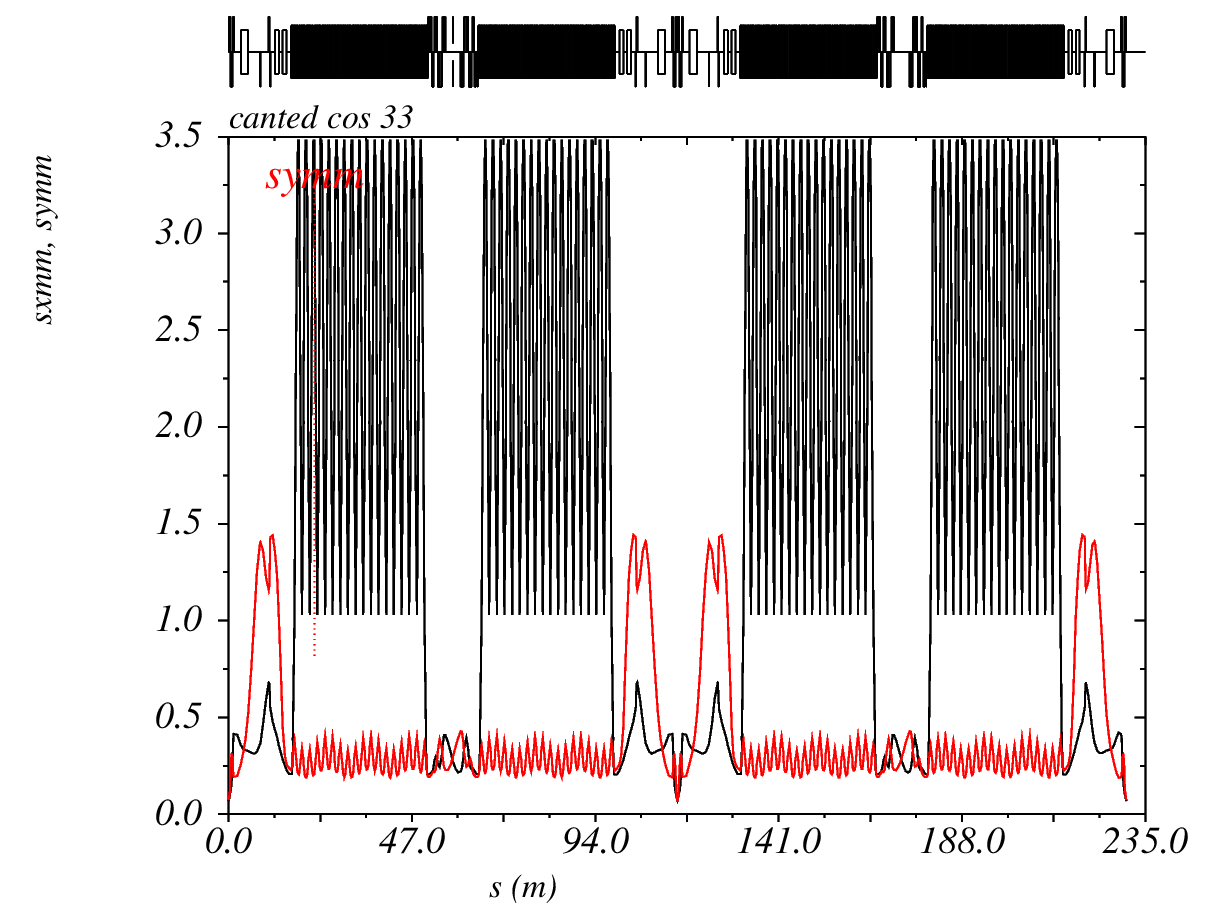}
  \caption{Beam size along the accumulator ring calculated as $\sigma = \sqrt{\epsilon\beta + \eta^2\delta^2}$ for 6\% energy spread and 20~$\pi$~nm~rad transverse emittance. Units in mm. The black line represents the horizontal plane while the red line represents the vertical. The beamsize at the target location is 60~$\mu$m in both planes.}\label{f:beamsize}
\end{figure}

Figure~\ref{f:beamsize} shows the expected beam size of a muon beam with a transverse emittance of 20~$\pi$~nm~rad and an energy spread of 6\%, calculated as $\sigma = \sqrt{\epsilon\beta + \eta^2\delta^2}$. One sigma of the beam relies well inside the good field region of $\pm$2~cm along the arcs, and it is much smaller than the 4~mm aperture at the inner triplet in the Interaction Region with magnets of 500~T/m. However, further studies are needed to determine if the stay clear criteria of 10$\sigma$ is required.\par
\begin{figure}[htb]
  \includegraphics[width=0.48\textwidth]{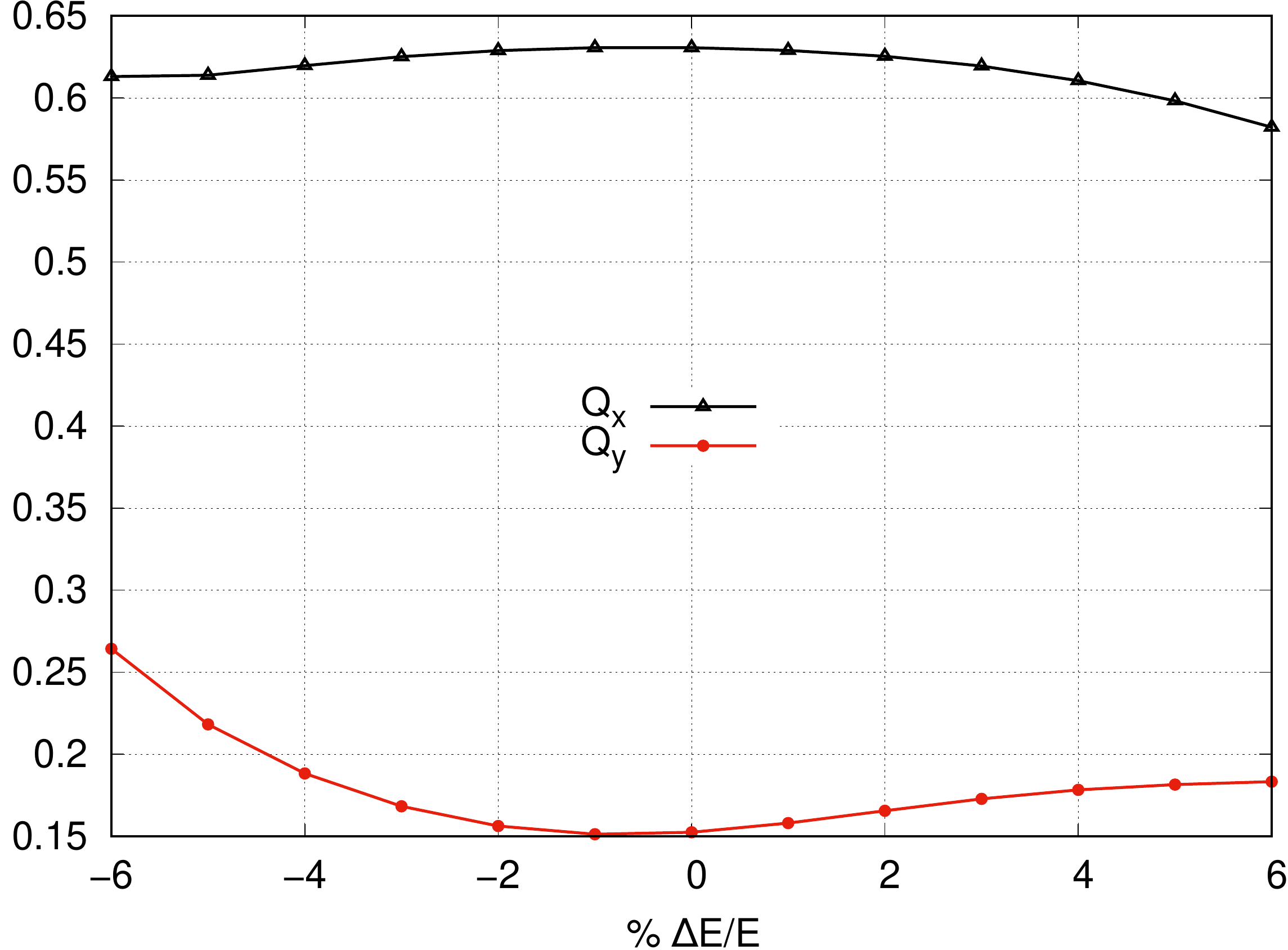}\\
  \includegraphics[width=0.48\textwidth]{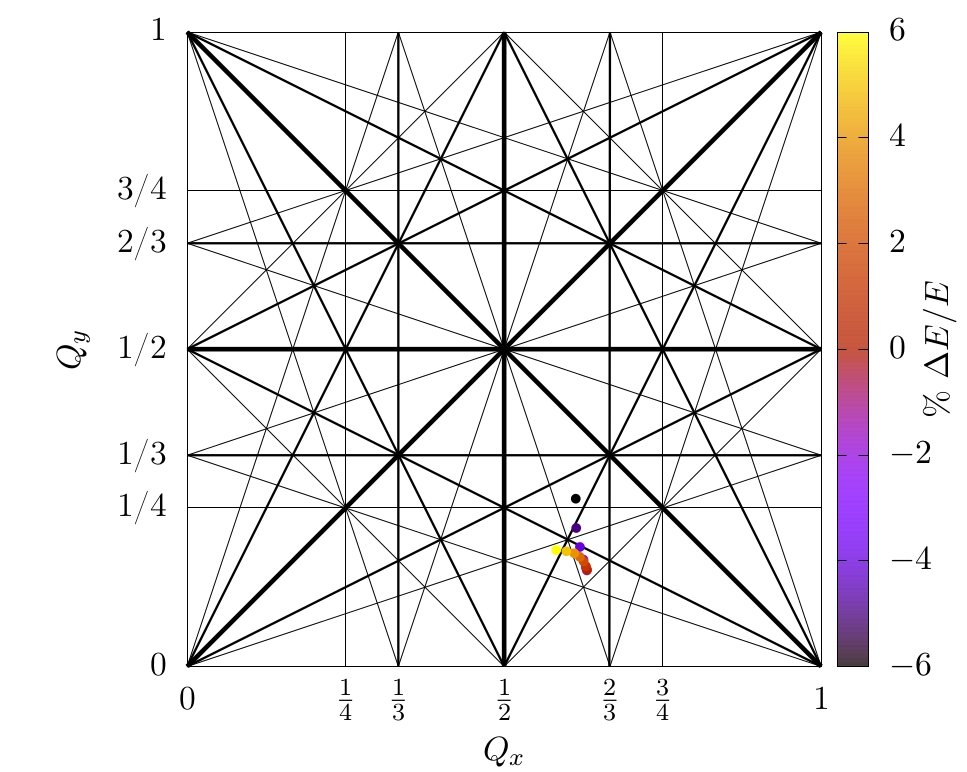}  
  \caption{Accumulator Ring tune (TOP) as a function of energy offset, (BOTTOM) footprint due to energy offsets.}\label{f:tunes}
\end{figure}

Figure~\ref{f:tunes} shows the tune footprint and the tune as a function of the energy for the accumulator. The particular tune chosen up to know is completely arbitrary, it must be moved to a region more distant from third and fourth order resonances, which we believe could be achieved by a rematch of the cell or increasing the number of cells in the arcs. However, the point of this plot is to show the behavior of the chromatic correction on the entire energy range.\par
\begin{figure}[htb]
  \includegraphics[width=0.48\textwidth]{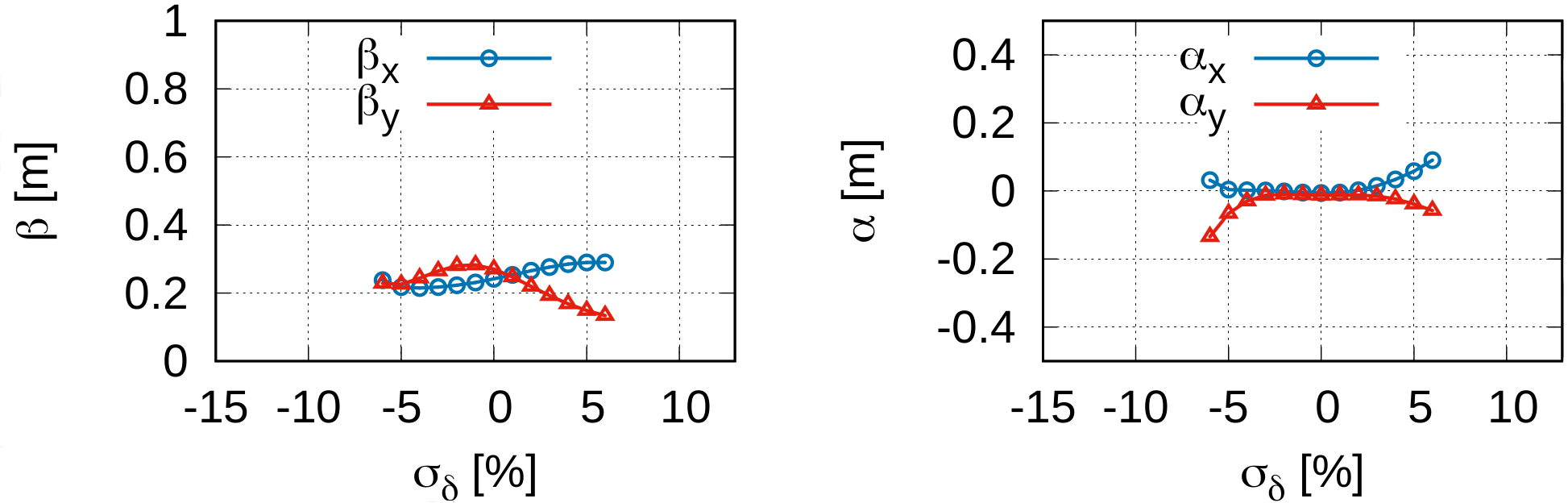}
  \caption{Accumulator Ring twiss $\beta^*_\mu$ and $\alpha^*_\mu$ vs energy offset ranging $\pm$6\% of the nominal energy. The optics effectively focuses the beam over the expected energy range.}\label{f:betastar}
\end{figure}

Figure~\ref{f:betastar} shows the variation of the twiss functions $\beta^*_\mu$ and $\alpha^*_\mu$ at the interaction point with the target. It shows nearly constant values along the expected energy range of $\pm$6\%. Further studies could be addressed to optimize these parameters, but the results from this first design are satisfactory.\par

\section{MUON ACCUMULATION STUDIES}\label{s:accumulation}
The muon transverse geometrical emittance, $\epsilon_{\mu}$,  for a thin target is
\begin{equation}
  \epsilon_{\mu} = \sigma_{e^+}\cdot\sigma'_{\mu}(E_{e^+})\label{eq:muemit},
\end{equation}
i.e. the product of the positron beam size $\sigma_{e^+}$ when crossing the target and the divergence of the produced muon beam $\sigma'_{\mu}$ from the kinematics of the target collision with a positron beam at some energy $E_{e^+}$. One could approximate the divergence to be 0.1 to 0.5~mrad for a positron beam at 43.72 to 45~GeV, respectively. Table~\ref{t:posEmusigmadelta} contains the main parameters of the muon production obtained from simulations with MUFASA~\cite{ciarma}, bench-marked with Geant4~\cite{geant}.\par
First, we calculate the muon beam size and emittance that matches the optics, and then will use those parameters to fix the requirements on the positron beam, as in~\cite{blancoprab2020}.\par
Given an energy acceptance of at least $\pm$5\% the muon energy and the results listed in the Table~\ref{t:posEmusigmadelta}, we can expect a beam divergence close to 0.3~mrad and energy spread close to $\pm$5\% from the kinematics of collision with an $e^+$ beam at $E_{e^+}=$43.8~GeV. The muon beam size at the target that matches the optics can be calculated as,
\begin{equation}
  \sigma_\mu = \beta_\mu^* \sigma'_\mu = 0.2\text{~m} \cdot 0.3\text{~mrad}=60~\mu\text{m}.
\end{equation}
Now, we can calculate the geometrical transverse emittance of the produced muon beam as
\begin{equation}
  \epsilon_\mu = \sigma_\mu\sigma'_\mu = 60~\mu\text{m}\cdot0.3\text{~mrad} = 18~\pi~\text{nm~rad}.
\end{equation}
We assume that the positron beam at $E_{e^+}=43.8$~GeV, has the same beam size given by Eq.~(\ref{eq:muemit}), $\sigma_{e^+}=60~\mu$m.  This value is above the minimum used in \cite{alesini2019positron} for simulations of power deposition in the target by a positron bunch population of $3\times10^{11}$  particles.\par
As a free parameter to match the optics $\beta^*_{e^+}$ we have the positron beam emittance. Thus,
\begin{equation}
\epsilon_{e^+}=\frac{\sigma_{e^+}^2}{\beta^*_{e^+}}=\frac{(60~\mu\text{m})^2}{0.5~\text{m}}=7.2~\pi~\text{nm~rad},
\end{equation}
which is similar to the 6~\pinmrad\; studied for the positron ring in~\cite{Boscolo:IPAC17-WEOBA3}.\par
Table~\ref{t:beamparams} summarizes the list of beam parameters obtained, where we also include the positron ring energy spread given in~\cite{Boscolo:IPAC17-WEOBA3}.\par
{\renewcommand{\arraystretch}{0.5}
  \begin{table}[htp]
    \caption{Positron and muon beam parameters matching the low beta $\beta^*_\mu=20$~cm of the interaction region design.}\label{t:beamparams}
    \begin{tabular}{ccccccc}\hline\hline
      Particle & Energy & $\epsilon$ & $\sigma$ & $\sigma'$ & \multicolumn{2}{c}{$\delta E$}\\
      & (GeV) & ($\pi$~nm~rad) & ($\mu$m) & (mrad) & (\%) & type\\\hline
      $e^+$ & 43.8 & \;\;7 & 60 & 0.12 & $\pm$0.1 & (Gaus)\\
      $\mu$\;\; & 21.9 &  18 & 60 & 0.30 & $\pm$5\;\;\;& (Flat)\\\hline\hline
    \end{tabular}
  \end{table}
}
{\renewcommand{\arraystretch}{0.5}
  \begin{table*}[htp]
    \caption{Muon production efficiency $eff$, divergence $\sigma'_\mu$ and energy spread $\delta E_\mu$ as a function of the positron beam energy $E_{e^+}$. The muon energy and divergence have a flat distribution. Efficiency has been estimated for 1\% of a radiation length of Beryllium~(3.5~mm of material). The muon beam divergence and energy spread are given by the collision kinematics.}\label{t:posEmusigmadelta}
  \begin{tabular}{lccccccc}\hline\hline
    $E_{e^+}$ & $eff$ & \multicolumn{2}{c}{$\sigma'_\mu$} &$E_\mu$&\multicolumn{2}{c}{$\delta E_{\mu}$} & $E_{\mu}(1\pm\delta E_{\mu})$\\
    (GeV) & ($10^{-8}\mu/e^+$) & (mrad$_{rms}$) & (mrad$_{max}$) & (GeV) & (\%$_{rms}$) & (\%$_{max}$) & Min/Max (GeV)\\\hline
    43.72 & \;\;2.3 & 0.07 & 0.16      & 21.860 & \;\;1.40 & \;\;2.9  & 21.20/22.50\\
    43.80 & \;\;3.0 & 0.14 & 0.26      & 21.900 & \;\;2.86 & \;\;4.9 & 20.82/22.98\\
    44.00 & \;\;4.6 & 0.23 & 0.42  & 22.000 & \;\;4.81 &   \;\;8.3 & 20.17/23.83\\
    44.50 & \;\;7.1 & 0.37 & 0.67  & 22.250 & \;\;7.85 &     13.6 & 19.44/25.56\\
    45.00 & \;\;8.7 & 0.47 & 0.83  & 22.500 & \;\;9.90 &     17.0 & 18.67/26.33\\
    46.00 & 10.9 & 0.62 & 1.11  & 23.000 &   12.91 &    22.6 & 17.80/28.19\\\hline\hline
  \end{tabular}
  \end{table*}
}
The muon longitudinal emittance comes from the product of the positron bunch length and the energy spread produced by the kinematics of the production. We assume here the previously studied parameters of a 6 or 27~km long positron ring in Refs.~\cite{eplusringopt,simone} with a bunch length of 3~mm, giving a longitudinal emittance at production in the order of $6.8$~$\pi$~mm~GeV (3$\times$0.10$\times$45/2~$\pi$~mm~GeV) and could be as low as 0.7~$\pi$~mm~GeV (3$\times$0.01$\times$43.72/2~$\pi$~mm~GeV). However, we remark that it is much smaller than the bunch length obtained from the optics studies, therefore the longitudinal emittance is determined by the accumulator.\par
With respect to the muon beam energy, the amount of synchrotron radiation along the accumulator is negligible while the energy loss from interacting with the target is small for a thin target and only becomes significant when passing several radiation lengths of material. The energy spread of the outgoing muon beam ranges from 1 to 10\% of half the positron beam energy in the range of 43.72 to 45~GeV, respectively (See Table~\ref{t:posEmusigmadelta}).\par
With respect to the muon life time at 22~GeV, we will be able to perform at most 520 turns around the accumulator. We will not go into details of muon decay and energy deposition in this article.\par
The muon pair production efficiency $eff$ has been defined in~\cite{NIM} as the ratio of muon pairs~($\mu^\pm$, or simply~$\mu$) per positron impinging on a fixed target. See Table~\ref{t:posEmusigmadelta}.\par
\subsection{Beam target interaction simulation}
In order to accelerate the calculation time, a dedicated fast Monte Carlo called MUFASA, written in CERN-ROOT~\cite{cernroot}, has been benched-marked with Geant4 and MDISim~\cite{Burkhardt:IPAC15-TUPTY031}, and has proved to be very useful and flexible to simulate the muon production and to study the dynamics of the stored muon beam which interacts at each turn with the target. The simulation starts by generating a positron beam population that interacts with the material of the target. Then, the outgoing positrons and produced muons are tracked through the lattice using MAD-X~PTC and the output distribution is saved to calculate positron and muon beam populations, emittances, beams divergence and beam size. We repeat the previous process over many turns until the muon population does not increase significantly.\par
\subsection{Simulation results}
We will concentrate in a positron beam at 43.8~GeV interacting with 1\%$X_0$ of Beryllium material, giving a production efficiency $eff=3\times10^{-8}$~$\mu$/$e^+$.\par
\begin{figure}[htb]
  \includegraphics[width=0.48\textwidth]{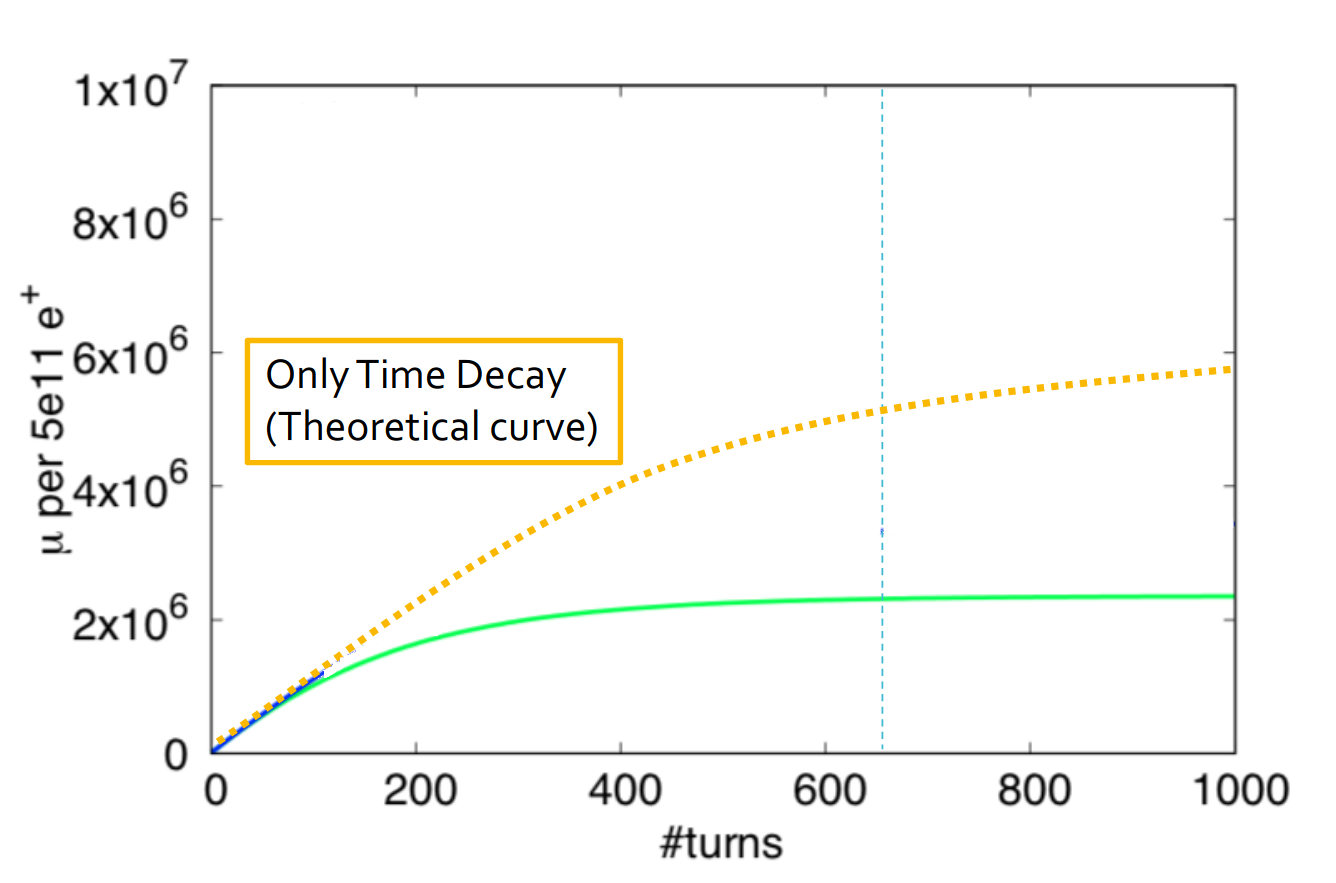}
  \caption{Muon accumulation for a positron beam at 43.8~GeV impinging on 0.01$X_0$ of Beryllium as target material. The population increases as expected in the first hundreds of turns, after that the muon beam population grows slowly because of particle losses produced by the multiple scattering with the target, exceeding the dynamic aperture.}\label{f:muonaccumulation}
\end{figure}

Figure~\ref{f:muonaccumulation} shows the muon accumulation over several hundreds of turns in the accumulator ring. A positron bunch population of $5\times10^{11}$ produces $1.5\times10^4$ muon pairs per interaction. The first hundred turns increase the population as expected to about $1.5\times10^6$, but after that the growth slows down because particles undergo multiple scattering interacting with the target leaving the dynamic aperture of the machine. After 400~turns the accumulation saturates, as new muon only replace the amount leaving the stable phase space.\par
\begin{figure}[htb]
  \includegraphics[width=0.45\textwidth]{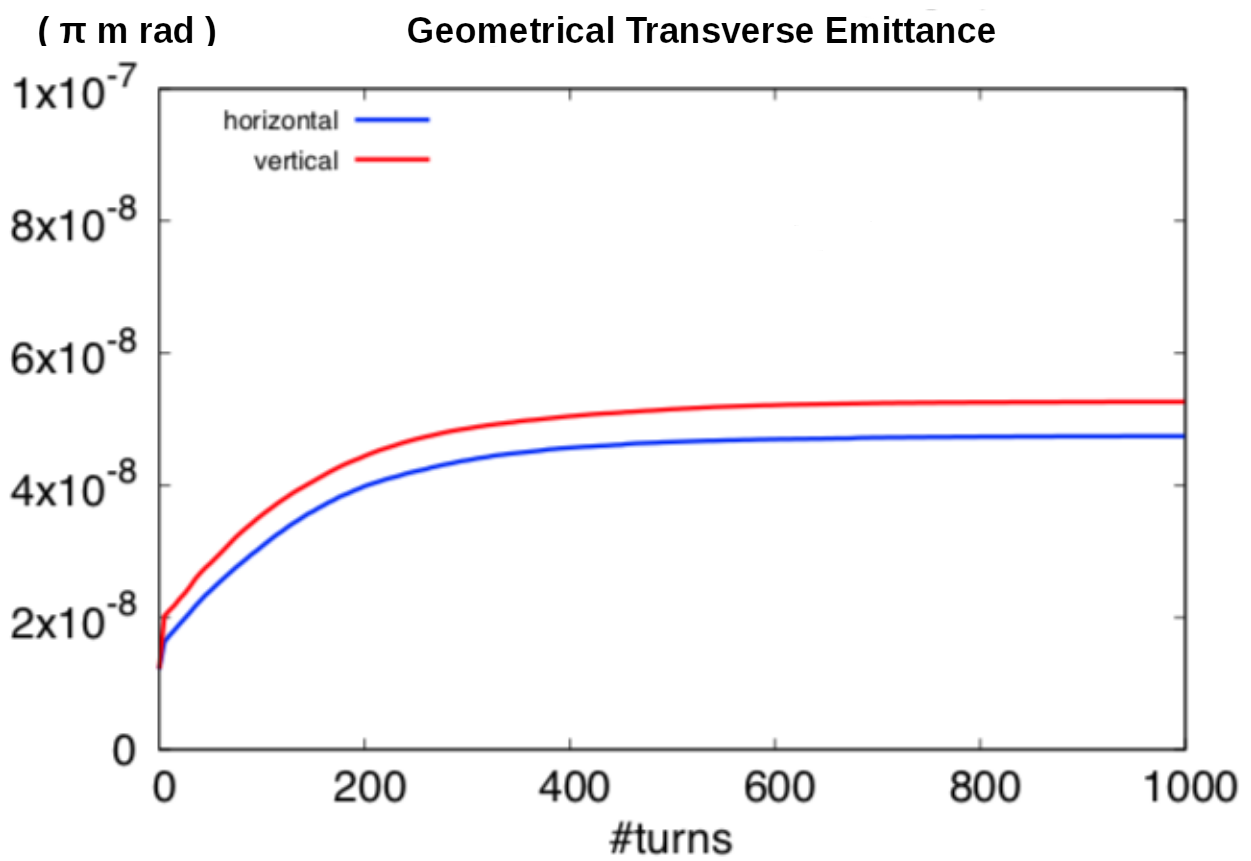}\\
  \includegraphics[width=0.45\textwidth]{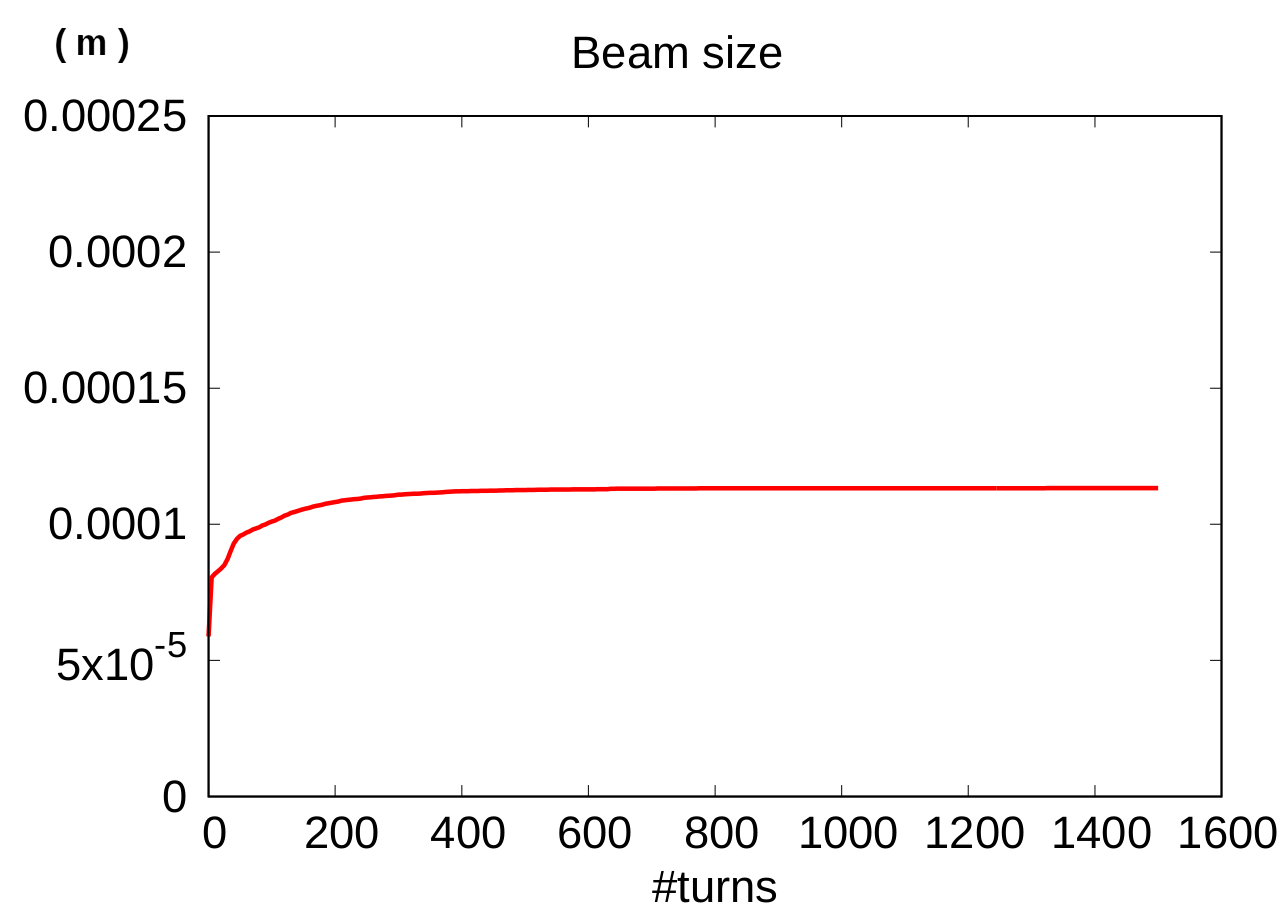}\\
  \includegraphics[width=0.45\textwidth]{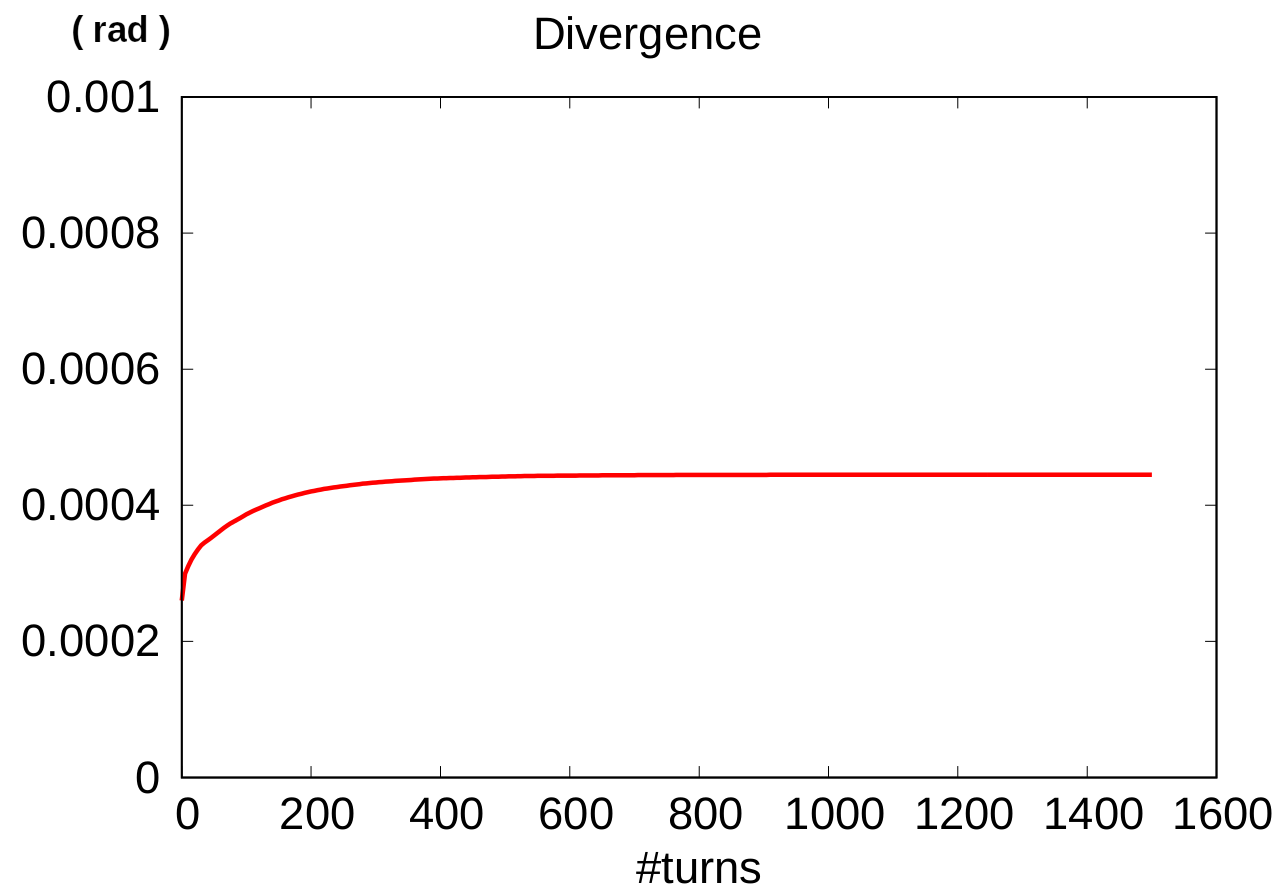}
  \caption{(TOP) Transverse muon beam emittance as a function of the accumulator turns for a $\beta^*_\mu=20$~cm. The emittance grows in the first hundreds of turns due to multiple scattering, after that, accumulation saturates due to the limiting dynamic aperture of the machine. (CENTER) Horizontal beam size (BOTTOM) Horizontal divergence.}\label{f:muonemittanceturns}
\end{figure}

Figure~\ref{f:muonemittanceturns} show the emittance growth as a function of the accumulator turns, as well as the beam dimensions in the horizontal plane. The emittance grows slowly and proportionally to the number of turns around the accumulator ring due to multiple scattering with the target. After a few hundred turns the emittance growth is limited by the dynamic aperture of the machine. The final normalized transverse emittance is approximately $10$~$\pi$~$\mu$m (200$\times$50~$\pi$~nm).\par
\begin{figure}[htb]
  \includegraphics[width=0.45\textwidth]{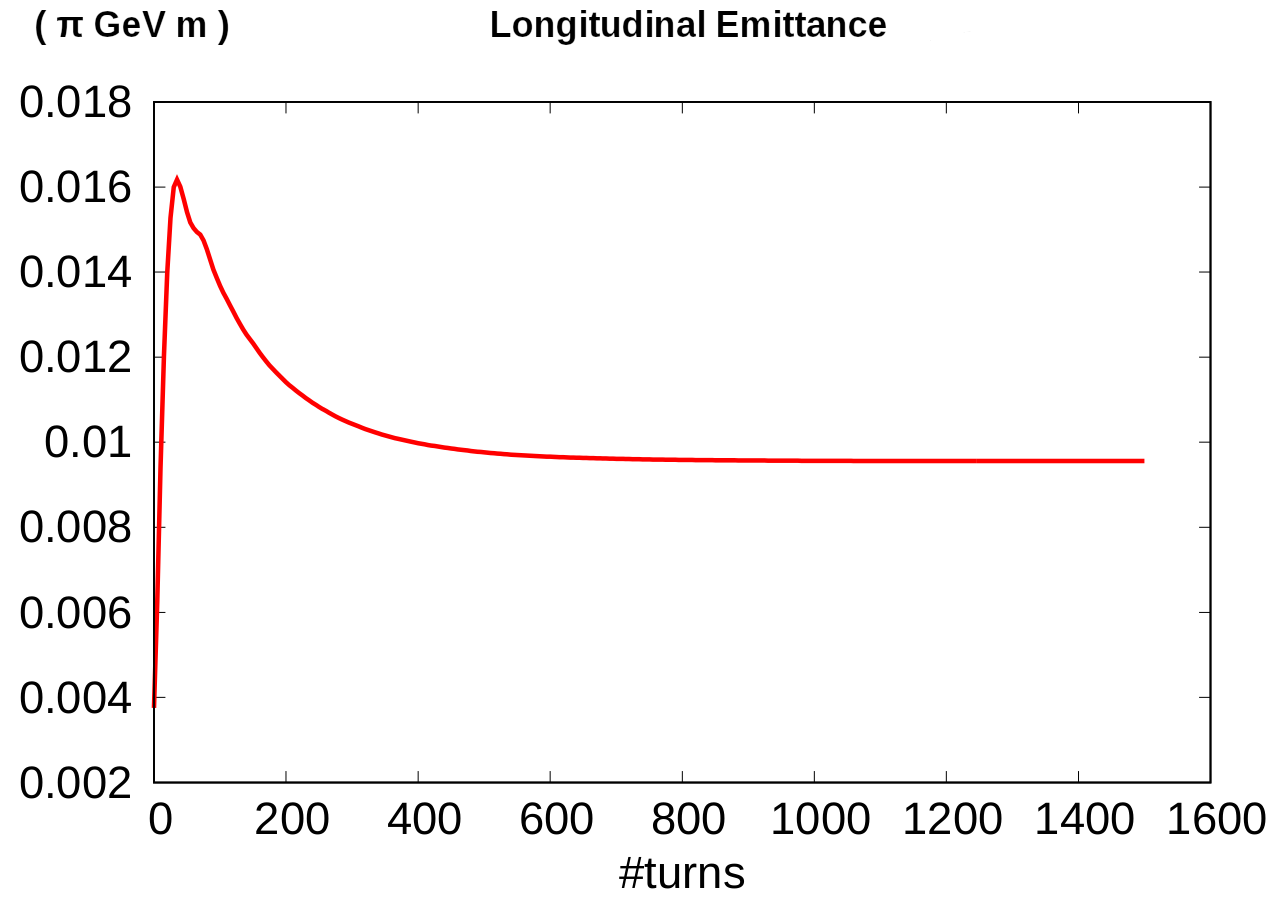}\\
  \includegraphics[width=0.45\textwidth]{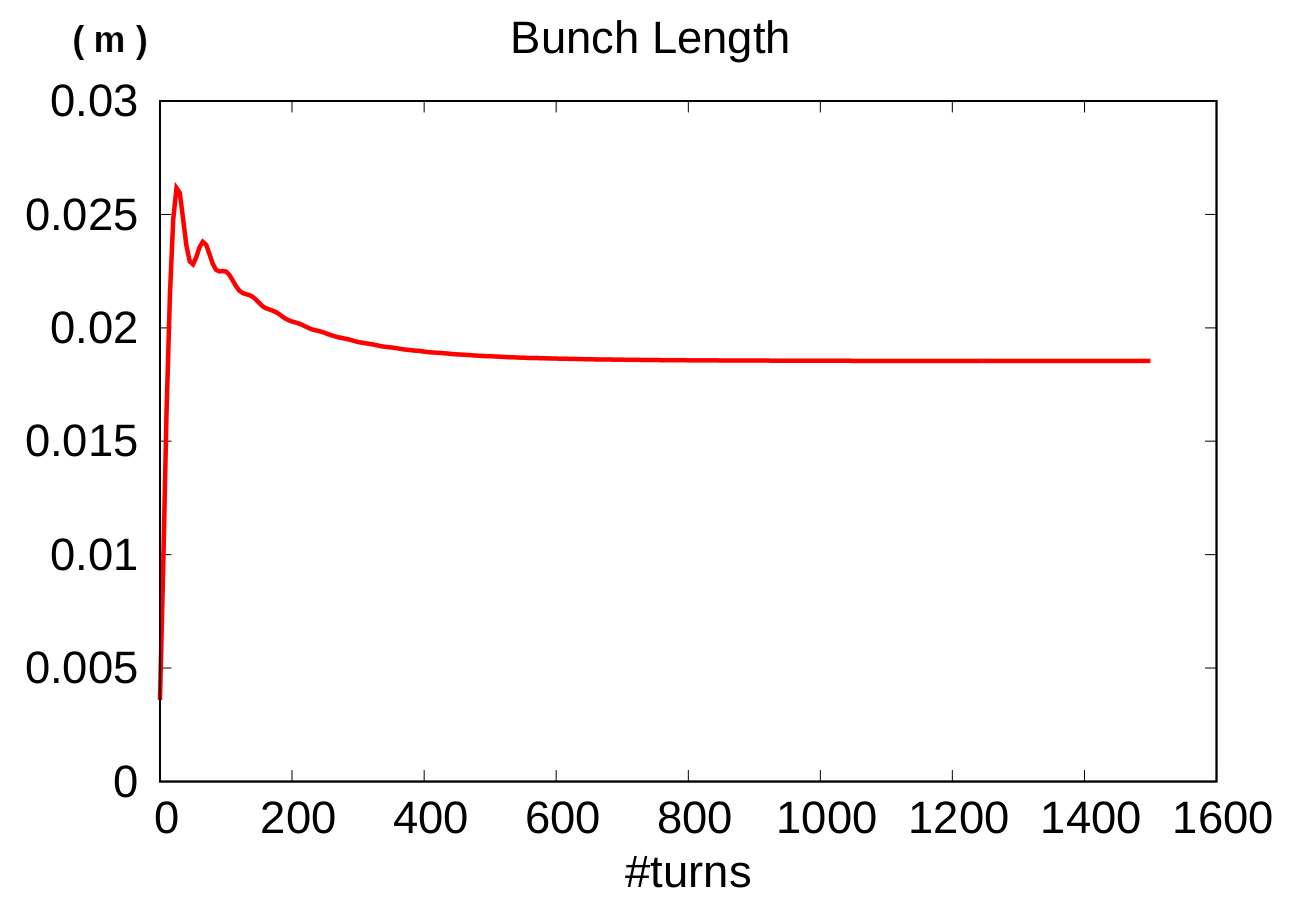}\\
  \includegraphics[width=0.45\textwidth]{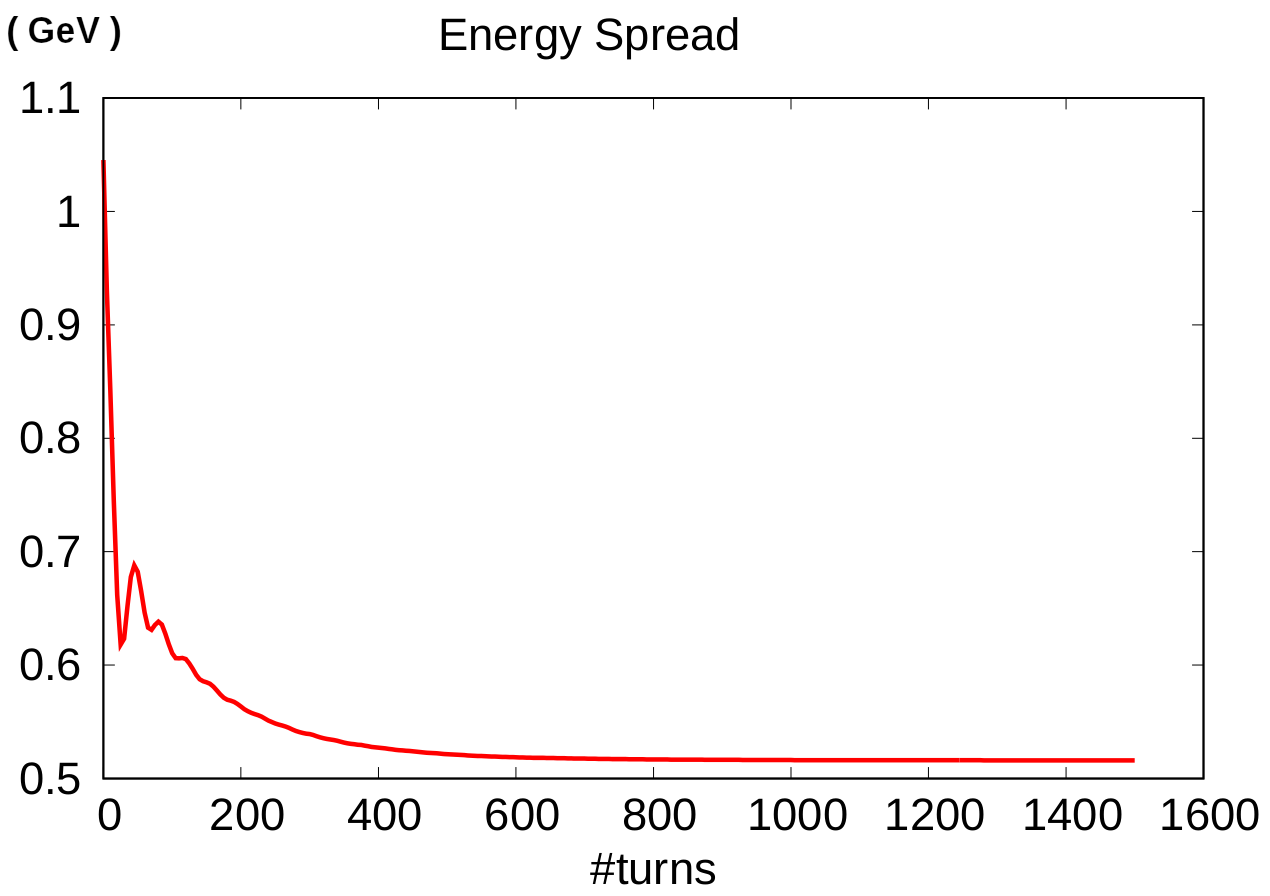}
  \caption{Muon beam (TOP) longitudinal emittance, (CENTER) bunch length and (BOTTOM) energy spread.}\label{f:bunchlength}
\end{figure}

Figure~\ref{f:bunchlength} shows the bunch length, energy spread and longitudinal emittance during the accumulation period. The final emittance is approximately 10~$\pi$~GeV~mm, given by the product of a 2~cm bunch length and 0.5~GeV of rms energy spread.\par
\begin{figure}[htb]
  \includegraphics[width=0.48\textwidth]{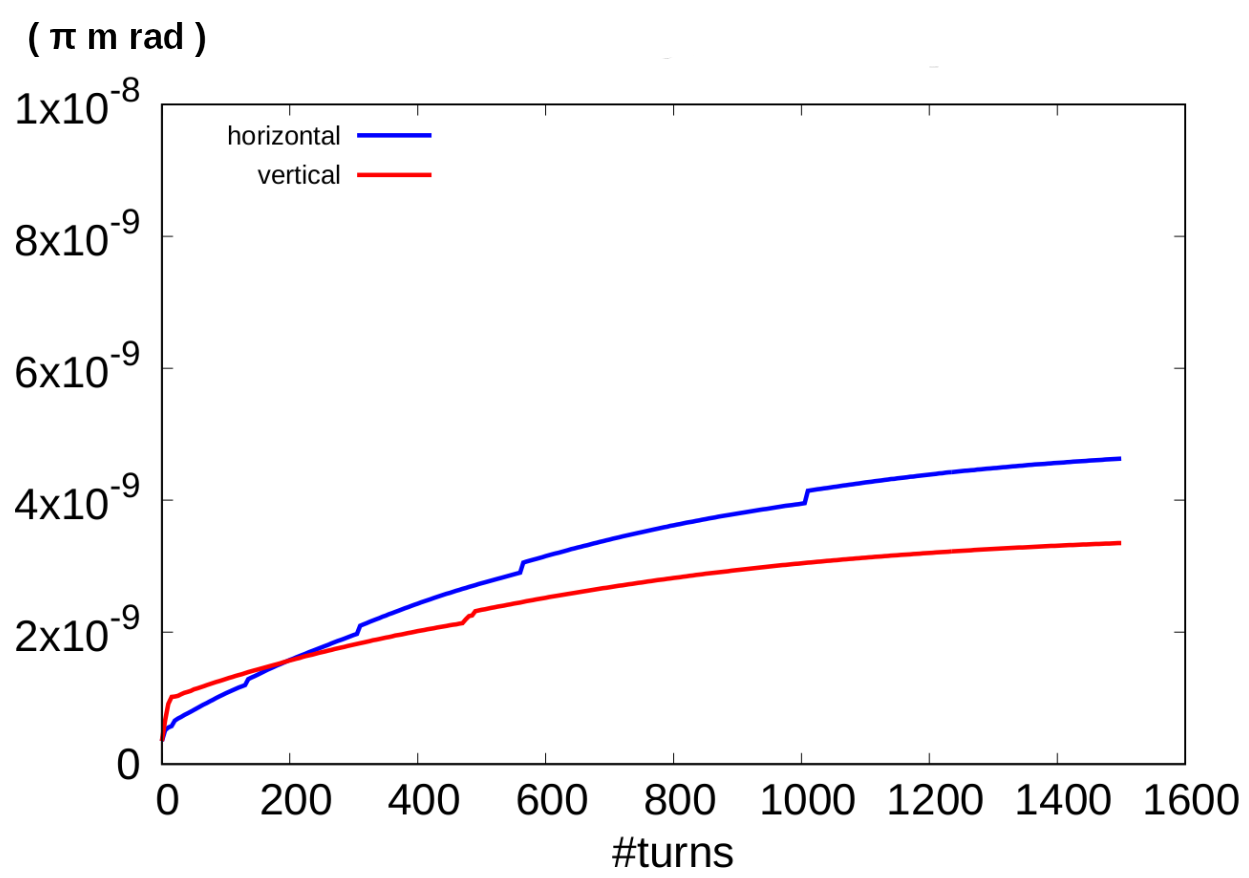}
  \caption{Muon beam emittance as a function of the accumulator turns for a $\beta^*_\mu=1$~cm. The emittance growth is reduced and the beam remains always inside the dynamic aperture of the lattice.}\label{f:turnsbetastar1cm}
\end{figure}

We have also tested the theoretical mitigation of multiple scattering with an ultra low $\beta^*_\mu=1$~cm. Figure~\ref{f:turnsbetastar1cm} shows the beam emittance for a $\beta^*_\mu=1$~cm, with the design shown in Section~\ref{ss:ir}. The emittance growth is much smaller due to the ultra-low $\beta^*_\mu$ and it always remains smaller than the dynamic aperture of the machine.

\subsection{Materials other than Beryllium}
One possible way to increase the muon production efficiency is to consider other materials for the target. The transverse and longitudinal emittance depend on the fraction of radiation length of material interacting with the muon beam. Therefore, using the same 1\%$X_0$ units for other materials, we can have the same beam dynamics while changing only the muon production efficiency. Table~\ref{tab:Ceff} shows a small list of possible targets with different lengths and efficiencies that have been foreseen.\par
In particular, we highlight the possibility to use 1\%$X_0$ of liquid Hydrogen~($H_2$) which will increase the muon population by a factor 2 with respect to Be. The target dimensions fit the 20~cm ($2 L^*$) available space among quadrupoles close to the interaction point. A possible small mismatch of the optics due to the target length seems negligible because the beta function grows just to $25$~cm at the entry point of the first quadrupole magnet.\par
Despite of the advantages of liquid Hydrogen regarding the muon production efficiency, there are materials that could become more practical solutions as they withstand better the power deposition or have dimensions that better suit the short space for beam--target interaction. Therefore, in Table~\ref{tab:Ceff} we include also Carbon, a carbon based composite C/C~A412~~\cite{maciariello}, liquid Lithium~(LLi), and a mix of 10\%~liquid Lithium and 90\%~Diamond Powder~(LLi-D)~\cite{PhysRevAccelBeams.23.051001}.
We note that the emittance obtained with these materials is equal to the case of Beryllium when using the same amount of radiation lengths of material.\par
\begin{figure}[htb]
  \includegraphics[width=0.48\textwidth]{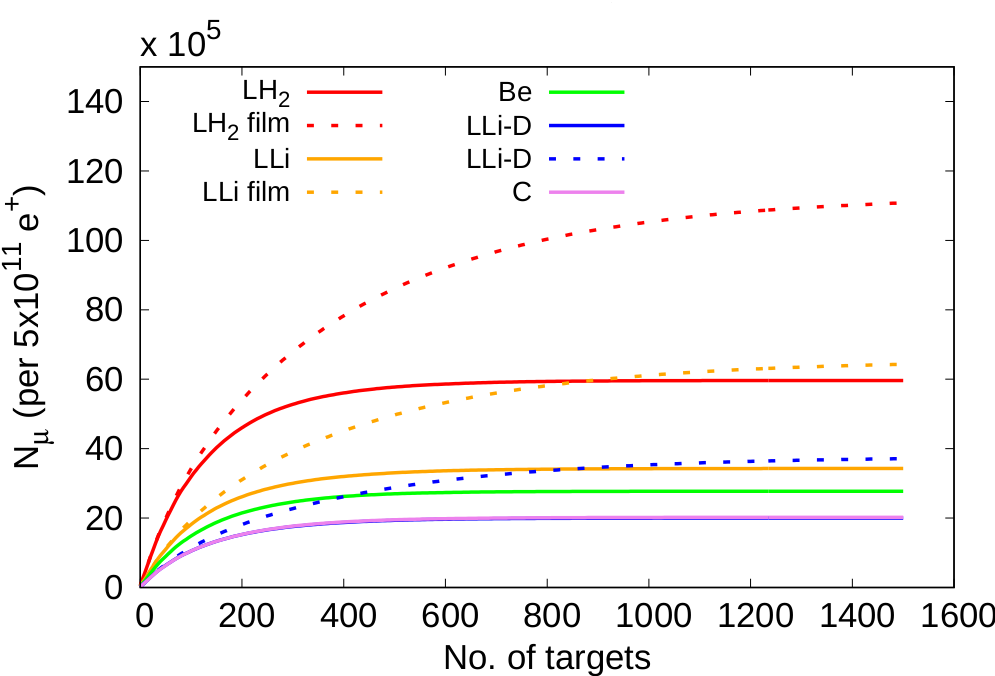}
  \caption{Muon beam accumulation with 1\% of a radiation length for different target materials interacting with a positron beam at 44~GeV with $5\times10^{11}$~$e^+$ per bunch.}\label{f:othermaterials}
\end{figure}
\begin{figure}[htb]
  \includegraphics[width=0.48\textwidth]{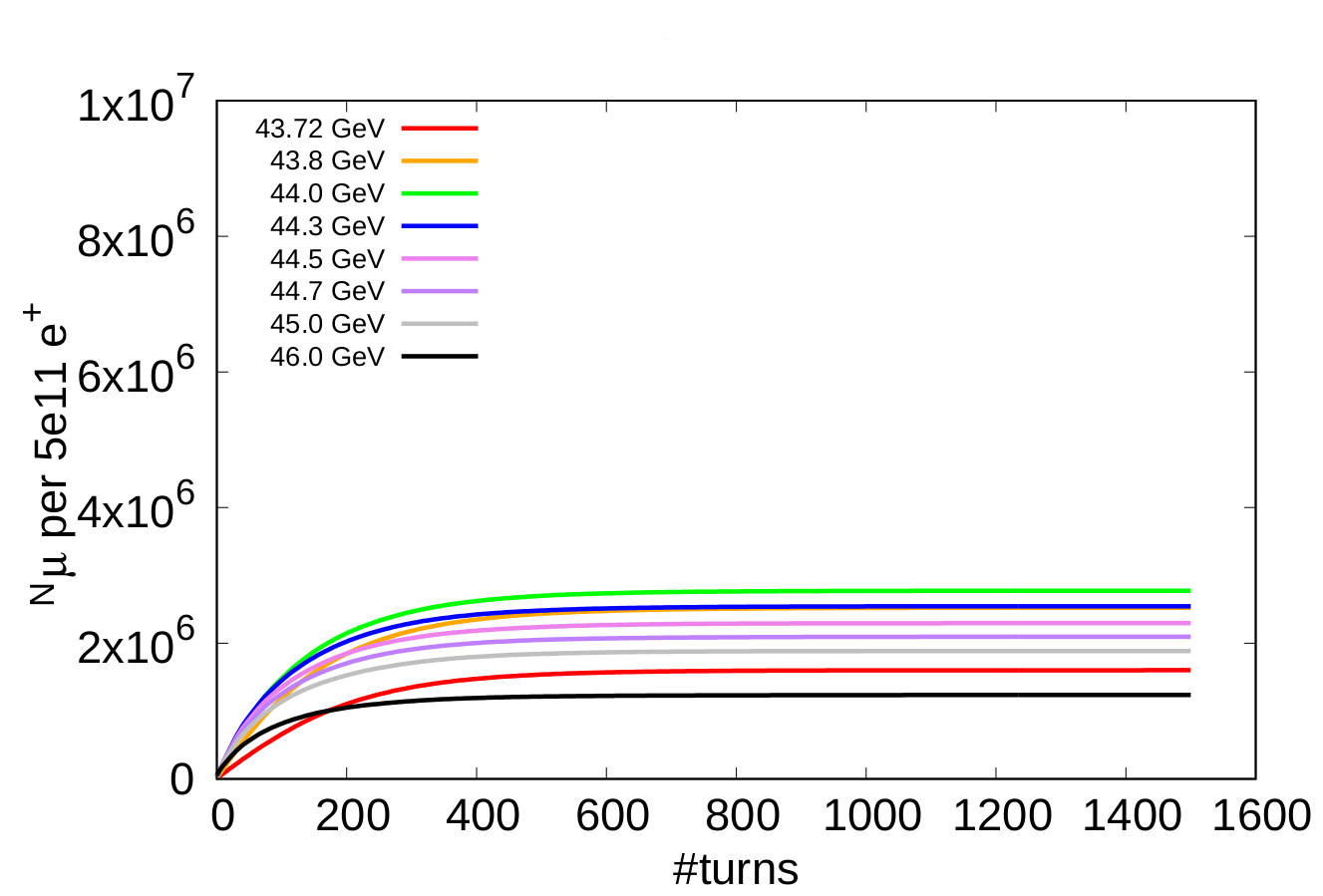}\\
  \includegraphics[width=0.48\textwidth]{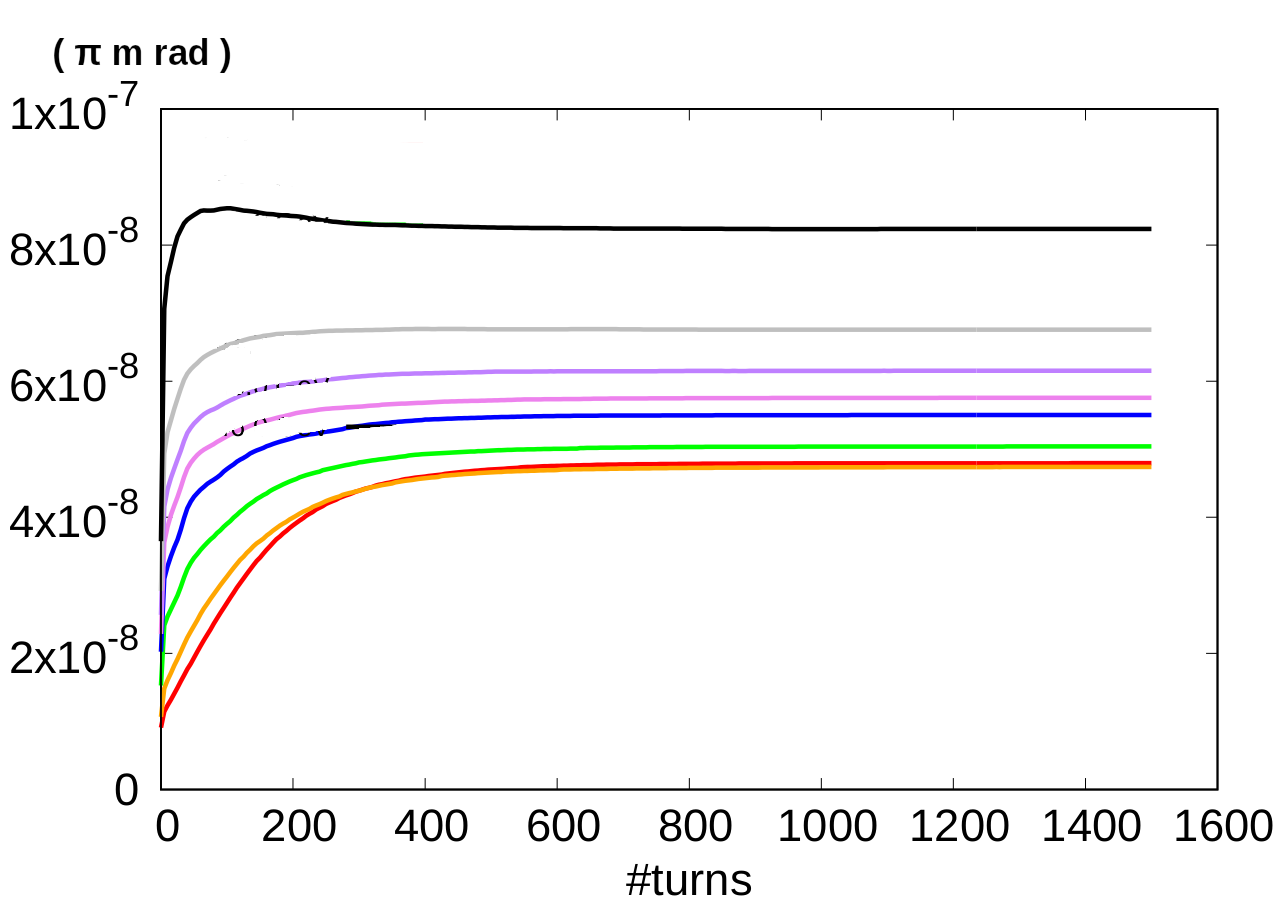}
  \caption{(TOP) Muon beam population as a function of the number of turns for various positron beam energies  on 1\%$X_0$ of Be. (BOTTOM) Transverse emittance of the accumulated beam.}\label{f:otherenergies}
\end{figure}

\begin{table}[th]
    \centering
    \caption{Muon production efficiency from a positron beam at 45~GeV interacting with Beryllium, liquid Hydrogen~(LH2), liquid Lithium~(LLi), a composition of 10\%~liquid Lithium and 90\%~Diamond powder~(LLi-D), Carbon and a Carbon composite~(C/C~A412). Density and length for  1\%X$_0$ are included.}
    \label{tab:Ceff}
    \begin{tabular}{lccc}\hline\hline
        Material & Density & 1\%$X_0$~Length &  $eff$  \\
        & (g cm$^{-3}$)&(mm)   & ($10^{-8}\mu/e^+$)\\\hline
LH2	& 0.071			& 88.8			& 18.9\\
LLi	& 0.534			& 15.5			& 10.9\\
Be	& 1.848			& \;\;3.5			& \;\;8.7\\
LLi-D 	& 3.221			& \;\;1.3		        & \;\;6.3\\
C	& 2.267			& \;\;1.8			& \;\;6.2\\
C/C A412 & 1.7\;\;\;\;  & \;\;2.5 & \;\;6.2 \\
        \hline\hline
    \end{tabular}
\end{table}
Figure~\ref{f:othermaterials} show the result of accumulation with other materials, in particular, we remark the results of using \emph{film} targets, as described in~\cite{PhysRevAccelBeams.23.051001}, that were created to reduce the impact of multiple scattering. Figure.~\ref{f:otherenergies} shows the effect of increasing the beam energy spread above the design specification of 43.8~GeV.\par

\section{Issues and Further design improvements}\label{s:further}
The biggest limit for accumulation in the design is the $\pm$5\% energy acceptance of the Interaction Region. It was already known from the beginning that this magnet configuration would have such performance because it corresponds to the symmetric solution of a first order apochromatic design. However, it seems possible to increase the momentum acceptance by designing a second order apochromatic Interaction Region. This kind of design will require the integration of MAD-X PTC transport maps with custom scripts to calculate the beam optics at several orders in the polynomial expansion. A similar approach is followed in Mapclass2~\cite{mapclass2} to calculate the final beamsize. While, a recent effort to do the symbolic polynomial calculation of the twiss beta functions dependence on energy is available in~\cite{ob1julia}.\par
The momentum compaction factor is in the order of $10^{-4}$ bringing the bunch length to about 2~cm. it will be beneficial to consider a reduction to $10^{-5}$ or even $10^{-6}$ as the energy acceptance grows. Further optimizations considering magnets below 20~T could reduce the bunch length. However, a more interesting idea is the exploration of a cell with more than three magnets and/or a vertical FFA~\cite{brooks} lattice that provides a zero momentum compaction factor at least to first order approximation.\par
On other aspects, we remark that the principle to use multiple IPs could be beneficial as it reduces the requirements of the dipole fields to close the ring, while the distance between IPs could, in principle, remain constant. Of course, this would mean several positron sources available, one per IP.\par
The good field region of a magnet has been assumed to be at least $\pm2$~cm. While this is a very common value for the FCC magnets, it would be beneficial to consider magnets with good field regions larger by a factor 10, i.e. $\pm20$~cm. This will allow to increase the dispersion in the arcs, reducing the sextupole strength and increasing the dynamic aperture.\par
The focal point has a $\beta^*_\mu=20$~cm over $\pm5$\% using quadrupole gradients of about 500~T/m. The beam emittance growth due to multiple scattering would benefit from a $\beta^*_\mu$ reduction, therefore an exploration of designs with higher gradients could be interesting. Previous work for CLIC points to gradients  close to 1~kT/m or even larger~\cite{silverstrov,modena}.\par
With a toy design achieving $\beta^*_\mu=1$~cm over $\pm2$\% energy spread, we have estimated that we will need gradients at least a factor 10 larger, i.e. in the order of 10 to 20~kT/m. This values are available with plasma lenses, e.g. the E-150 experiment~\cite{plasma3,plasma,plasma2} at the SLAC Final Focus Test Beam Facility~(FFTB), reports gradients in the order of $10^6$~T/m for electron a positron beams of $1.5\times10^{10}$ particles per pulse, energy of 28.5~GeV, with a horizontal and vertical beamsize in the order of 10~$\mu$m, with a bunch length of $600$~$\mu$m and normalized emittance of about 50~$\mu$m crossing a gas jet of $N_2$ or $H_2$. These positron and electron beam parameters are similar to the current muon beam parameters studied for LEMMA at the beginning of the accumulation cycle.\par
\section{Conclusions}\label{s:conclusion}
The design of the two identical muon accumulator rings with FFA technology for the LEMMA project has been presented.\par
This 231~m long optics design for the accumulator is composed by high energy acceptance arcs based on Fixed Field Alternating Gradient~(FFA) cells, two Interaction Regions~(IRs) achieving a low $\beta^*_\mu$ of 20~cm over $\pm5$\% energy spread, an extraction and a RF section with low contribution to chromaticity.\par
Due to the energy acceptance and the kinematics of muon production, this design is apt to accumulate muons at 21.9~GeV.\par
The Interaction Region has been conceived to combine and then separate three beams ($\mu^+$ and $\mu^-$ at 21.9~GeV, and $e^+$ at twice the muon beam energy) reducing the amount of synchrotron radiation coming from the positron beam, and focalizing the muon beams at the Interaction Point~(IP) where $\beta^*_\mu=20$~cm. The IP marks the location for a thin target of 1\% of a radiation length of material. The distance between the nearests quads around the IP is 20~cm (2$L^*$).\par
The arc magnets superimpose dipole, quadrupole and sextupole components that, in principle, could be realized with a canted cosine type of magnet.\par
The highest quadrupole gradient of 500~T/m, like in the CLIC Final Focus, is used in the IR to focalize the beams. All other magnets in the design have smaller gradients in the range of 100 to 250~T/m with a good field region of $\pm1$ to $\pm2$~cm and a peak magnetic field below 15~T, similar to the FCC magnet parameters.\par
The RF and the extraction section have been designed to minimize the contribution to the ring chromaticity while giving enough space for a kicker or radiofrequency cavities of few hundred MV.\par
Due to the complexity of the requirements to get a high quality muon beam, Table~\ref{t:designs} summarizes the FFA design exploration and intermediate results on the most relevant optics parameters. It reflects the reduction of the energy acceptance due to the requirement of low $\beta^*_\mu$ at the target location. For completeness, Table~\ref{t:muacc} gives a longer list of parameters of the accumulator design.\par
Simulations of the beam--target interaction with MUFASA show that 100 bunches of $5\times10^{11}$~$e^+$ each at 43.8~GeV impinging on 3.5~mm of Beryllium produce a final population of about $10^6$ muons at 21.9~GeV with a normalized transverse emittance of 10~$\pi$~$\mu$m and longitudinal emittance of 10~$\pi$~GeV~mm.\par
The number of accumulation cycles is limited because multiple scattering with the target increases the muon beams emittance leading to particles losses when they populate the dynamic aperture of the machine. The initial normalized transverse beam emittance is 5~$\pi$~$\mu$m and grows due to multiple scattering with the target in few hundreds of turns to the final 10~$\pi$~$\mu$m. In order to further mitigate this effect we will need to produce a smaller $\beta^*_\mu$ using quadrupole gradients above 500~T/m.\par
The muon population is small because the $\pm5$\% energy acceptance of the IR limits the muon production efficiency (in the order of $10^{-8}$ muon pairs per positron), therefore, the accumulation studies point to the need of a redesign of the Interaction Region for a larger energy acceptance. One of the possibilities is to create a second order apochromatic IR that could accept an energy spread of $\pm10$\%.\par
Alternatively, one could consider different materials for the target. We show the results of accumulation for few interesting ones, from where we would like to highlight a \emph{film} of liquid Hydrogen~(LH$_2$) that allows to increase the production efficiency and reduces the multiple scattering with the target reaching a population of $10^7$ muons in a thousand turns.\par
The arc magnets could also be reoptimized for larger dynamic aperture given that good field regions of more than $\pm2$~cm in superconducting canted cosine magnets are possible.\par
In addition, the accumulation opens up the possibility to quantitatively study the multiparticle interaction of three beams at two different energies, one is the high intensity low emittance positron beam and the other two are a growing population of muons.\par

\section{Acknowledgments}
This work has been financially supported by the Istituto Nazionale di Fisica Nucleare~(INFN), Italy, Commissione Scientifica Nazionale~5, Ricerca Tecnologica -- Bando n.~20069.\par
The authors would like to thank Manuela Boscolo, Mario Antonelli, Susanna Guiducci, Alessandro Variola, Marica Biagini and Francesco Collamati from INFN, and Pantaleo Raimondi and Simone Liuzzo from ESRF for useful discussions on the muon beam production.\par

\newpage
\bibliographystyle{unsrt}

\newpage

\newpage
{\renewcommand{\arraystretch}{0.5}
  \begin{table*}[htp]
    \caption{Lattice design summary. The \emph{FFA cell} and \emph{Low $\alpha_c$} designs refer to rings with high energy acceptance and without IP. The \emph{Zero $\eta_x$} design refers to the ring with 4 point where the dispersion is cancelled, it is a first approach to the inclusion of the insertions in the ring. The \emph{+IR} and \emph{1+1* IPs} designs refer to the inclusion of all ring sections considering 1 or 2 IPs.}\label{t:designs}
  \begin{tabular}{lcc|cccc|c}\hline\hline
    Parameter & Unit & Requirement & FFA cell&Low $\alpha_C$&+zero $\eta_x$ & +IR & 1+1* IPs\\\hline
    Energy Acceptance & \% & $\pm$20 & $>\pm20$ & $\pm10$ & $\pm9$ & $\pm5$ &  $\pm5$\\
    IPs & & 1 & -- & -- & 1 & 1 & 2\\
    Circumference/IP & m & 60 & 98~& 98 & 98 & 230 & 230/2\\
    $\beta^*_\mu$ & m & 0.01 & -- & 2.2 & 2.2 & 0.2 & 0.2\\
    Bunch length & mm & 3 & $\sim10^3$ & $\sim$150 & 100 & 20 & 20\\\hline
  \end{tabular}
  \end{table*}
}

{\renewcommand{\arraystretch}{0.5}
  \begin{table*}[htp]
    \caption{Accumulator Ring Parameters.}\label{t:muacc}
    \begin{tabular}{lccc}\hline\hline\\
      Parameter & Unit & Requirement &  FFA design\\\hline
      Energy & GeV & 22.5 & 21.9\\
      Relativistic Gamma Factor & -- & 212.95 & 207.272\\
      Length & m & 60 & 230\\
      Revolution Frequency & MHz & 5 & 1.30275 \\
      Revolution Time & $\mu$s & 0.2 & 0.7676 \\
      Energy Loss per Turn & MeV & -- & $3\times10^{-6}$~(S.R.), $\sim10$~(thin target)\\
      Energy Acceptance & \% & $\pm$20 & $\pm$5\\
      Number of Bunches & -- & 1 & 1\\
      Bunch Population & -- & $10^9$ & $2\times10^6$ (Be), $10^7$ (LH$_2$ film,  $E_{e^+}=44$~GeV) \\
      Normalized Transverse Emittance & $\pi$~$\mu$m & 0.04 & 5~(at production), 10~(end of accumulation)\\
      Geometrical Transverse Emittance & $\pi$~nm & 0.2 & 25~(at production), 50~(end of accumulation)\\
      Longitudinal Emittance & $\pi$~GeV~mm & -- & 10~(end of accumulation)\\
      Number of IPs & -- & 1 & 1+1* \\
      Cycles of accumulation & -- & 1000 & $<400$~(Be), $<1000$~(LH$_2$ film) \\
      Nat. Chrom. x/y & -- & -- & -26.8 / -29.4\\
      Qx/Qy/Qs & -- & -- & 25.6306 / 10.1525 / 0.0137\\
      $\beta^*_\mu$ at the IP (target location) & cm & 1 & 20\\
      Distance from IP to first magnet, $L^*$ & cm & -- & 10\\
      $\alpha_C$ & -- & very small & $3\times10^{-4}$\\
      Bunch length & mm & 3 & 20\\
      Straight Sections & -- & -- & 4 (2 IPs, RF, extraction)\\\hline
      Arc Magnets Peak Field & T & -- & 15\\
      Type of magnets & -- & -- & Canted Cosine Theta\\
      Highest Dipole Component & T & -- & 12\\
      Highest Quadrupole Component & T/m & -- & 240\\
      Highest Sextupole Component & T/m$^2$ & -- & 11000\\
      Arc Magnet Good Field Region & cm & -- & $\pm$2\\\hline
      IR Magnets Inner Triplet Peak Field & T & -- & 2\\
      Type of magnets & -- & -- & Dedicated function\\
      Highest Quadrupole Gradient & T/m & -- & 500\\
      Aperture Radius & mm & -- & 4\\\hline
      IR Magnets Second Triplet Peak Field & T & -- & 5\\
      Type of magnets & -- & -- & Dedicated function/superconductor\\
      Highest Quadrupole Gradient & T/m & -- & 200\\\hline
      Insertion Magnets Peak Field & T & -- & 5\\
      Type of magnets & -- & -- & Dedicated function/superconductor\\
      Highest Quadrupole Gradient & T/m & -- & 200\\\hline
      RF Voltage & MV & -- & 150\\
      RF Harmonic & -- & -- & 600\\
      RF Frequency & MHz & -- & 782\\
      RF Drift Space & m & -- & 4\\\hline
      Extraction Kicker Drift Space & m & -- & 4\\\hline
    \end{tabular}
  \end{table*}
}
\newpage
\par

\newpage

\end{document}